\documentclass{cernyrep}

\usepackage{color}
\newcommand{\aq}[1]{}
\newcommand{\tq}[1]{}
\newcommand{\ced}[1]{#1}
\def\rmi{\mathrm{i}} 
\def\rmd{\mathrm{d}} 
\def\rme{\mathrm{e}} 
\def\rmp{\mathrm{p}} 
\def\rmn{\mathrm{n}} 
\def\qku{\mathrm{u}} 
\def\qkd{\mathrm{d}} 
\def\qkc{\mathrm{c}} 
\def\qks{\mathrm{s}} 
\def\qkt{\mathrm{t}} 
\def\qkb{\mathrm{b}} 
\def\th13{\theta_{13}}
\def\nutau{\nu_\tau}
\def\anutau{\bar{\nu}_\tau}
\def\numu{\nu_\mu}
\def\anumu{\bar{\nu}_\mu}
\def\nue{\nu_\rme}
\def\anue{\bar{\nu}_\rme}
\def\nux{\nu_x}
\def\JPCS{\textit{J. Phys. Conf. Ser.}}

\def\PRL{\textit{Phys. Rev. Lett.}}

\providecommand{\Rule}[2][0mm]{\rule[#1]{0mm}{#2}} 

\begin{document}

\title{Neutrino Physics}

\author{I.~Gil-Botella}

\institute{Centro de Investigaciones Energ\'eticas, Medioambientales y
  Tecnol\'ogicas, Madrid, Spain}

\maketitle

\begin{abstract}
The fundamental properties of neutrinos are reviewed in these
lectures. The first part is focused on the basic characteristics of
neutrinos in the Standard Model and how neutrinos are
detected. Neutrino masses and oscillations are introduced and a summary
of the most important experimental results on neutrino oscillations
to date is provided. Then, present and future experimental proposals are
discussed, including new precision reactor and accelerator
experiments. Finally, different approaches for measuring the neutrino
mass and the nature (Majorana or Dirac) of neutrinos are
reviewed. The detection of neutrinos from supernovae explosions and
the information that this measurement can provide are also summarized
at the end.

\end{abstract}

\section{Introduction}

The last 20 years have been a revolution for neutrino physics. The
observation of neutrino oscillations has established that neutrinos
have masses and this implies physics beyond the Standard Model. This
fact has a clear impact not only on particle physics, but also on astroparticle
physics and cosmology.

Nevertheless, neutrinos are still quite unknown particles. At the
moment we know that there are three light neutrinos, although some
theoretical models propose the existence of sterile neutrinos (not
interacting weakly with matter). Neutrinos are much lighter than their
charged leptonic partners and they interact very weakly with
matter. In addition, during the last 13 years, neutrino experiments
have proved that neutrinos have mass, contrary to the zero-neutrino-mass
hypothesis of the Standard Model. Neutrinos oscillate when they
propagate through space. During the past few years the solar neutrino
problem has been solved and the solar and atmospheric oscillation
parameters have been confirmed using artificial sources. A period of
precision measurements in neutrino physics has started.

However, many fundamental questions still remain unanswered: What is the value of the neutrino masses? Are neutrinos Majorana or Dirac particles? What is the mass hierarchy? What is the value of the neutrino oscillation parameters, in particular,
$\theta_{13}$ and $\theta_{23}$ (if this is maximal or not)? Is there
CP-violation in the leptonic sector? Are there more than three
neutrinos? What is the relation with the quark sector? Can neutrinos
be related to leptogenesis? Why are neutrinos much lighter than other
fermions?~\ldots\ In summary, there are many aspects of neutrinos still
unknown.\aq{I've rewritten these in the form of actual short questions.}

The history of neutrinos began with the investigation of beta
decays. During the early decades of the past century, radioactivity was
explored and the nuclear beta decay was observed. In this process, a
radioactive nucleus emits an electron and increases its positive
charge by one unit to become the nucleus of another element. The beta
decay was studied and, because of the energy conservation, the electron
should always carry away the same amount of energy. A line in the
energy spectrum was expected.

However, in 1914, Chadwick showed that electrons follow a continuous
spectrum of energies up to the expected value. Some of the energy
released in the decay appeared to be lost. To explain this observation,
only two solutions seemed possible: either energy is not conserved
(preference of Bohr) or an additional undetectable particle carrying
away the additional energy was emitted (preference of Pauli). To solve
the energy crisis, in 1930 Pauli wrote his famous letter explaining
that he invented a desperate remedy to save the energy conservation
law. There could exist in the nuclei electrically neutral particles
that were emitted in beta decays and able to cross the detectors
without leaving any trace. These particles (which he wished to call
neutrons) were carrying all the missing energy. These particles have
spin 1/2 and obey the exclusion principle. The mass should be the same
order of magnitude as the electron mass.

Later on, in 1932, Chadwick discovered the neutron; and, in 1934, Fermi
took Pauli's idea and on its basis developed a theory of beta
decay. Fermi named this particle ``neutrino''. The weak force is so weak
that the probability of inverse beta decay was calculated to be close
to zero. The possibility to detect a neutrino seemed null.

However, the development of very intense sources of neutrinos (fission
bombs and fission reactors) changed the prospect. In 1951, Reines
thought about using an intense burst of antineutrinos from bombs in an
experiment designed to detect them. At the end, they decided to use
fission reactors as sources, in particular the Hanford reactor. In
collaboration with Cowan at the Los Alamos Scientific Laboratory, they
began the ``Poltergeist Project''. They chose the inverse beta decay on
protons to detect the free neutrino. The detection principle was a
coincident measurement of the 511~keV photons associated with positron
annihilation and a neutron capture reaction a few microseconds
later. The idea was to build a large detector filled with liquid
scintillator loaded with Cd to increase the probability of capturing a
neutron. The process releases 9~MeV gammas a few microseconds later
than the positron detection. This delayed coincidence provides a
powerful means to discriminate the signature of the inverse beta decay
from background noise. The 300~litre neutrino detector was read by 90
two-inch photomultipliers (PMTs) and surrounded by homemade
boron--paraffin shielding intermixed with lead to stop reactor neutrons
and gamma rays from entering the detector and producing unwanted
background. The expected rate for delayed coincidences from
neutrino-induced events was 0.1--0.3 counts per minute. However, the
delayed-coincidence background, present whether or not the reactor was
on, was about 5 counts per minute, many times higher than the expected
signal rate. The background was due to cosmic rays entering the
detector. The small increase observed when the reactor was on was not
sufficient. The results of this first experiment were not conclusive.

Nevertheless, after the unsuccessful trial, they redesigned the
experiment to better distinguish events induced by cosmic rays and
those initiated by reactor neutrinos. Two large flat plastic target
tanks were filled with water. The protons in the water provided the
target for inverse beta decay. Cadmium chloride dissolved in the water
provided the Cd nuclei that would capture the neutrons. The target
tanks were sandwiched between three large scintillator detectors
having 110 PMTs to collect scintillation light and produce electronic
signals. With this detector, neutrinos were detected for the first
time in 1956 by Reines and Cowan using the nuclear reactor neutrinos
from the Savannah River Plant in South Carolina~\cite{reines}. Several
tests confirmed that the signal was due to reactor antineutrinos. The
experiment was also able to provide a measurement of the cross-section
for inverse beta decay. This detection was rewarded with the Nobel
Prize in 1995.

Other important historical facts related to neutrinos were the
detection of muon neutrinos in 1962, the detection of solar neutrinos
by Davis in 1970, the discovery of neutral current neutrino
interactions in 1973 with a bubble chamber experiment in a $\numu$ beam
at CERN, the detection of neutrinos from a supernova type-II explosion in
1987 with large underground neutrino detectors, and the determination
at \ced{the Large Electron--Positron Collider (LEP)} of three light neutrinos by measuring the total decay width of the Z resonance.

\section{Neutrinos in the Standard Model}

In the Standard Model (SM) of particle physics, fermions come in three
families. Among them, neutrinos are the less known particles. We know that
they have zero electric charge and they only interact via weak
interactions.

The SM is based on the gauge group $\mbox{SU}(3)_C \times \mbox{SU}(2)_L$
$\times$ \mbox{U}(1)$_Y$ that is spontaneously broken to the subgroup
$\mbox{SU}(3)_C \times \mbox{U}(1)_{EM}$.
All the fermions of the SM are representations of this group with
the quantum numbers indicated in Table~\ref{tab:reps}, where the
family structure is shown. Neutrinos are the partners of the charged
leptons. They form left-handed weak isospin doublets under the \mbox{SU}(2)
gauge symmetry. In the SM, neutrinos are strictly massless. They do
not carry electromagnetic or colour charge but only the weak
charge. They are extremely weakly interacting.

\begin{table}[ht]
\caption[]{Fermionic representations in the Standard Model.}
\label{tab:reps}
\[
\begin{array}{@{}cc|ccc@{}}
\hline\hline
\Rule[-1em]{2.5em}
L_L(1,2,-\frac{1}{2})
& Q_L(3,2,\frac{1}{6})~~
& ~~E_R(1,1,-1)
& U_R(3,1,\frac{2}{3})
& D_R(3,1,-\frac{1}{3}) \\\hline
\Rule[-2em]{4em}
\begin{pmatrix}\nu_\rme  \\ \rme \end{pmatrix}_{L}
 & \begin{pmatrix} \qku \\ \qkd \end{pmatrix}_{L}
  & \rme_R
   & \qku_R
    & \qkd_R                                        \\
\Rule[-2em]{4em}
\begin{pmatrix}\nu_\mu \\ \mu\end{pmatrix}_{L}
 & \begin{pmatrix} \qkc  \\ \qks \end{pmatrix}_{L}
  & \mu_R
   & \qkc_R
    & \qks_R                                        \\
\Rule[-2em]{4em}
\begin{pmatrix} \nu_\tau \\ \tau\end{pmatrix}_{L}
 & \begin{pmatrix} \qkt \\ \qkb \end{pmatrix}_{L}
  & \tau_R
   & \qkt_R
    & \qkb_R                                        \\ \hline
\end{array}
\]
\end{table}

Under \ced{charge, parity and time reversal symmetry (CPT)}\aq{Given CPT in full on first occurrence. OK?} conservation, for any left-handed fermion there exists a
right-handed antiparticle with opposite charge. But the right-handed
particle state may not exist. This is precisely what happens with
neutrinos in the SM. Since, when the SM was postulated, neutrino masses
were compatible with zero, neutrinos were postulated to be Weyl
fermions: the left-handed particle was the neutrino and the
right-handed antiparticle was the antineutrino.

A neutrino of a flavour $l$ is defined by the charged-current (CC)
interaction with the corresponding charged lepton $l$. For example, the
muon neutrino always comes with the charged muon. The CC
interactions between neutrinos and their corresponding charged leptons
are given by
\begin{equation}
-\mathcal{L}_{\rm CC} =\frac{g}{\sqrt{2}}\sum_l{\bar{\nu}_{Ll}\gamma}^{\mu}l_{\bar{L}}W^+_{\mu}
+ {\rm h.c.}
\label{eq:a1}
\end{equation}
The SM neutrinos also have neutral-current (NC) interactions, as
indicated in
\begin{equation}
-\mathcal{L}_{\rm NC} =\frac{g}{2\cos\theta_W}\sum_l{\bar{\nu}_{Ll}\gamma}^{\mu}\nu_{Ll}
Z^0_{\mu}.
\label{eq:a2}
\end{equation}
From this equation, one can determine the decay width of the Z$^0$ boson
into neutrinos, which is proportional to the number of light
left-handed neutrinos.

We know thanks to neutrinos that there are exactly three families in
the SM. An extra SM family with quarks and charged leptons so heavy
that they remain unobserved would also have massless neutrinos that
would have been produced in Z decay, modifying its width, which has
been measured at LEP with impressive precision. The combined result
from the four LEP experiments is $N_{\nu} = 2.984 \pm 0.008$~\cite{PDG}.

The SM presents an accidental global symmetry. This is a consequence of
the gauge symmetry and the representations of the physical states. The
total lepton number given by $L = L_\rme + L_\mu + L_\tau$ is conserved.

Other neutrino properties summarized in the Particle Data
Book~\cite{PDG} are upper limits on neutrino masses, on neutrino decay
processes and on the neutrino magnetic moment.

\section{Neutrino interactions and detection}

Neutrinos are produced copiously in natural sources: in the burning of
stars, in the interaction of cosmic rays, in the Earth's radioactivity,
in supernova explosions and even as relics of the Big Bang. In the
laboratory, neutrinos are produced in nuclear reactors and particle
accelerators.

The neutrino energies expand through a huge range: from $10^3$~eV to
$10^{15}$~eV. In the low-energy range there are neutrinos from double-beta decay, geoneutrinos, nuclear reactors, supernovas and the
Sun. Artificial neutrinos from particle accelerators, beta beams or
neutrino factories have energies in the medium range. Atmospheric
neutrinos extend from medium to high energies, while neutrinos coming
from extragalactic sources can reach very high energies.

\begin{figure}[ht]
\begin{center}
\includegraphics[width=8cm]{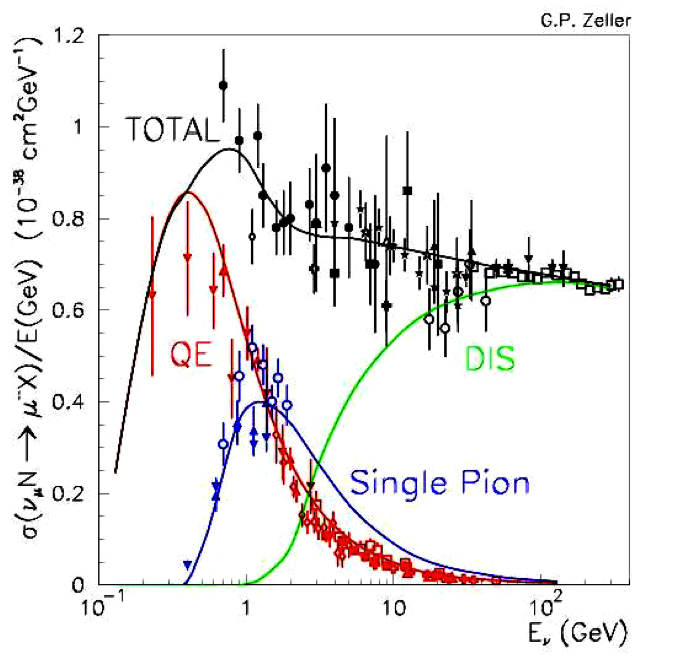}
\caption{Neutrino interaction cross-sections.}
\label{fig:nu_int}
\end{center}
\end{figure}

Another important ingredient for neutrino detection is the neutrino
interaction cross-section. Neutrino cross-sections are not equally
well known in the whole range (Fig.~\ref{fig:nu_int}). For neutrino
energies lower than 100~MeV, cross-sections are well known because the
interaction processes are dominated by the inverse beta decay, elastic
scattering, and CC and NC interactions with nuclei. These interactions
are theoretically better known than determined in experiments. For
neutrino energies above 100~GeV, up to $10^7$~GeV (ultrahigh
energies), they are also accurately known. However, in the
intermediate range, critical for atmospheric and accelerator
experiments with neutrino energies around 1~GeV, cross-sections are
poorly known (with uncertainties of 20--40\%) due to the
complexity of the processes like quasi-elastic (QE) scattering, single pion
production, deep inelastic scattering (DIS), and nuclear effects, form
factors, etc.

MINER$\nu$A~\cite{minerva} (Main Injector Experiment for $\nu$-A) is a detector designed to precisely study neutrino--nucleus interactions in the 1--10~GeV range
in the NuMI high-intensity neutrino beam at Fermilab. This experiment
will improve our knowledge of neutrino cross-sections at low energy
and study the $A$ dependence in neutrino interactions. These data will
be important to reduce the systematic errors in long-baseline neutrino
oscillation experiments.

They will study four main reaction channels: QE, resonance production,
deep inelastic scattering, and coherent neutrino--nucleus reactions (CC
and NC coherent single pion production). The MINER$\nu$A detector is a
fine-grained tracking calorimeter with a fully active
solid-scintillator tracker. The active detector are solid-scintillator
strips of triangular cross-section providing a spatial resolution of
2.5~mm. The scintillation light due to a charged particle is collected
by a wavelength-shifting optical fibre located at the centre of each
strip and routed to PMTs. The detectors are hexagonal modules
containing one or two active planes. After the tracker region there is
the electromagnetic calorimeter (ECAL) and the hadronic calorimeter
(HCAL) to contain forward-going particles. Both calorimeters also
surround the inner detector to contain particles with high transverse
momentum. At the back of the detector they use the MINOS Near detector
as a muon spectrometer to measure the energy and charge of
muons. Before the tracker there is an area of the nuclear targets
(liquid He, carbon, iron, lead and water) interleaved with tracking
planes. Just before, there is a veto wall and a steel shield.

The MINER$\nu$A collaboration built a first prototype of 24 full-size modules that was first commissioned with cosmic rays in 2008--2009 and then moved underground
into the NuMI beam upstream of the MINOS Near detector with an iron
target prototype and a veto wall. It \ced{began operating} in summer 2009. The
complete detector was finished in March 2010. Modules of four
types (120 in total) -- nuclear target, tracker, ECAL and HCAL -- were built with a total mass of $\sim$~200~ton. MINER$\nu$A has taken data in the low-energy (peak at
3~GeV) antineutrino beam since November 2009 with 55\% of the full
detector. In March 2010 they took data in low-energy neutrino mode
until September 2010. \ced{After November 2010 they took antineutrino data,}
turning again to low-energy neutrino mode in spring 2011. In summer 2012
Fermilab will switch to the medium-energy beam (peak at 6~GeV) for
NO$\nu$A, and MINER$\nu$A will continue to take data.

Different technologies have been used by past and present neutrino
detectors. {\it Radiochemical techniques} were used by the first solar neutrino
experiments like Homestake, SAGE and GALLEX. They use the interaction
of neutrinos with Cl or Ga isotopes, producing Ar or Ge, and developed
methods to extract these isotopes using different solutions. They were
not real-time detectors. At present, one of the most common
technologies exploited is that used by {\it Cerenkov detectors} like
Super-Kamiokande, SNO, MiniBooNE, Antares, IceCube, etc. They detect
the Cerenkov light of the charged leptons produced by neutrinos using
PMTs. The pattern of the detected rings allows electrons to be distinguished
from muons. This is the best technique for low rates and
low-multiplicity events with energies below 1~GeV and also very high
energies. A different technique used by some neutrino accelerator
experiments like MINOS, MINER$\nu$A and NO$\nu$A is {\it tracking
  calorimetry}. They use alternating planes of absorber material (such as
lead) with detector planes for tracking (essentially liquid or plastic
scintillators read by PMTs). This is appropriate for high-rate and
high-multiplicity events with energies around 1~GeV.
Another type of detectors are the {\it unsegmented scintillator
calorimeters} like KamLAND, Borexino and \ced{Double Chooz}\aq{It appears that usually you refer to the CHOOZ (all capitals) experiment and the Double Chooz (initial capitals) experiment. This seems also to be the case with a Google search, so I've left this}. They provide large light yields at MeV energies. This is very convenient for the
detection of reactor and solar neutrinos.
\ced{\textit{Liquid argon time projection chambers} (LAr TPCs)}\aq{Given in full on first occurrence. OK?} like ICARUS have high granularity and are
potentially good for large masses.
 Finally we have the {\it emulsion technique}, which is in fashion again
with the OPERA experiment. This is the only technique providing the micrometre-level
spatial resolution needed, for example, to detect tau neutrinos.

\section{Massive neutrinos}

As already mentioned, there are only upper limits to neutrino masses. The direct limits come
from the precise measurement of the endpoint of the lepton energy
spectrum in weak decays, which gets modified if neutrinos are
massive.

The SM predicts that neutrinos are precisely massless. In
order to add a mass to the neutrino, the SM has to be extended. The SM
gauge invariance does not imply lepton number symmetry. Total lepton
number can or cannot be a symmetry depending on the neutrino nature.

Neutrino masses can be easily accommodated in the SM. A massive
fermion necessarily has two states of helicity. The mass is the
strength of the coupling between the two helicity states. To introduce
such a coupling in the SM for the neutrinos, we need to identify the
neutrino right-handed states, which in the SM are absent. There are
two ways to proceed:
\begin{enumerate}
\item
We introduce a right-handed neutrino coupled to the matter just
through the neutrino masses and impose lepton number conservation
({\it Dirac neutrinos})
\begin{equation}
\mathcal{L}=\mathcal{L}_{\rm SM} - M_{\nu}\overline{\nu_R}\nu_L + {\rm h.c.}
\label{eq:a3}
\end{equation}
\item
We do not impose lepton number conservation and we identify the
right-handed state with the antiparticle of the left-handed state
({\it Majorana neutrinos})
\begin{equation}
\mathcal{L}=\mathcal{L}_{\rm SM} - \tfrac{1}{2} M_{\nu}\overline{\nu_{L}^C}\nu_L + {\rm h.c.}
\label{eq:a4}
\end{equation}
\end{enumerate}

In the first case, we enlarge the SM by adding a set of three
right-handed neutrino states, which would be singlets under $\mbox{SU}(3)
\times \mbox{SU}(2) \times U_Y(1)$, but coupled to matter just through
the neutrino masses. This coupling has to be of the Yukawa type to
preserve the gauge symmetry. Masses are proportional to the vacuum
expectation value of the Higgs field, like for the remaining
fermions. One important consequence of this is a new hierarchical
problem: Why are neutrinos much lighter than the remaining leptons?

In the second case, Majorana identified the right-handed state with
the antiparticle of the left-handed state. $C$ is the operator of charge
conjugation in spinor space that connects particle and
anti\-particles:
\begin{eqnarray}
\nu_R \rightarrow (\nu_L)^c = C \bar{\nu}^T_L = C \gamma_0 \nu_L^* .
\end{eqnarray}
The Majorana neutrino masses are of the form
\begin{eqnarray}
m_\nu = \alpha_\nu \frac{v^2}{\Lambda}.
\end{eqnarray}
If $\Lambda$ is much higher than the electroweak scale $v$, a strong
hierarchy between neutrino and charged lepton masses arises
naturally. A Majorana mass violates the conservation of all charges
carried by the fermion, including global charges as lepton number.

The simplest example to explain the origin of the scale $\Lambda$ in
the Majorana masses is the famous {\it see-saw mechanism}
\cite{see-saw}.
In this case, the scale of the mass eigenvalues is much higher than
the scale of electroweak symmetry breaking ($\Lambda \gg v$). The
Majorana effective interaction results from the interchange of very
heavy right-handed Majorana neutrinos. The new physics scale is simply
related to the masses of the heavy Majorana neutrinos and the Yukawa
couplings.


Neutrino masses imply neutrino mixing, as happens in the quark
sector. Majorana and Dirac possibilities differ in the number of
observable phases. The real physical parameters are the mass eigenstates
and the mixing angles, while the imaginary parameters are CP-violating
phases. In the case of three families, there are three mixing angles
and one phase for the Dirac case or three phases for the Majorana
case.

\section{Neutrino oscillations in vacuum and matter}

If neutrinos have masses and mix, there can be neutrino flavour
change. Oscillations appear because of the misalignment between the
neutrino interaction eigenstates and the propagation mass
eigenstates.

The neutrino flavour eigenstates, $\nu_\alpha$, produced in a
weak interaction are linear combinations of the mass
eigenstates $\nu_j$:
\begin{eqnarray}
|\nu_\alpha \rangle = \sum_{j=1}^n  U_{\alpha j}^* |\nu_j \rangle ,
\end{eqnarray}
where $n$ is the number of light neutrino species and $U$ is the
\ced{Pontecorvo--Maki--Nakagawa--Sakata (PMNS)}\aq{Given in full. OK?} mixing matrix.

A standard parametrization of the mixing matrix is given by
\begin{equation}
U_\text{PMNS} =
\begin{pmatrix}
1 & 0 & 0           \\
0 & c_{23} & s_{23} \\
0 & -s_{23} & c_{23}
\end{pmatrix}
\begin{pmatrix}
c_{13} & 0 & s_{13} \rme^{\rmi  \delta} \\
0 & 1 & 0           \\
-s_{13} \rme^{\rmi  \delta}& 0  & c_{13}
\end{pmatrix}
\begin{pmatrix}
c_{12} & s_{12} & 0  \\
-s_{12} & c_{12}  & 0 \\\
0 & 0 & 1
\end{pmatrix}
\begin{pmatrix}
  1 & 0 & 0             \\
  0 & \rme^{\rmi  \alpha_1} & 0\\
0 & 0 & \rme^{\rmi  \alpha_2}
\end{pmatrix} ,
\label{mns}
\end{equation}
where $c_{ij}=\cos\theta_{ij}$ and $s_{ij}=\sin\theta_{ij}$,
with $\theta_{12}$, $\theta_{13}$ and $\theta_{23}$ the three mixing
angles, $\delta$ is the Dirac CP-violating phase, and $\alpha_1$ and
$\alpha_2$ are the Majorana phases, not accessible by oscillation
experiments.

After travelling a distance $L$ (or, equivalently for
relativistic neutrinos, time $t$), a neutrino originally produced with a
flavour $\alpha$ evolves as follows:
\begin{eqnarray}
|\nu_\alpha(t) \rangle = \sum_{j=1}^n  U_{\alpha j}^* |\nu_j(t)
\rangle .
\end{eqnarray}

Using the standard approximation that the neutrino state is a plane
wave $|\nu_j(t) \rangle = \rme^{-\rmi  E_j t} |\nu_j(0) \rangle$, that neutrinos
are relativistic with
\begin{eqnarray}
E_j = \sqrt{p_j^2 + m_j^2} \approx p + \frac{m_j^2}{2E}
\end{eqnarray}
and the orthogonality relation $\langle \nu_i(0)|\nu_j(0)
\rangle = \delta_{ij}$, the transition probability between $\nu_\alpha$
and $\nu_\beta$ is
\begin{eqnarray}
P(\nu_\alpha{\rightarrow}\nu_\beta) = |\langle \nu_\beta
|\nu_\alpha(t)\rangle|^2 &=& \left| \sum_{j=1}^n \sum_{k=1}^n U_{\alpha
    j}^* U_{\beta k} \langle \nu_k |\nu_j(t)\rangle \right|^2 \nonumber \\
&\approx& \sum_{j,k} U^*_{\alpha j} U_{\beta j} U_{\alpha k} U^*_{\beta k} \, \rme^{{-\rmi} {\Delta m^2_{jk} L}/{2E}} ,
\end{eqnarray}
with $\Delta m^2_{jk} = m_j^2 - m_k^2$. The probability for
flavour transition is a periodic function of the distance between the
source and the detector.

Dominant oscillations are well described by effective two-flavour
oscillations. The three-flavour oscillation neutrino effects are
suppressed because of the small value of $\theta_{13}$ and the
hierarchy between the two mass splittings,
$\Delta m^2_{21} \ll \Delta m^2_{32}$. In most cases
the problem can be reduced to two-flavour oscillations.

In the simplest case of two-family mixing, the mixing matrix depends
on just one mixing angle and there is only one mass square
difference. The probability that a neutrino $\nu_\alpha$ of
energy $E_\nu$ oscillates into a neutrino $\nu_\beta$ after travelling
a distance $L$ is given by
\begin{eqnarray}
P(\nu_\alpha{\rightarrow}\nu_\beta) =
\sin^2 2\theta \sin^2\!\left(\frac{\Delta m^2 L}{4 E_\nu}\right), \qquad\quad
\alpha\neq\beta .
\label{eq:wk}
\end{eqnarray}
The probability is the same for neutrinos and antineutrinos, since
there are no imaginary entries in the mixing matrix.

The transition probability has an oscillatory behaviour with a period
determined by the oscillation length ($L_{\rm osc}$), which is
proportional to the neutrino energy and inversely proportional to the
neutrino mass square difference, and an amplitude proportional to the
mixing angle. Hence the name ``neutrino oscillations'':
\begin{eqnarray}
L_{\rm osc} = \frac{4 \pi E_\nu}{\Delta m^2}.
\end{eqnarray}
If $L \gg L_{\rm osc}$, the oscillating phase goes through many cycles
before detection and is averaged to 1/2.

Experimentally, the free parameters are the source--detector distance and the
neutrino energy. In order to be sensitive to a given value of $\Delta m^2$, the
experiment has to be set up with $E/L \approx \Delta m^2$. For
example, to measure $\theta_{23}$ and $\Delta m^2_{32}$ parameters,
one should look for an $L/E$ of around 500~km/GeV (which is the case for
atmospheric neutrinos). To measure $\theta_{12}$ and
$\Delta m^2_{21}$ parameters $L/E$ should be around 15\,000~km/GeV (solar
neutrinos case).


In the most general case of three neutrino families, the oscillation
probability can be rewritten in one term conserving CP and another
term violating CP, as follows:
\begin{eqnarray}
P(\nu_\alpha{\rightarrow}\nu_\beta) = \delta_{\alpha\beta}
-4  \sum_{i<j}^n {\rm{Re}}[J^{\alpha\beta}_{ij}]
\sin^2\!\left(\frac{\Delta m^2_{ij} L}{4 E_\nu}\right)
\pm 2  \sum_{i<j}^n {\rm{Im}}[J^{\alpha\beta}_{ij}]
\sin\!\left(\frac{\Delta m^2_{ij}L}{2 E_\nu}\right),
\label{eq:prob}
\end{eqnarray}
with $J^{\alpha\beta}_{ij} \equiv U_{\alpha i} U_{\beta
i}^* U_{\alpha j}^* U_{\beta j}$. The two terms have opposite sign
for neutrinos and antineutrinos. By comparing neutrino and
antineutrino oscillation probabilities, we could test the violation of
CP.


From the experimental point of view, to measure neutrino oscillations,
we need to compute or to measure the flavour composition and the flux
and energy spectra of the produced neutrinos (near data) and also the interaction
cross-section at their energies. After propagation of neutrinos through a
distance $L$, we need to measure the flavour composition and energy
spectrum (far data) with a detector. By comparing predictions with observations
or near/far data, we can measure neutrino oscillations and determine
the oscillation parameters.

When neutrinos propagate in matter, the interactions with the medium
affect their properties. The amplitude of this propagation
is modified due to coherent forward scattering on electrons and
nucleons. Different flavours have different interactions. The effect of
the medium can be described by an effective potential that depends on
the density and composition of the matter.

The effective potential for the evolution of $\nue$ in a
medium with electrons, protons and neutrons due to its CC interactions
is given by\aq{What is $G_F$ in the following equation?}
\begin{equation}
V_{\text{CC}} = \pm \sqrt{2} G_F n_\rme ,
\end{equation}
where $n_\rme$ is the electron number density and $G_F$ is the Fermi constant. The effective
potential has different sign for neutrinos and antineutrinos.

For example, the matter potential at the Earth's core is $\sim 10^{-13}$~eV while at the solar core it is $\sim 10^{-12}$~eV. In spite of these tiny values, these effects are
non-negligible in neutrino oscillations.

For $\numu$ and $\nutau$, the potential due to CC interactions is zero,
since neither muons nor taus are present in the medium. The effective
potential for any active neutrino due to neutral current
interactions in a neutral medium can be written as
\begin{equation}
V_{\text{NC}} = \mp \frac{\sqrt{2}}{2} \, G_F n_\rmn
\end{equation}
where $n_\rmn$ is the number density of neutrons.

In general, the electron number density in the medium changes along
the neutrino trajectory and so does the effective potential. We can
describe neutrino oscillations in a medium as in vacuum but with an
effective mass matrix (${\tilde M}_\nu^2$)  that depends on the
neutrino energy and the matter density, as follows:
\begin{equation}
{\tilde M}_\nu^2 = M_\nu^2 \pm 4 E V_{\rm m} ,
\end{equation}
with
\begin{eqnarray}
V_{\rm m} = \begin{pmatrix}
       V_\rme = V_{\text{CC}} + V_{\text{NC}}
         & 0
           & 0\\
       0
         &  V_{\mu} = V_{\text{NC}}
           & 0 \\
       0
         & 0
           & V_{\tau} = V_{\text{NC}}
\end{pmatrix} .
\end{eqnarray}

In the case of two flavours, the mixing angle and effective masses in
matter can be written as
\begin{equation}
\tan 2\theta_{\rm m} = \frac{\Delta m^2 \sin 2\theta}{\Delta m^2 \cos
  2\theta - A}
\end{equation}
and
\begin{equation}
\mu^2_{1,2}(x) = \frac{m_1^2 + m_2^2}{2} + E (V_\alpha + V_\beta) \mp
\frac{1}{2} \sqrt{(\Delta m^2 \cos 2\theta - A)^2 + (\Delta m^2 \sin 2\theta)^2}.
\end{equation}
They depend on the matter density and neutrino energy. The  $-$ (+) sign
corresponds to neutrinos (antineutrinos). The quantity $A$ is defined as $A
\equiv 2 E (V_\alpha - V_\beta)$, the potential difference factor
between $\alpha$ and $\beta$ flavours. Depending on the sign of $A$, the
mixing angle in matter can be larger or smaller than in vacuum. For
constant potential, the mixing angle and effective masses are constant
along the neutrino evolution.

Matter effects are important when the potential difference factor $A$ is
comparable to the mass difference term $\Delta m^2 \cos 2\theta$. The
oscillation amplitude has a resonance when the neutrino energy
satisfies this relation:
\begin{equation}
A_R= \Delta m^2 \cos 2\theta .
\end{equation}
Even if the mixing angle in vacuum is very small, we will have
maximal mixing at the resonance condition. The resonance happens for
neutrinos or antineutrinos but not for both, and depends on the sign of
$\Delta m^2 \cos 2\theta$.

The value of the mixing angle in matter changes if the density is
changing along the neutrino trajectory. The mixing angle $\theta_{\rm m}$
changes sign at $A_R$. For $A \gg A_R$,
we have $\theta_{\rm m} = \pi/2$. For $A = A_R$, $\theta_{\rm m} = \pi/4$.

For a neutrino system that is travelling across a monotonically
varying matter potential, the dominant flavour component of a given mass
eigenstate changes when crossing the region with $A = A_R$. This
phenomenon is known as {\it level crossing}. For constant or sufficiently slowly
varying matter potential, the instantaneous mass eigenstates
behave approximately as energy eigenstates and they do not mix in the
evolution. This is the {\it adiabatic transition} approximation.

The \ced{Mikheyev--Smirnov--Wolfenstein (MSW)}\aq{Given in full. OK?}
effect~\cite{msw} describes the adiabatic flavour neutrino conversion in a
medium with varying density. We can consider the propagation of a
two-family neutrino system in the matter density of the Sun. The solar
density decreases monotonically with the distance to the centre of the
Sun.
The eigenstates in matter can be written as
\begin{eqnarray}
|\nu_1^{\rm m}\rangle &=& |\nu_\rme\rangle \cos\theta_{\rm m} - |\nu_\mu\rangle \sin\theta_{\rm m} ,\\
|\nu_2^{\rm m}\rangle &=& |\nu_\rme\rangle \sin\theta_{\rm m} + |\nu_\mu\rangle \cos\theta_{\rm m} .
\label{eq:eigenmat}
\end{eqnarray}

Neutrinos are produced close to the centre where the electron density ($n_\rme(0)$)
is very large. The potential is much larger than the resonance potential
\begin{equation}
2 E \sqrt{2} G_F n_\rme(0) \gg \Delta m^2 \cos 2\theta ,
\end{equation}
and therefore the mixing angle in matter is $\theta_{\rm m} = \pi/2$. In
this case, the electron neutrino is mostly the second mass eigenstate
($\nu_\rme \approx \nu_2^{\rm m}$).

\begin{figure}[ht]
\begin{center}
\includegraphics[width=7cm]{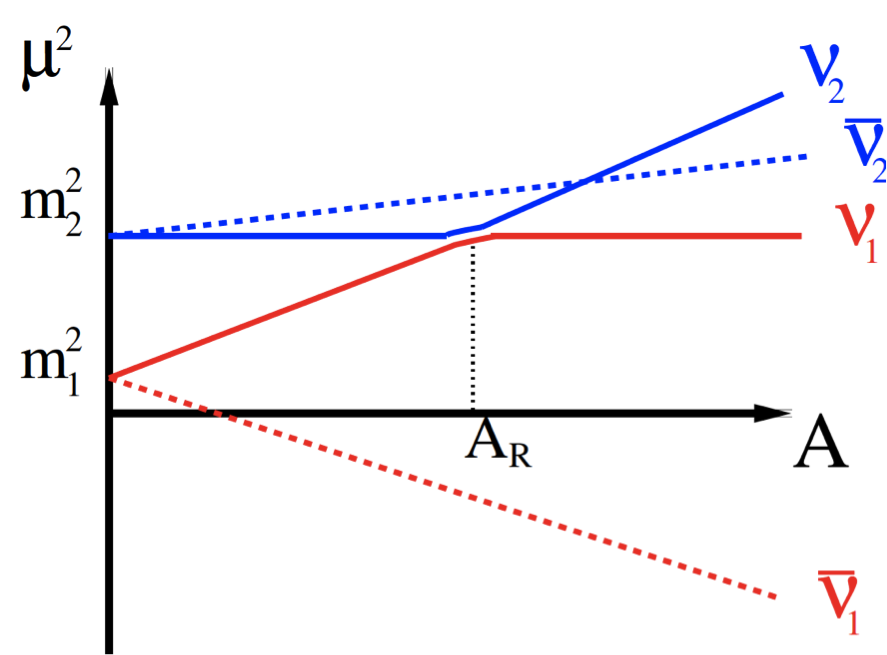}
\caption{Effective masses acquired in the medium by a system of two
  massive neutrinos as a function of the potential $A$.}
\label{fig:msw}
\end{center}
\end{figure}

When neutrinos exit the Sun, the matter density falls to zero and the
effective mixing angle is the one in vacuum, $\theta_{\rm m} = \theta$. If
$\theta$ is small, the eigenstate $\nu_2^{\rm m}$ is mostly $\nu_\mu$. There
is maximum conversion $\nue \rightarrow \numu$ if the adiabatic approximation is
correct (Fig.~\ref{fig:msw}). This is the MSW effect. There is a level crossing in the
absence of mixing. As we will explain later, the deficit of electron
neutrinos coming from the Sun has been interpreted in terms of an MSW
effect in neutrino propagation in the Sun.

\section{Experimental results from neutrino oscillation experiments}

Over the years, neutrino experiments have provided spectacular
evidence for neutrino oscillations. There are essentially three
pieces of evidence: one provided by solar and reactor neutrinos, a
second by atmospheric and accelerator neutrinos, and a
third by the \ced{Liquid Scintillator Neutrino Detector (LSND)}\aq{I've corrected the acronym to LSND throughout and given it in full here on first occurrence. OK?} experiment. They correspond to three values of
mass-squared differences of different orders of magnitude. There is no
consistent explanation of all three signals based on oscillations
among the three known neutrinos, since there are only two independent
mass-squared differences.

In the next sections, I will describe these experimental
results in detail.

\subsection{Solar neutrinos}

Solar electron neutrinos are produced in thermonuclear reactions
happening in the Sun through two main chains, the \ced{proton--proton (pp) chain} and the \ced{carbon--nitrogen--oxygen (CNO) cycle.}\aq{Given both pp and CNO in full on first occurrence. OK?} There are five reactions that produce $\nue$ in the pp chain and three in the CNO cycle.

\begin{figure}[ht]
\begin{center}
\includegraphics[width=9cm]{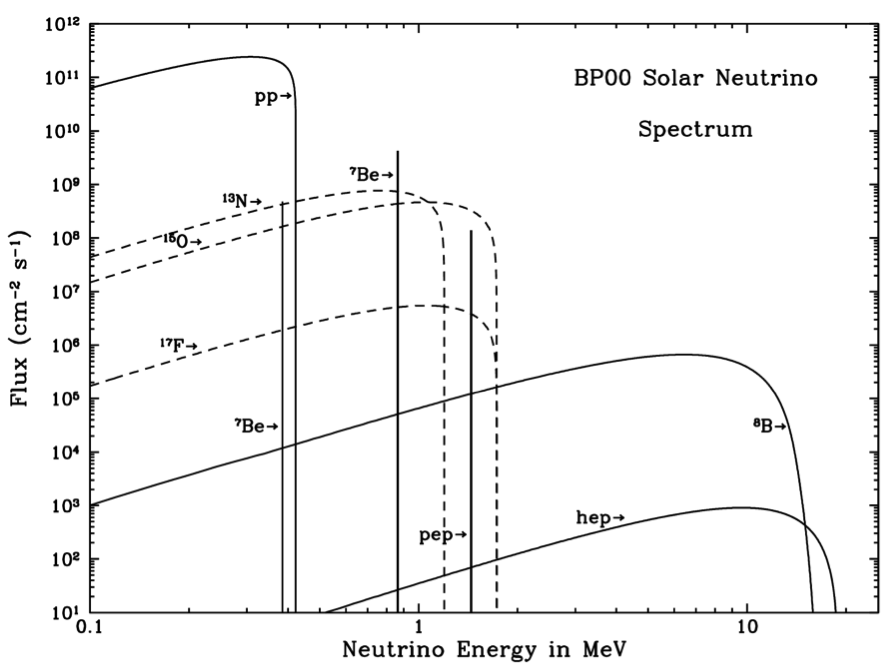}
\caption{Neutrino fluxes from the pp chain reactions and the CNO cycle
reactions as a function of the neutrino energy.}
\label{fig:bahcall}
\end{center}
\end{figure}

Figure~\ref{fig:bahcall} shows the solar neutrino spectrum as predicted
by Bahcall~\cite{bahcall} from the eight reactions.

The standard solar model (SSM) is the theoretical model describing the
evolution of the Sun and allows one to predict the spectra and the fluxes
of all the solar neutrino sources. As a consequence, solar neutrinos
provide a unique probe for studying both the nuclear fusion reactions
that power the Sun and the fundamental properties of neutrinos.

The first indication of oscillations happened in the 1970s by measuring
the solar neutrino flux. Radiochemical experiments were trying to
understand the energy production mechanism in the Sun and they found a
huge difference between what they measured and what was expected from
solar models.

The Davis experiment was installed in the Homestake mine in South
Dakota~\cite{homestake}. They
built a 615~ton tank of perchloroethylene C$_2$Cl$_4$ to measure the
$\nue$ interaction with Cl, which gives a radioactive isotope $^{37}$Ar that
can be extracted and counted.
\begin{equation}
\nu_\rme + \text{$^{37}$Cl} \rightarrow \text{$^{37}$Ar} + \rme^-
\end{equation}
The energy threshold for this reaction is 0.814~MeV, so the relevant
fluxes are the $^7$Be and $^8$B neutrinos. The $^{37}$Ar produced is
extracted radiochemically every three months approximately and the
number of $^{37}$Ar decays is measured in a proportional counter.

In the 1990s, other radiochemical experiments like GALLEX/GNO
\cite{gallex} in Italy and SAGE~\cite{sage} in Russia tried to measure the
solar neutrinos using a $^{71}$Ga target and extracting Ge isotopes.
\begin{equation}
\nu_\rme + \text{$^{71}$Ga} \rightarrow \text{$^{71}$Ge} + \rme^-
\end{equation}
This reaction has a very low energy threshold ($E_\nu > 0.233$~MeV)
and a large cross-section for the lower-energy pp neutrinos. The
extraction of $^{71}$Ge takes place every 3--4 weeks. The GALLEX programme was
completed in the autumn of 1997 and its successor GNO started taking
data in spring 1998.

All the radiochemical neutrino experiments found a solar neutrino flux
much lower (between 30\% and 50\%) than the predicted value. They could
provide neither information on the directionality nor the energy of the
neutrinos.

The Kamiokande experiment~\cite{kamioka} pioneered a new technique to observe solar
neutrinos using water Cerenkov detectors. This was a real-time
experiment and provided information on the directionality and the energy
of neutrinos by measuring the electrons scattered from the water by
the elastic reaction
\begin{equation}
\nu_\rme + \rme^- \rightarrow \nu_\rme + \rme^-
\end{equation}
producing Cerenkov light, which is detected by
photomultipliers. The threshold for this type of experiment is much
higher and they are only able to measure the $^8$B neutrinos.

\begin{figure}[ht]
\begin{center}
 \includegraphics[width=9cm]{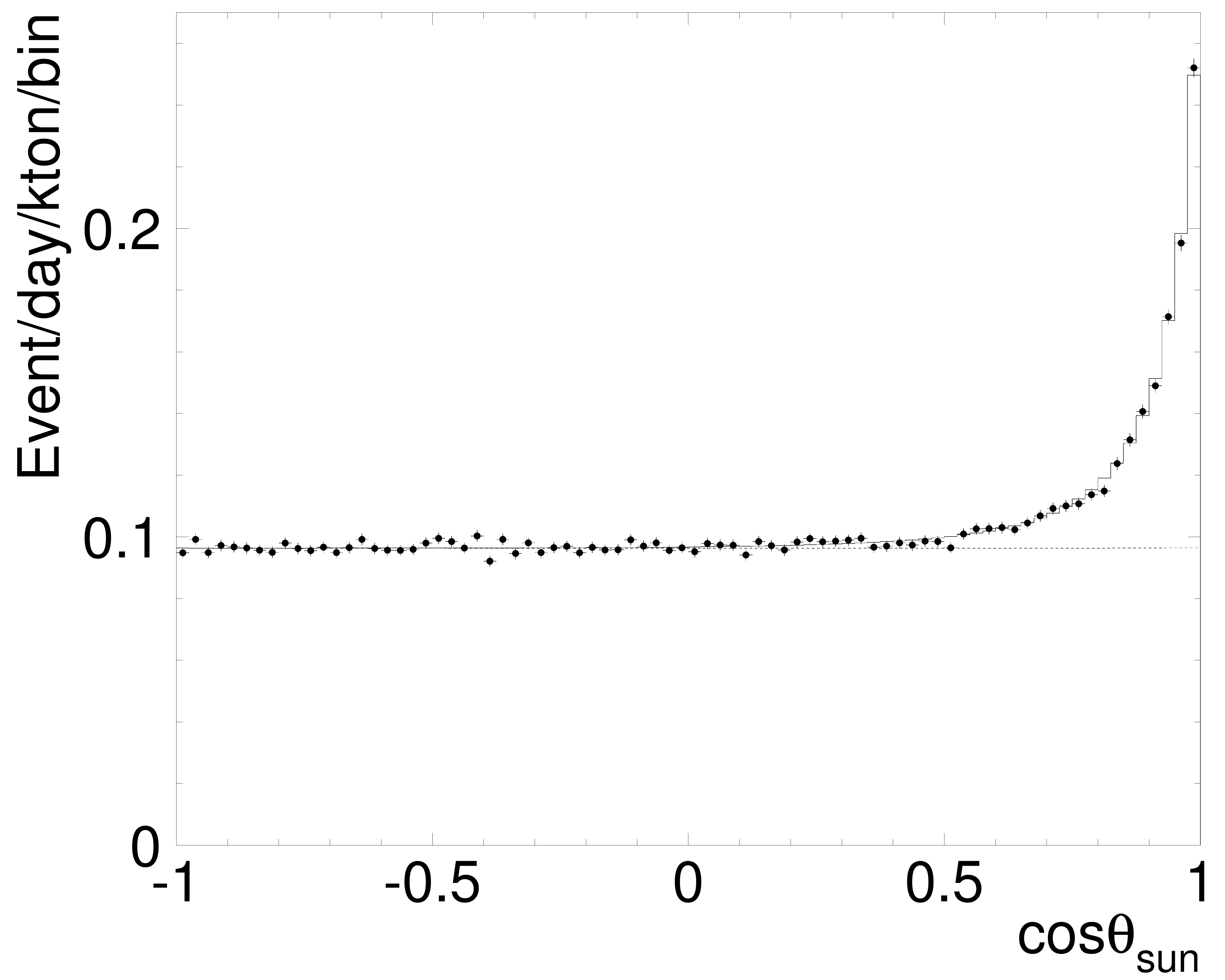}
\caption{Solar neutrino direction as a function of the zenith angle of
  the Sun in Super-Kamiokande phase I (from Ref.~\cite{sk_solar}).}
\label{fig:solar_sk}
\end{center}
\end{figure}

Later on, the Super-Kamiokande (SK) experiment~\cite{sk_solar}, with 50~kton of water,
measured the solar neutrinos with unprecedented precision in the
energy region 5--20~MeV. Figure~\ref{fig:solar_sk} shows the reconstructed direction of
the incoming neutrinos correlated to the Sun direction as measured by
SK during the first phase of operation (1996--2001).

In 2001 the SNO experiment showed clear evidence that solar
neutrinos oscillate. This allowed the solar model predictions to be studied
independently of the neutrino properties.

SNO~\cite{sno} is a Cerenkov detector made of 1~kton of heavy water (D$_2$O) located
underground in the Sudbury mine in Canada and is able to detect $^8$B
solar neutrinos via three different reactions:
\begin{itemize}
\item CC interactions on deuterons in which only electron neutrinos
  participate
\begin{equation}
\nu_\rme  + \rmd \rightarrow \rmp + \rmp +  \rme^{-}
\end{equation}
\item elastic scattering (ES) sensitive to other neutrino flavours but dominated by electron  neutrinos
\begin{equation}
\nu_x + \rme^- \rightarrow \nu_x + \rme^-
\end{equation}
\item NC interactions with equal sensitivity to all flavours and an
  energy threshold of 2.2~MeV
\begin{equation}
\nu_x  + \rmd \rightarrow \rmp + \rmn +  \nu_x
\end{equation}
\end{itemize}

\begin{figure}[ht]
\begin{center}
\includegraphics[width=8cm]{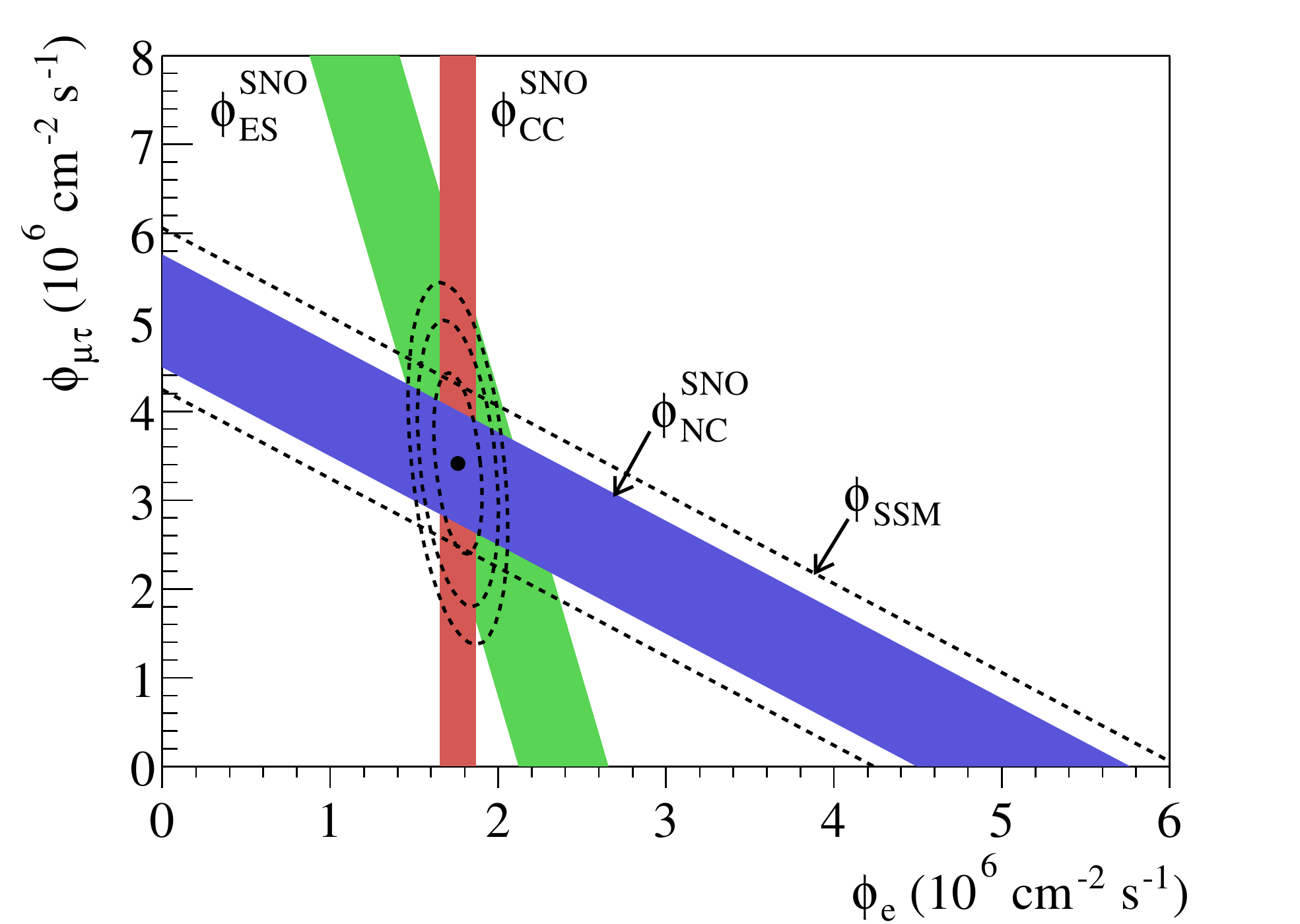}
\caption{Flux of $^8$B solar neutrinos that are $\mu$ or $\tau$
  flavour versus flux of electron neutrinos deduced from the three neutrino
  reactions in SNO (from Ref.~\cite{sno-I}).}
\label{fig:sno}
\end{center}
\end{figure}

In the case of no oscillations, the neutrino fluxes from the three
interactions should be equal since there are only electron neutrinos
coming from the Sun. However, Fig.~\ref{fig:sno} shows the neutrino
fluxes measured by the three reactions by SNO.
The flux of non-electron neutrinos ($\phi_{\mu\tau}$) is plotted as a
function of the electron neutrino flux ($\phi_\rme$). The NC events give
a measure of the total solar neutrino flux and it is in good
agreement with the SSM theoretical predictions. SNO can test if the deficit
of solar $\nue$ is due to changes in the flavour composition of the
solar neutrino beam, since the ratio CC/NC compares the number of
$\nue$ interactions with those from all active flavours. This comparison
is independent of the overall flux normalization.

The SNO detector operated during 1999--2006 in three phases with different
 detection techniques to detect NC neutrons: phase~I, in pure heavy
 water; phase~II, 2000~kg of salt were dissolved in the heavy water,
 increasing the neutron capture cross-section; and phase~III, the salt was
 removed and ultra-pure $^3$He counters were deployed into the SNO
 detector. SNO finished data taking in November 2006.

From the latest results including the SNO-III phase, the ratio between
the CC and NC events is~\cite{sno_ratio}
\begin{equation}
\frac{\phi_{\rm CC}}{\phi_{\rm NC}} = 0.301 \pm 0.033.
\end{equation}
This result provides clear evidence for solar neutrino oscillations
independently of the solar model.

\begin{figure}[ht]
\begin{center}
\includegraphics[width=0.6\linewidth]{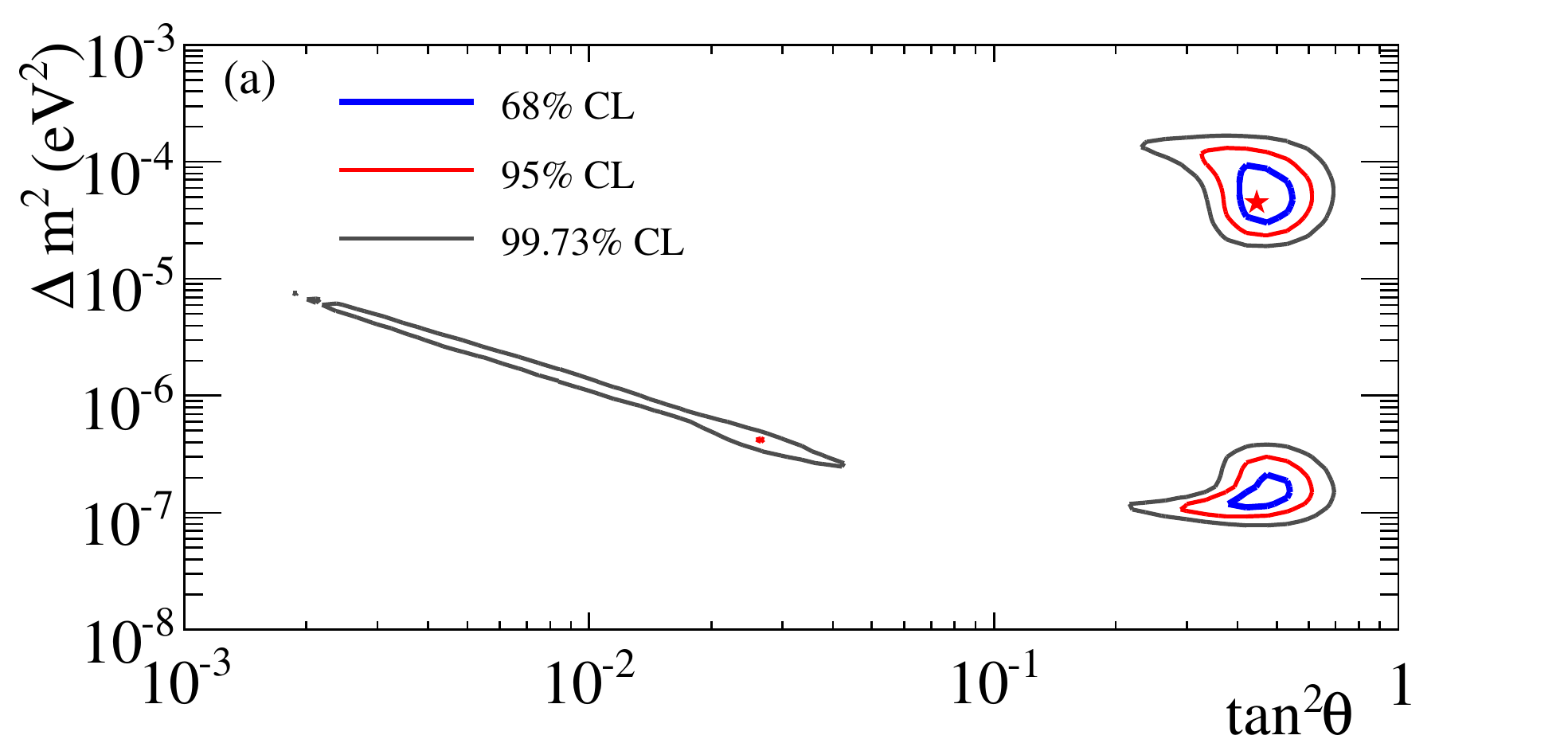}
\includegraphics[width=0.3\linewidth]{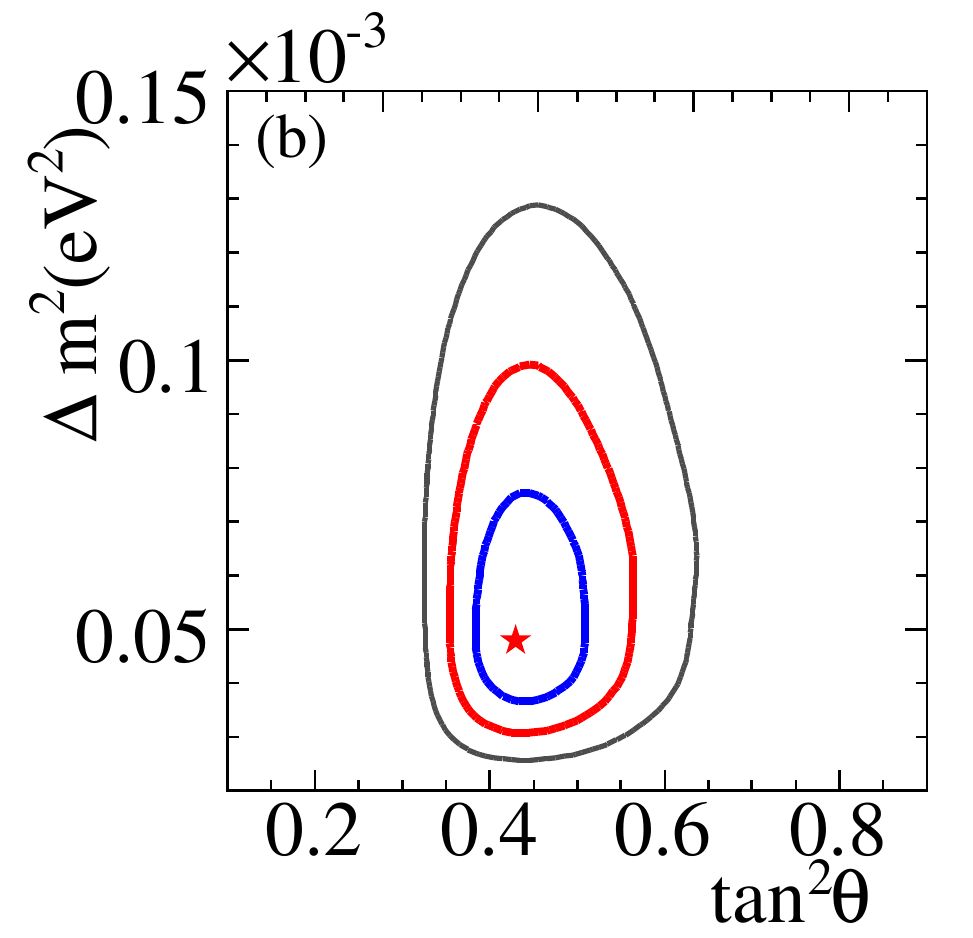}
\caption{Allowed oscillation parameters from the analysis of SNO
  neutrino data (left) and from the global analysis of all solar neutrino
  data (right) in terms of neutrino oscillations (from
  Ref.~\cite{sno_ratio}).}
\label{fig:sno_region}
\end{center}
\end{figure}

From the SNO results, it is possible to constrain the neutrino mixing
parameters. Figure~\ref{fig:sno_region} shows the allowed regions of
parameters from SNO data (left) and from the global analysis including
data from all the solar experiments (right). Of all the possible
solutions, only the one at the largest mixing angle and mass-squared
difference survives, the famous \ced{large-mixing-angle (LMA)}\aq{Given in full on first occurrence. OK?} solution, for which matter effects
in the Sun are important.

\begin{figure}[ht]
\begin{center}
\includegraphics[width=8cm]{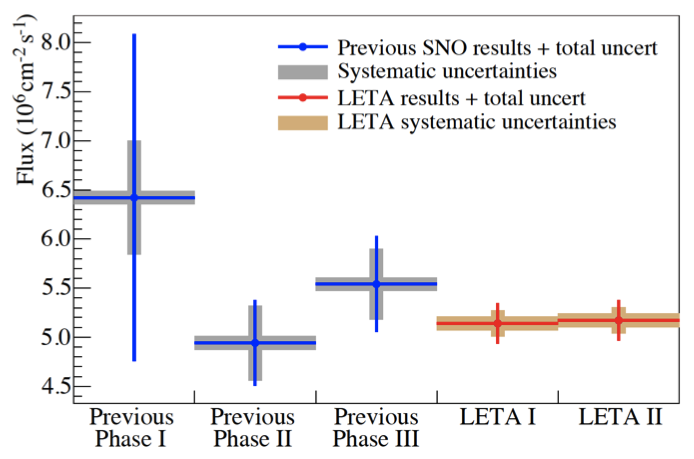}
\caption{Total $^8$B neutrino flux results using the NC reaction from
  both unconstrained signal extraction fits (LETA) in comparison to
  unconstrained fit results from previous SNO analyses (from
  Ref.~\cite{sno_new}).}
\label{fig:sno_new}
\end{center}
\end{figure}

The phase I and II data from SNO have been reanalysed (see Fig.\ \ref{fig:sno_new})~\cite{sno_new} with a lower effective electron kinetic energy threshold (3.5~MeV). The total uncertainty on the flux of $^8$B solar neutrinos has been reduced by
more than a factor of~2 \ced{compared to} the best previous SNO results.

One of the most important results in the last few years has come from the
Borexino detector in the Gran Sasso laboratory. Borexino
\cite{borexino} is a 300~kton ultra-pure liquid scintillator
detector using the elastic scattering on electrons to measure the low-energy flux and spectrum of solar neutrinos. The main goal is the
measurement of the monochromatic $^7$Be solar neutrinos at 0.862~MeV. Thanks to its excellent radiopurity, Borexino also measures $^8$B
neutrinos with an energy threshold of only 3~MeV. This is the lowest
energy threshold ever reached in real-time experiments.

Before Borexino, radiochemical experiments measured the very low energy
range (where oscillations happen essentially in vacuum) while SNO and
SK measured the $^8$B part of the spectrum. Borexino has measured the
$^7$Be spectrum and provided a confirmation of the MSW--LMA model. This
is the first direct measurement of the survival probability for solar
electron neutrinos in the transition region between matter-enhanced
and vacuum-driven oscillations~\cite{borexino_results1}.

\begin{figure}[ht]
\begin{center}
\includegraphics[width=0.4\linewidth]{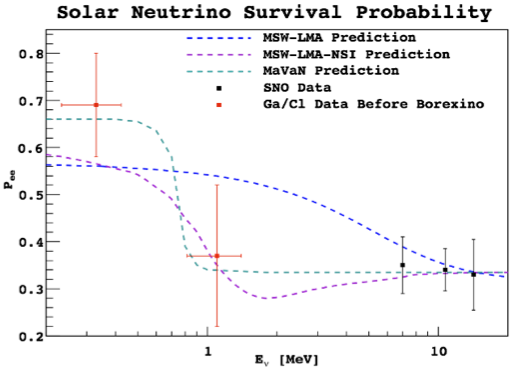}
\includegraphics[width=0.4\linewidth]{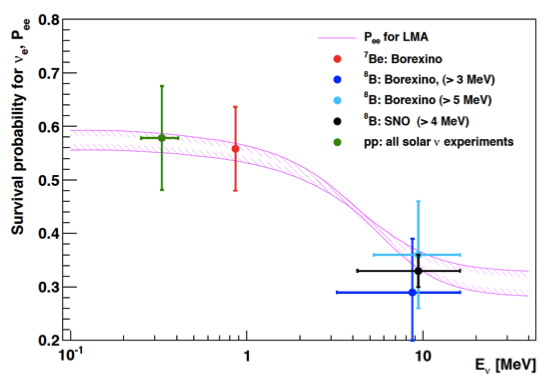}
\caption{Comparison of solar neutrino fluxes as a function of the
  energy measured by several solar neutrino experiments before (left) and
  after (right) Borexino data (from Ref.~\cite{borexino_results2}).}
\label{fig:borexino_results}
\end{center}
\end{figure}

A prediction of the MSW--LMA model is that neutrino oscillations are
dominated by vacuum oscillations at low energies ($<1$~MeV) and by
resonant matter-enhanced oscillations taking place in the Sun's core
at high energies ($>5$~MeV). A measurement of the survival probability as a
function of the neutrino energy is very important to confirm the
MSW--LMA solution. Figure~\ref{fig:borexino_results} shows the survival
probability ($P_{\rm ee}$) before (left) and after (right) including the
Borexino data and the fit assuming LMA oscillations. The MSW--LMA model
is confirmed at 4.2$\sigma$ level. For the first time the same
apparatus can measure two different oscillation regions predicted by
the MSW--LMA model.

``Geoneutrinos'' are electron antineutrinos produced by beta decays of
the nuclei in the decay chains of $^{238}$U and $^{232}$Th. Geoneutrinos are
direct messengers of the abundances and distribution of radioactive
elements within our planet. By measuring their flux and spectrum, it is
possible to reveal the distribution of long-lived radioactivity in the
Earth and to assess the radiogenic contribution to the total heat
balance of the Earth. As these radioactive isotopes beta-decay, they
produce antineutrinos. So, measuring these antineutrinos may serve as
a cross-check of the radiogenic heat production rate.

KamLAND is the first detector to conduct an investigation on
geoneutrinos~\cite{kamland_geo}. In 2005 they provided the first
experimental indication for geoneutrinos. Borexino has also been able to measure
geoneutrinos at 4.2$\sigma$~\cite{borex_geo}. Both detectors use the
inverse beta decay to detect geoneutrinos.

\subsection{Atmospheric neutrinos}

Atmospheric neutrinos are produced in the collision of primary cosmic
rays (typically protons) with nuclei in the upper atmosphere. This
creates a shower of hadrons, mostly pions. The pions decay to a muon
and a muon neutrino. The muons decay to an electron, another muon
neutrino, and an electron neutrino. Based on this simple kinematic
chain, one predicts a flux ratio of \ced{two muon neutrinos to one electron
neutrino.}\aq{Is this what you meant?}

The first experiment proving neutrino
oscillations without ambiguities was the Super-Kamiokande experiment located 1000~m
underground in the Kamioka mine in Japan in 1998. This 50~kton water
Cerenkov detector (22.5~kton fiducial mass) measured the atmospheric
neutrinos produced by cosmic-ray collisions with the
atmosphere. Two muon neutrinos are produced per one electron neutrino
from the pion decay with energies between 0.1 and 100~GeV.
More than 11\,000 20-inch PMTs covering 40\% of the surface detect the
Cerenkov light coming from the neutrino CC interactions.

\begin{figure}[ht]
\begin{center}
\includegraphics[width=8cm]{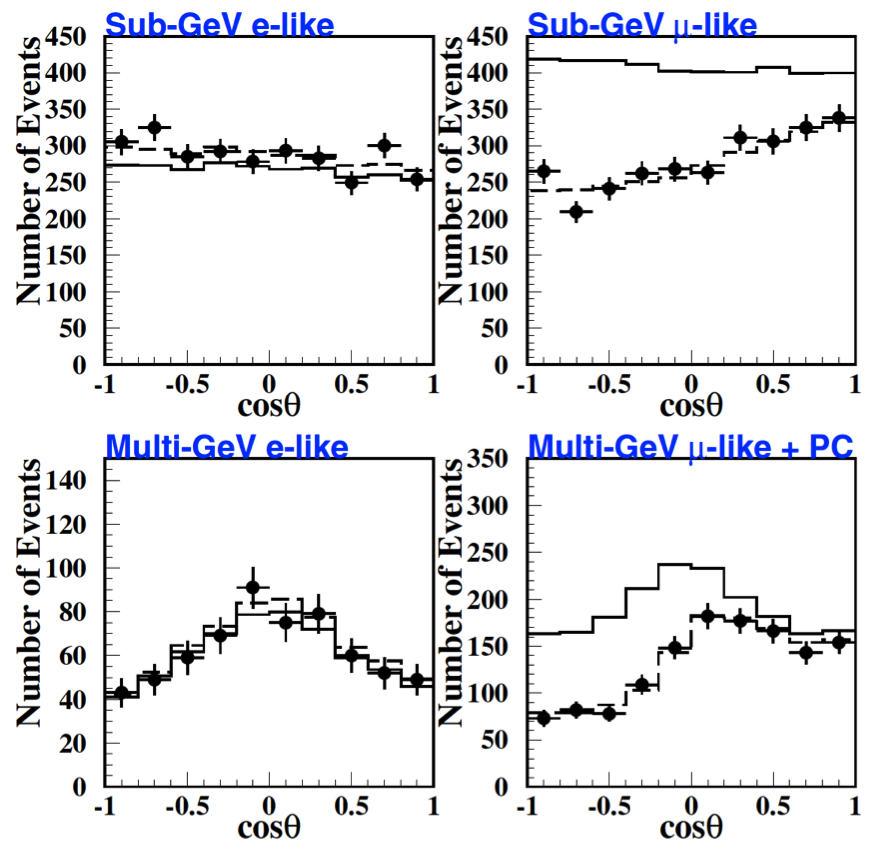}
\caption{Zenith angle distribution of SK data. Dots, solid line and
  dashed line correspond to data, Monte Carlo without oscillation, and
  Monte Carlo with best-fit oscillation parameters.}
\label{fig:sk_atm}
\end{center}
\end{figure}

By measuring the number of events of each type, as a function of
energy and direction, we can find out if neutrino oscillations are
affecting the results. SK has shown a big deficit of muon neutrinos
\ced{dependent on}\aq{Sense OK?} the energy and at distances compatible with neutrino
oscillations. The distributions of electrons and muons as a function
of the azimuthal angle show an asymmetry between upward and downward
muon neutrinos (Fig.~\ref{fig:sk_atm}). The muon neutrinos traversing
the Earth present a clear deficit, which is not the case for downward
muon neutrinos or electron neutrinos. This deficit is compatible
with a $\numu$--$\nutau$ oscillation~\cite{sk_osc}.

On 12 November 2001, about 6600 of the photomultiplier tubes in the
Super-Kamiokande detector imploded, apparently in a chain reaction due
to a shock wave. The detector was partially restored by
redistributing the photomultiplier tubes that did not implode. In
2005 they reinstalled 6000 PMTs and they called the new phase SK-III.

The zenith angle two-flavour analysis of the data before the SK PMT implosion
(SK-I and SK-II) has been updated~\cite{sk_oscpar} and allowed to better
constrain the $\Delta m^2_{32}$ and $\theta_{23}$ oscillation
parameters.

\begin{figure}[ht]
\begin{center}
\includegraphics[width=8cm]{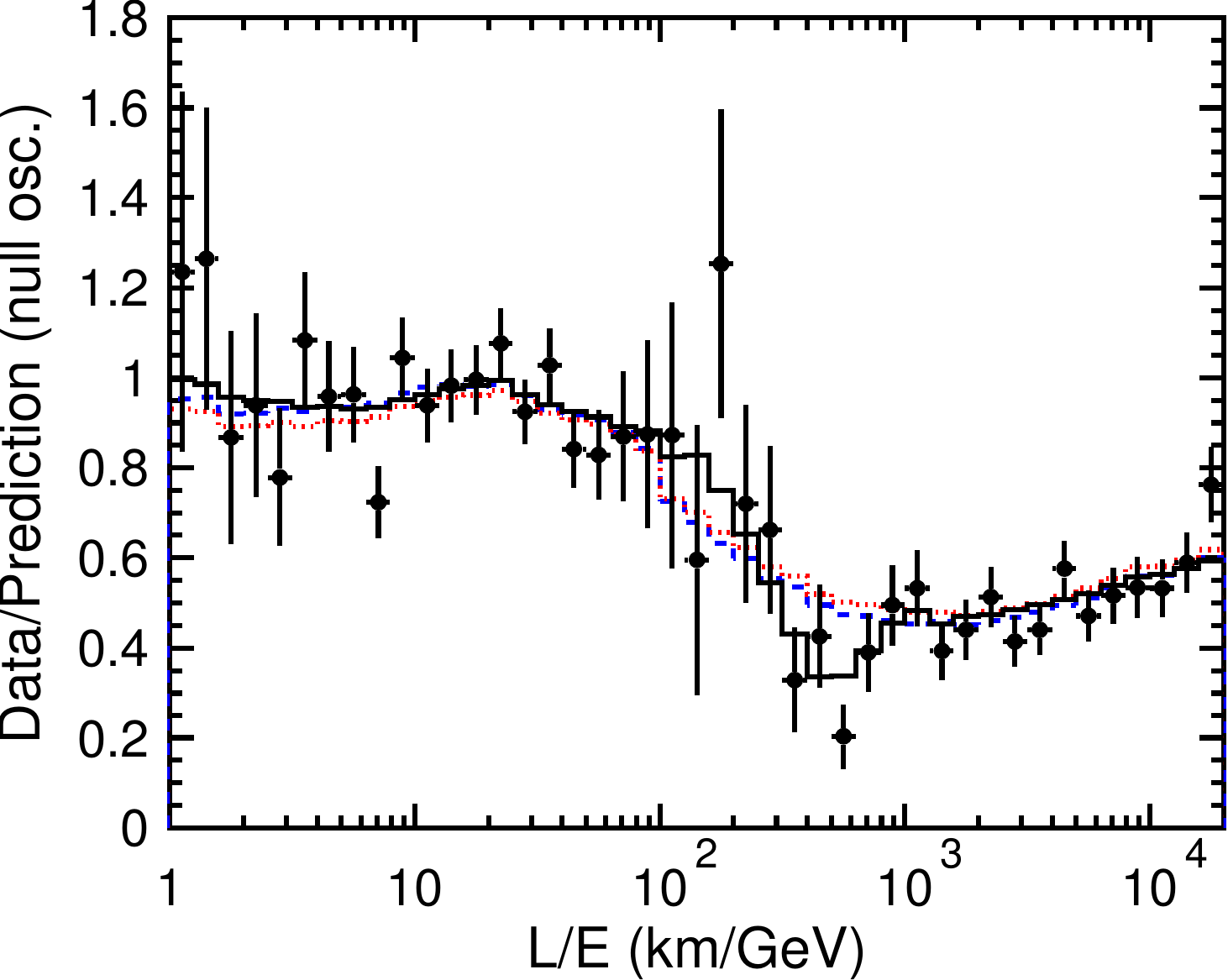}
\caption{Ratio of the data to the non-oscillated Monte Carlo events
  (points) with the best-fit expectation for two-flavour
  $\numu \rightarrow \nutau$ oscillation
  analysis (solid line) as a function of $L/E$ (from Ref.~\cite{sk_dip}).}
\label{fig:sk_dip}
\end{center}
\end{figure}

SK was also able to observe the expected dip in the $L/E$ spectrum due
to oscillations (Fig.~\ref{fig:sk_dip}). Other hypotheses have been
excluded at 4.1$\sigma$ and 5$\sigma$ levels.

The latest zenith angle and $L/E$ analysis results from SK-I, II and III
data~\cite{sk_latest} are consistent and provide the most stringent
limit on $\sin^2\theta_{23}$.

\subsection{Reactor neutrinos}

Reactor neutrinos have also played a crucial role in neutrino
oscillations. They have helped to understand the solar anomaly and
they have provided unique information on the $\theta_{13}$ mixing
angle, still unknown.

Nuclear reactors are the major source of human-generated
neutrinos. They are very intense, pure and isotropic sources of
antineutrinos coming from the beta decay of the neutron-rich fission
fragments. The four main isotopes contributing to the antineutrino
flux are $^{235}$U, $^{238}$U, $^{239}$Pu and $^{241}$Pu.

On average, each fission cycle produces $\sim$~200~MeV and six
antineutrinos. For typical modern commercial light-water reactors with
thermal power of the order of 3~GWth, the typical yield is \mbox{$\sim 6
\times 10^{20}$} antineutrinos per core per second. But not all these
neutrinos can be detected.


\begin{figure}[ht]
\begin{center}
\includegraphics[width=8cm]{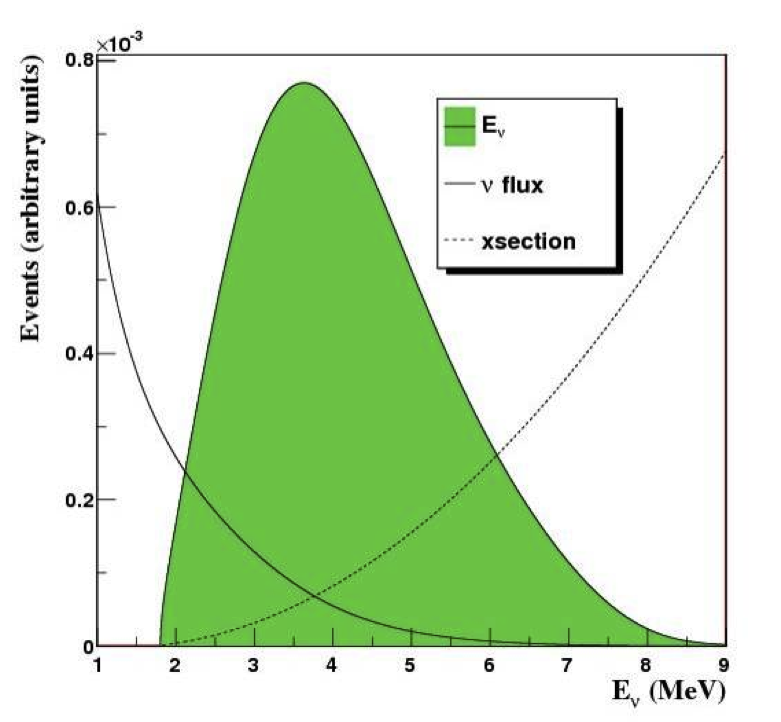}
\caption{Typical energy spectrum of antineutrinos from nuclear reactors.}
\label{fig:anue_spect}
\end{center}
\end{figure}

The observed neutrino spectrum will be the product of the reactor
neutrino flux and the inverse beta decay cross-section, as shown in
Fig.~\ref{fig:anue_spect}. The inverse beta decay (Eq.~(\ref{eq:ibd}))
has an energy threshold of 1.8~MeV and only about 1.5~$\anue$/fission can
be detected (25\% of the total).
\begin{equation}
\bar{\nu}_\rme + \rmp \rightarrow \rme^+ + \rmn
\label{eq:ibd}
\end{equation}
Past reactor experiments were looking for the disappearance of reactor
$\anue$ with the goal of solving the atmospheric problem at short
baselines. All of them found negative results. The most sensitive of
these experiments was the CHOOZ experiment.

CHOOZ~\cite{chooz_result} was looking for the disappearance of electron
antineutrinos from the CHOOZ nuclear power plant in France in the
1990s. CHOOZ was a quite simple liquid scintillator detector doped with 0.1\%
Gd located 1.05~km away from the reactors. It was hosted in a
cylindrical pit 7~m in diameter and height. The cylindrical steel tank
was surrounded by a 75~cm thick low-radioactivity sand contained in an
acrylic vessel and covered by cast iron. The target was 5~ton 0.1\% Gd-loaded liquid scintillator contained in a transparent acrylic
vessel. A 17~ton non-Gd-loaded liquid scintillation region contained
192 eight-inch PMTs. A muon veto region was read by two rings of 24 eight-inch PMTs.

This experiment has strongly influenced the present and upcoming
reactor experiments. They had the unique opportunity to have
both reactors off and periods with only one of the reactors on. This
allowed a good measurement of the backgrounds.

\begin{figure}[ht]
\begin{center}
\includegraphics[width=8cm]{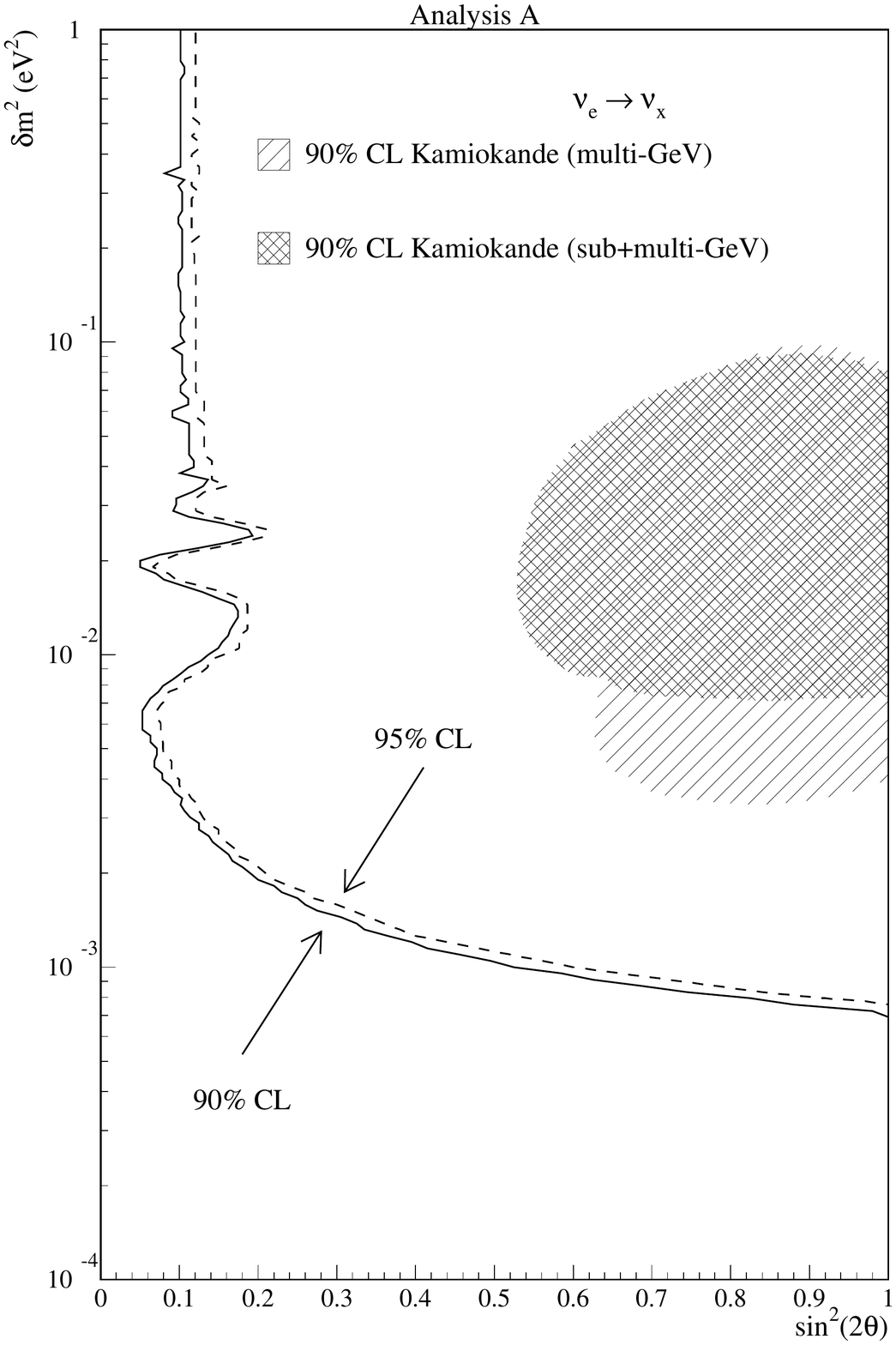}
\caption{Exclusion region in the parameter space by the CHOOZ data
  (from Ref.~\cite{chooz_result}).}
\label{fig:chooz_excl}
\end{center}
\end{figure}

The ratio of measured to expected events was $1.01 \pm 2.8\%$~(stat.)
$\pm\ 2.7\%$~(sys.). No evidence for $\nue \rightarrow
\numu$ oscillations at the $10^{-3}$ scale was found. Despite the negative
result, this experiment was very sensitive to the $\nue \rightarrow
\nutau$ oscillation.
They have not observed the disappearance of electron antineutrinos but
they could exclude a region in the parameter space
(Fig.~\ref{fig:chooz_excl}). The upper bound obtained is
$\sin^2(2\theta_{13}) < 0.12\mbox{--}0.2$ at 90\% confidence level (CL), depending on the value of $\Delta m^2_{32}$.

The CHOOZ constraint is also relevant to the global interpretation of
the solar and atmospheric neutrino data in the framework of
three-neutrino mixing.

The neutrino oscillation in the solar range has been confirmed with
reactor neutrinos by the KamLAND long-baseline reactor experiment
\cite{kamland}. KamLAND is a 1~kton liquid scintillator detector located at the
Kamioka mine in Japan (at the old Kamiokande site) at an average
distance of $L_{0} = 180$~km from 55 nuclear reactors at a depth of 2700~mwe \ced{(metres of water equivalent).}\aq{Given `unit' in full on first occurrence. OK?} They were looking for the disappearance of electron antineutrinos
at $E/L \sim 10^{-5}$~eV$^2$, in the oscillation range indicated by the
solar data. They started taking data in 2002 and finished in 2007.

The liquid scintillator is contained in a 13~m diameter spherical
nylon balloon surrounded by oil in a 18~m diameter spherical
stainless-steel vessel. This holds the 1879 PMTs with a photocathode
coverage of 34\%. A cylinder filled with water surrounds the previous
volumes, being a Cerenkov veto against backgrounds (cosmic muons, gamma
rays and neutrinos from the surrounding rock). This is the largest
scintillator detector ever constructed. Neutrinos are detected through
the inverse beta decay reaction (Eq.~(\ref{eq:ibd})), the neutrons
being captured in protons, giving photons of 2.22~MeV.

\begin{figure}[ht]
\begin{center}
\includegraphics[width=0.4\linewidth]{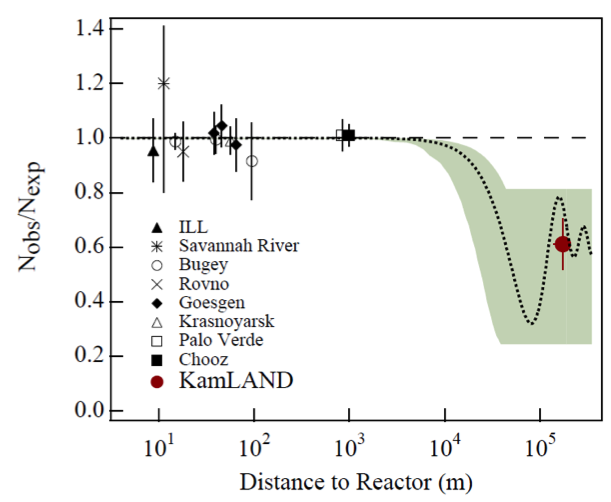}
\includegraphics[width=0.5\linewidth]{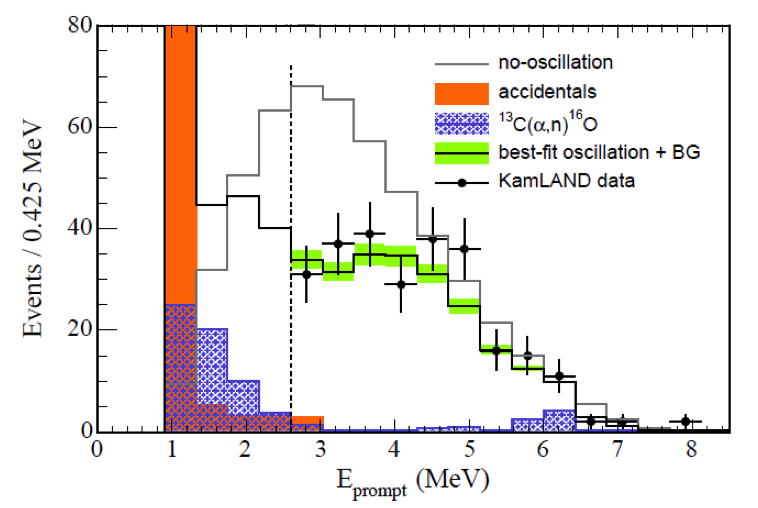}
\caption{Evidence for $\anue$ disappearance (left) and spectral
  distortion (right) measured by KamLAND.}
\label{fig:kamland_results}
\end{center}
\end{figure}

They reported the first evidence for the disappearance of reactor electron antineutrinos
in 2002~\cite{kamland_1}. In Fig.~\ref{fig:kamland_results} we can see the
ratio of observed over expected events (without oscillations) as a
function of the distance. The deficit measured by KamLAND ($R = 0.611
\pm 0.085$~(stat.) $\pm\ 0.041$~(syst.) for $\anue > 3.5$~MeV) is
compared with previous unsuccessful reactor experiments. This was
consistent with the LMA region.

KamLAND presented the first evidence of spectral distortion in
2004~\cite{kamland_2}. Figure~\ref{fig:kamland_results} shows data compared with the
non-oscillation
scenario and with the best-fit oscillation spectrum as a function of
the prompt event energy ($E_{\rm prompt} \approx E_{\bar{\nu}_\rme} + m_\rmp +
m_\rmn$). The shaded band indicates the systematic error in the best-fit
reactor spectrum above 2.6~MeV. The observed energy spectrum disagrees
with the expected spectral shape in the absence of neutrino
oscillation at 99.6\% significance and prefers the distortion expected
from the oscillation effects.

KamLAND has presented new results~\cite{kamland_3} with more statistics and a lower
energy threshold (0.9~MeV compared to  2.6~MeV). They have enlarged
the fiducial volume and performed a campaign to purify the
liquid scintillator. They have reduced the systematic uncertainty in
the number of target protons and background up to 4.1--4.5\%. The
significance of spectral distortion is now $>5\sigma$.

\begin{figure}[ht]
\begin{center}
\includegraphics[width=7cm, angle=-90]{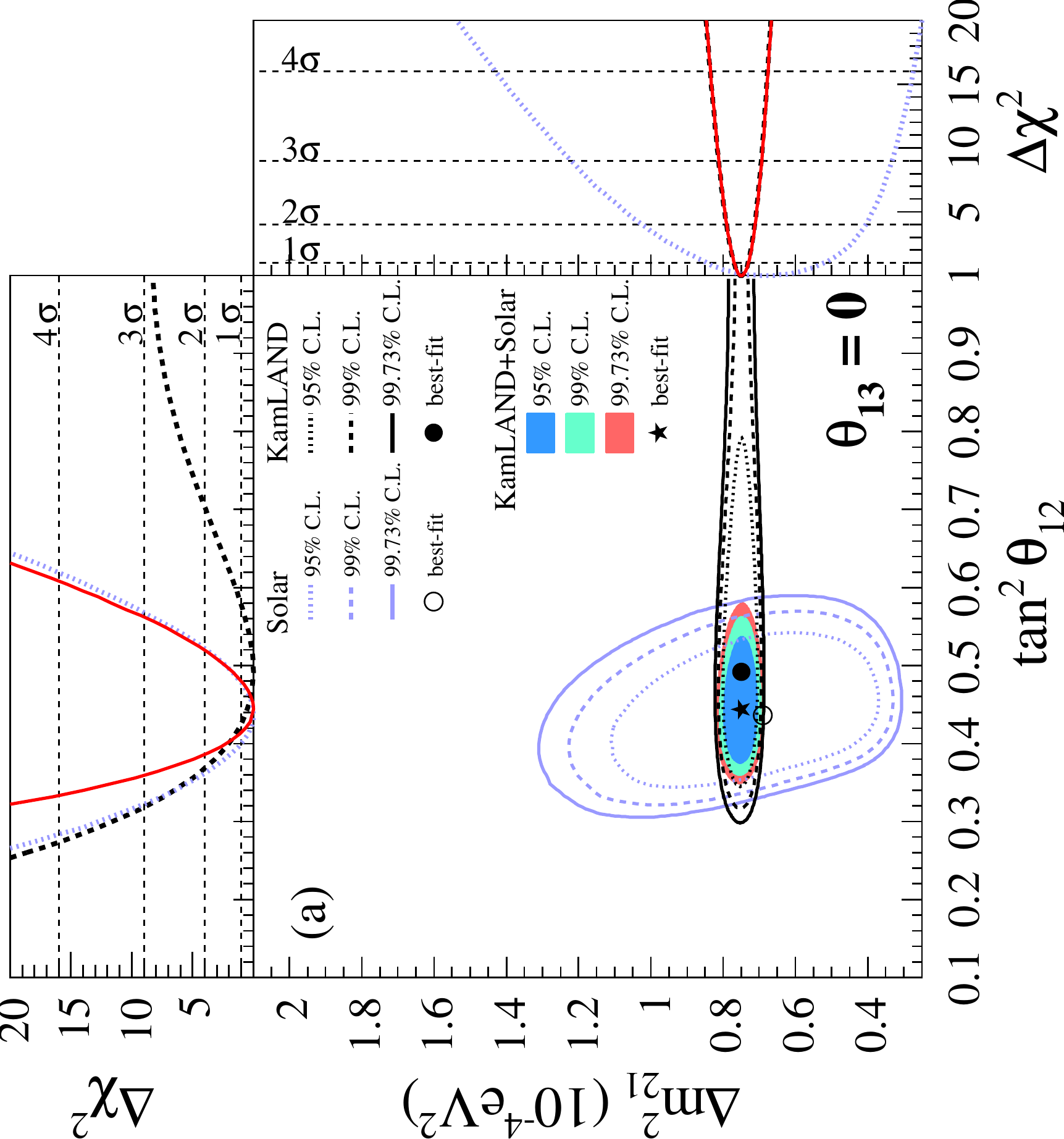}
\caption{Two-flavour neutrino oscillation analysis including solar and
  KamLAND data (from Ref.~\cite{kamland_3}).}
\label{fig:kamland_3}
\end{center}
\end{figure}

The KamLAND results can be interpreted in terms of
$\anue$ oscillations. Figure~\ref{fig:kamland_3} shows the  allowed contours in the
oscillation parameter space for solar and  KamLAND data from the
two-flavour oscillation analysis (assuming $\theta_{13}$ = 0). The
solar region is in agreement with the KamLAND data. The $\Delta
m^2_{21}$ parameter is strongly determined by the KamLAND experiment.

\begin{figure}[ht]
\begin{center}
\includegraphics[width=6cm, angle=-90]{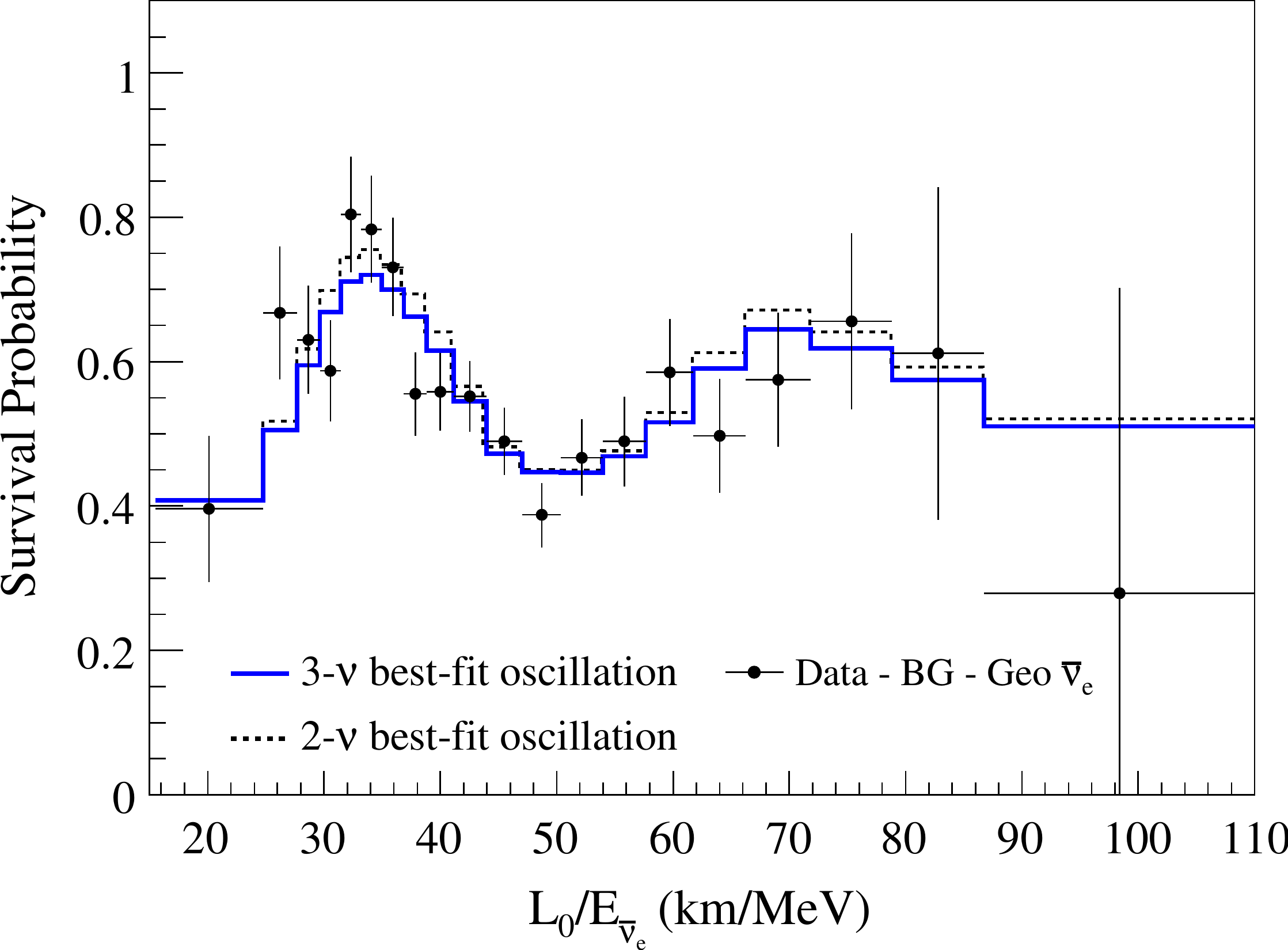}
\caption{Ratio of the observed $\anue$ spectrum to the expectation for
  non-oscillation versus $L_0/E$ for the KamLAND data (from
  Ref.~\cite{kamland_3}).}
\label{fig:kamland_4}
\end{center}
\end{figure}

The ratio of the background-subtracted neutrino spectrum to
non-oscillation expectations as a function of $L_{0}/E_{\nu}$ is shown in
Fig.~\ref{fig:kamland_4}. We can clearly see the oscillation periods over
almost two full cycles. The oscillatory signature is distorted because
the reactor sources are distributed across multiple baselines.

In summary, KamLAND confirmed neutrino oscillation, providing the most
precise value of  $\Delta m^2_{21}$ to date and improving the
precision of $\tan^2\theta_{12}$ in combination with solar data. The
indication of an excess of low-energy antineutrinos consistent with an
interpretation as geoneutrinos persists. The scientific goals of the
KamLAND experiment are now expanded towards solar neutrino
detection and neutrino-less double-beta decay detection using enriched
Xe.

\subsection{Accelerator neutrinos}

Accelerator neutrinos \ced{not only have} confirmed neutrino oscillations in the
atmospheric region \ced{and in addition proved the appearance of flavours but also have}
opened new questions in neutrino physics.\aq{Is this construction now OK as changed?}

Neutrinos are produced from the collision of a proton beam with a target,
producing pions and kaons. Then, they are focused and decay,
giving muons, electrons and neutrinos. Muons and electrons are
absorbed and the surviving particles are 98\% muon neutrinos and
around 2\% electron antineutrinos.

There are two types of searches that can be undertaken at
accelerators: disappearance searches with experiments like K2K and
MINOS, with not enough energy to produce the lepton in the CC reaction,
and appearance searches with experiments like MiniBooNE and OPERA, with
enough energy to produce the lepton. These experiments are mainly
focused on the measurement of $\Delta m^2_{32}$ and $\theta_{23}$ and
they have very limited sensitivity to $\theta_{13}$.

\begin{figure}[ht]
\begin{center}
\includegraphics[width=0.4\linewidth]{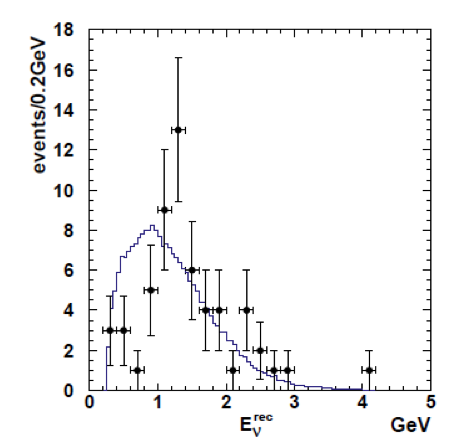}
\includegraphics[width=0.4\linewidth]{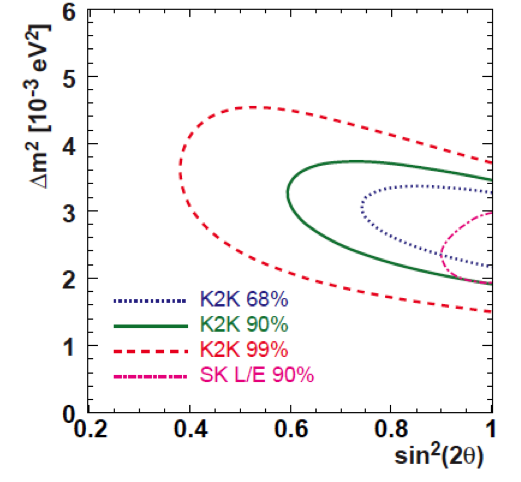}
\caption{(left) Distribution of $\numu$ events in K2K as a function of the
  reconstructed neutrino energy and (right) allowed regions from the
  analysis of K2K data compared to the $L/E$ SK analysis.}
\label{fig:k2k}
\end{center}
\end{figure}

The first accelerator-based long-baseline neutrino oscillation
experiment was K2K~\cite{k2k} starting in 1999 and running until
2004. They looked for muon neutrino disappearance using a beam
provided by KEK and detecting the oscillated neutrinos 250~km away
with the SK detector. The comparison between near and far detectors
allowed the measurement of 112 events, whereas 158 were expected, and a clear
distortion of the energy spectrum (Fig.~\ref{fig:k2k}). The best-fit
parameters are compatible with the SK atmospheric oscillation results.

More recently, the MINOS long-baseline experiment~\cite{minos} has
presented a positive result on
neutrino oscillations. MINOS is composed of two similar magnetized
steel/scintillator calorimeters to look for the disappearance of muon
neutrinos from the NUMI beam at Fermilab. The 1.5~kton near detector
is located near the source at Fermilab and the 5~kton far detector is
placed 735~km away in the Soudan mine.

\begin{figure}[ht]
\begin{center}
\includegraphics[width=0.43\linewidth]{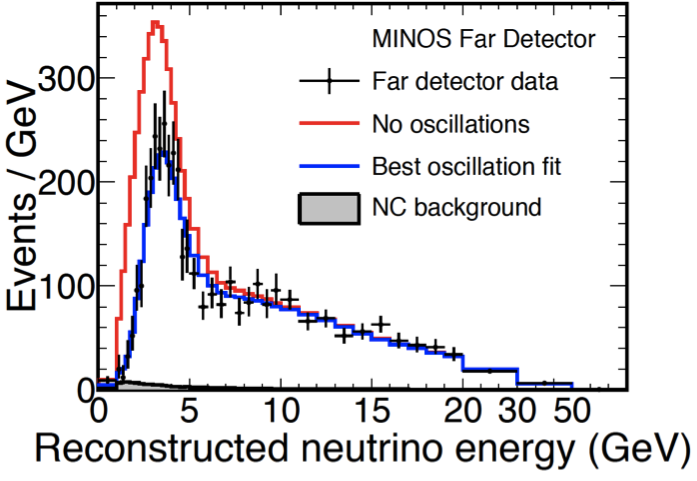}
\includegraphics[width=0.35\linewidth]{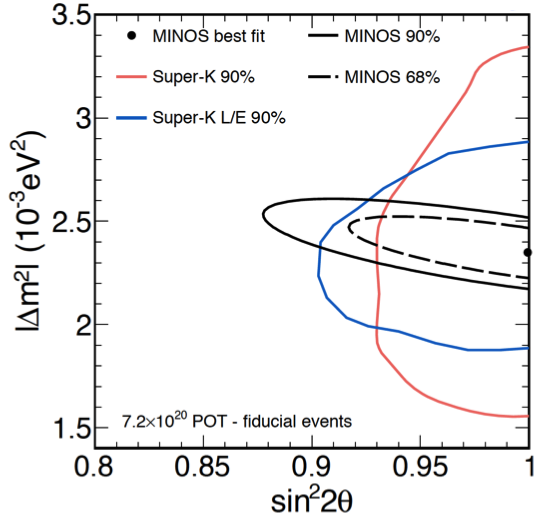}
\caption{(left) Distribution of the neutrino energy at the far MINOS detector
  compared to the non-oscillation case and (right) the corresponding
  allowed regions in the oscillation parameter space (from
  Ref.~\cite{minos_nus}).}
\label{fig:minos_nus}
\end{center}
\end{figure}

Figure~\ref{fig:minos_nus} shows the MINOS far detector data with a
significant deficit compared to the non-oscillation case and very good
agreement with the $\numu$--$\nutau$ oscillation scenario. The allowed
region in the parameter space is shown in this plot together with the
results from SK $L/E$ analysis and K2K. The measurement of $\Delta
m^2_{32}$ is dominated by MINOS, while the angle $\theta_{23}$ is
essentially determined by SK.

\begin{figure}[ht]
\begin{center}
\includegraphics[width=8cm]{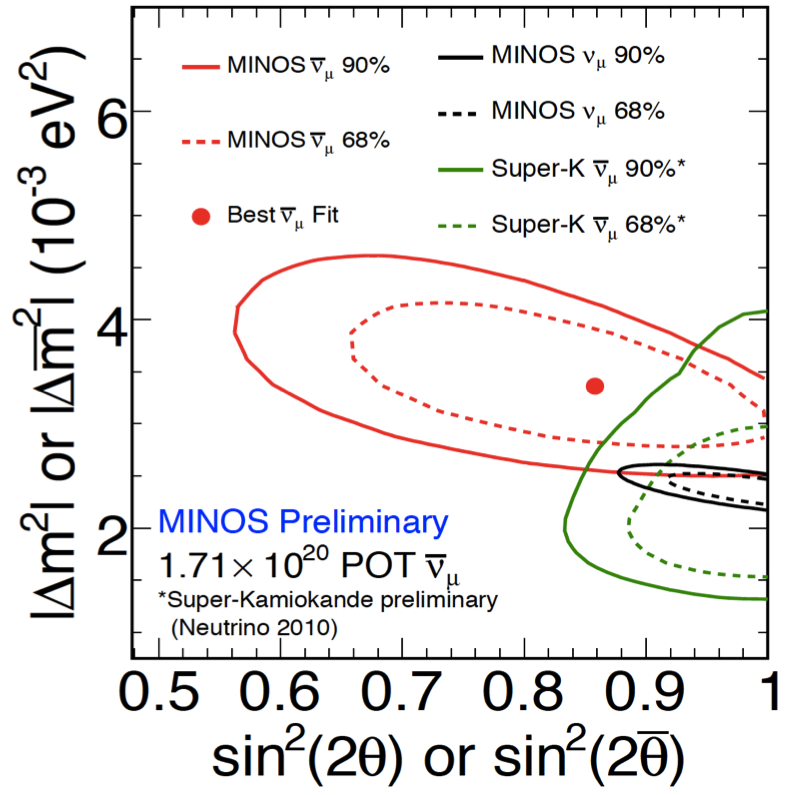}
\caption{Allowed regions in the oscillation parameter space for
  neutrino and antineutrino data (from Ref.~\cite{minos_anus}).}
\label{fig:minos_anus}
\end{center}
\end{figure}

A similar study to that discussed previously has been performed on the
antineutrino dataset~\cite{minos_anus}. MINOS is also able to distinguish between muon
neutrinos and antineutrinos. A total of $1.7 \times 10^{20}$ \ced{protons on target (POT)}\aq{Given in full. OK?}
were accumulated between September 2009 and March 2010. The
reconstructed energy spectrum of $\anumu$ CC events at the far
detector shows a deficit in the low-energy region. The best-fit
oscillation parameters to $\bar{\nu}$ data are shown in
Fig.~\ref{fig:minos_anus}. We see the corresponding contours for
neutrino and antineutrino oscillations. Antineutrinos favour a
slightly higher $\Delta m^2$ than neutrino data, which could violate CPT.
 Anyway, the results are compatible at 2$\sigma$ and more data are being
 taken to understand if this is a statistical fluctuation or
 not. Matter effects cannot explain this discrepancy (too small
 effect).

\begin{figure}[ht]
\begin{center}
 \includegraphics[width=0.4\linewidth]{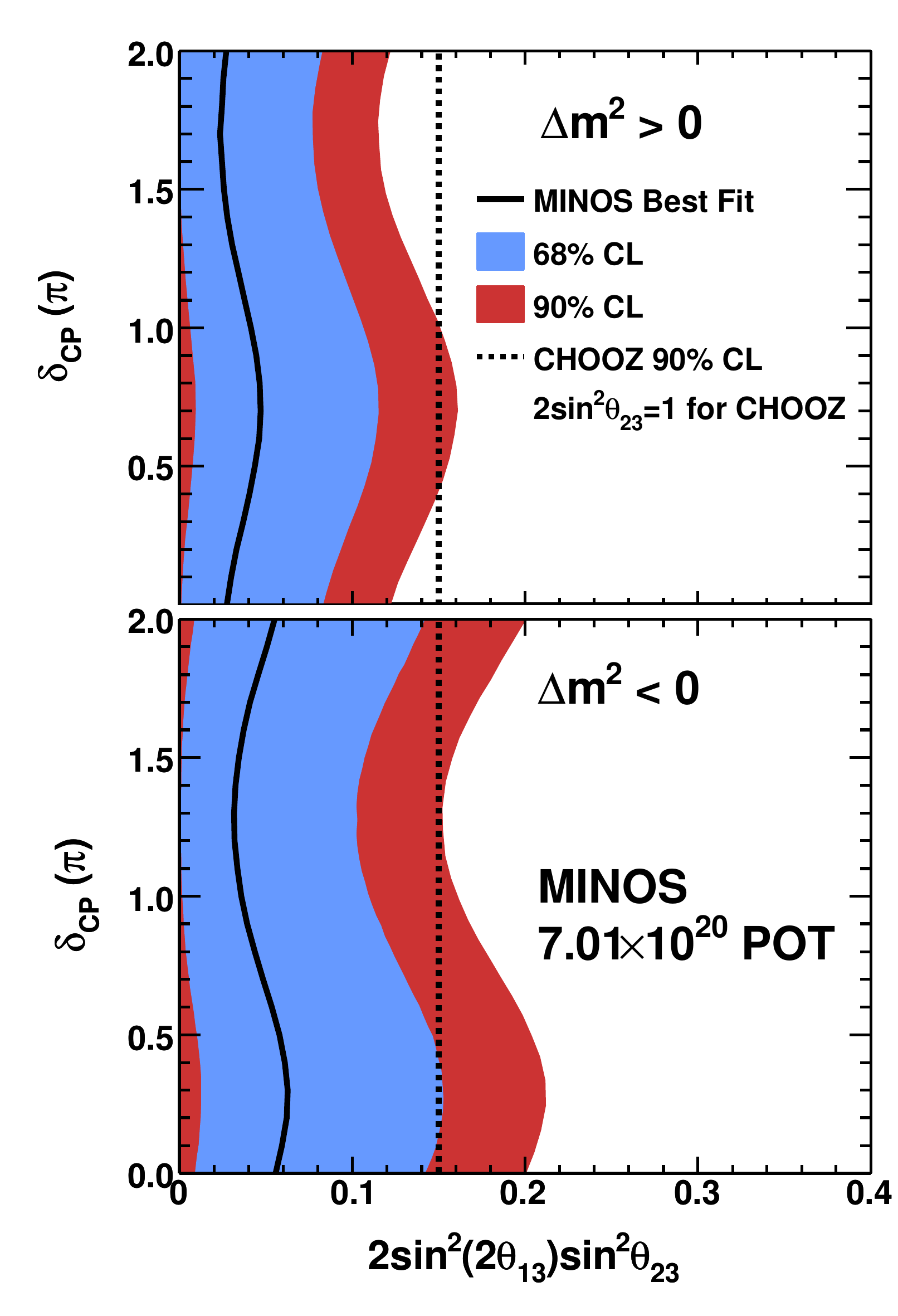}
\caption{Values of $2\sin^2(2\theta_{13})\sin^2\theta_{23}$ and
  $\delta_{\rm CP}$ that produce a number of candidate events in the far
  MINOS detector consistent with the observation for (top) the normal
  hierarchy and (bottom) the inverted hierarchy. Black lines are the
best fit and red (blue) regions show the 90\% (68\%) CL intervals
(from Ref.~\cite{minos_th13}).}
\label{fig:minos_th13}
\end{center}
\end{figure}

A subdominant transition $\numu \rightarrow \nue$ would be expected if
$\theta_{13} \neq 0$. MINOS was optimized for muon identification, and
thus the reconstruction of electromagnetic showers is difficult. They use an
artificial neural network technique for this analysis. Recent results
looking for $\nue$ appearance have shown a very small excess of data
(0.7$\sigma$ over the expected background)~\cite{minos_th13}. This
measurement is also consistent with no $\nue$ appearance. A limit has
been set around the CHOOZ value. Since MINOS is sensitive to matter effects,
they have different limits depending on the sign of $\Delta m^2$
(Fig.~\ref{fig:minos_th13}).

\begin{figure}[ht]
\begin{center}
\includegraphics[width=6cm]{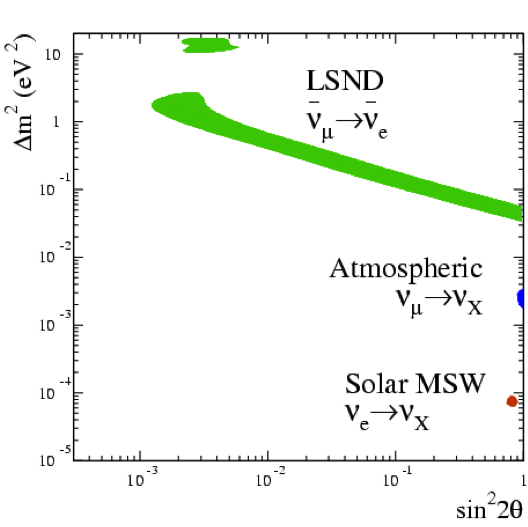}
\caption{Allowed regions in the parameter space including atmospheric,
  solar and LSND data.}
\label{fig:lnsd}
\end{center}
\end{figure}

Among the short-baseline accelerator neutrino experiments, LSND
is the first experiment that
claimed the observation of neutrino oscillation appearance~\cite{lsnd}. They were
taken data from 1993 to 1998 looking for the appearance of electron
antineutrinos in a muon antineutrino beam produced at Los Alamos
National Laboratory. The detector was a tank filled with 167~ton of
dilute liquid scintillator, located about 30~m from the neutrino
source. The experiment observed an excess of events above the MC
predictions (at 3.8$\sigma$) that could be interpreted in terms
of $\anumu \rightarrow \anue$ oscillations. The corresponding $\Delta
m^2$ is in the range shown in Fig.~\ref{fig:lnsd}.
These results created a huge controversy because they are not compatible
with atmospheric and solar oscillations since they cannot be explained
assuming three-flavour oscillations.

The region of parameter space that is favoured by the LSND
observations has been partly tested by other experiments like KARMEN
\cite{karmen} with negative results on neutrino oscillations. KARMEN
excluded part of the LSND region. Another experiment is needed to
definitively confirm this excess or not.

This is the case of the MiniBooNE experiment, designed to test the
neutrino oscillation interpretation of the LSND signal. This
experiment was proposed in 1997 and started running in 2002. MiniBooNE
\cite{miniboone} is an 800~ton mineral oil Cerenkov detector placed at
540~m from the neutrino source and uses the $\numu$ beam produced by the
Booster Neutrino Beamline at Fermilab. The $L/E$ baseline is similar to
the LSND, but the baseline and neutrino energies are one order of
magnitude higher. Therefore, MiniBooNE systematic errors are
completely different. They also have higher statistics and they are
taking data in both neutrino and antineutrino modes.

\begin{figure}[ht]
\begin{center}
\includegraphics[width=0.5\linewidth]{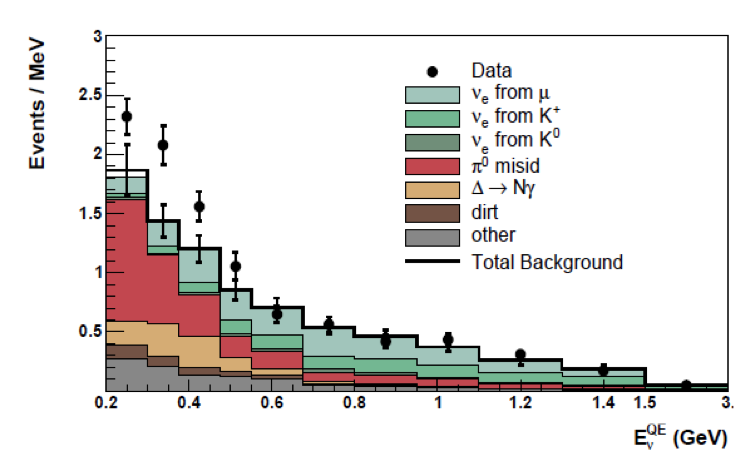}
\includegraphics[width=0.37\linewidth]{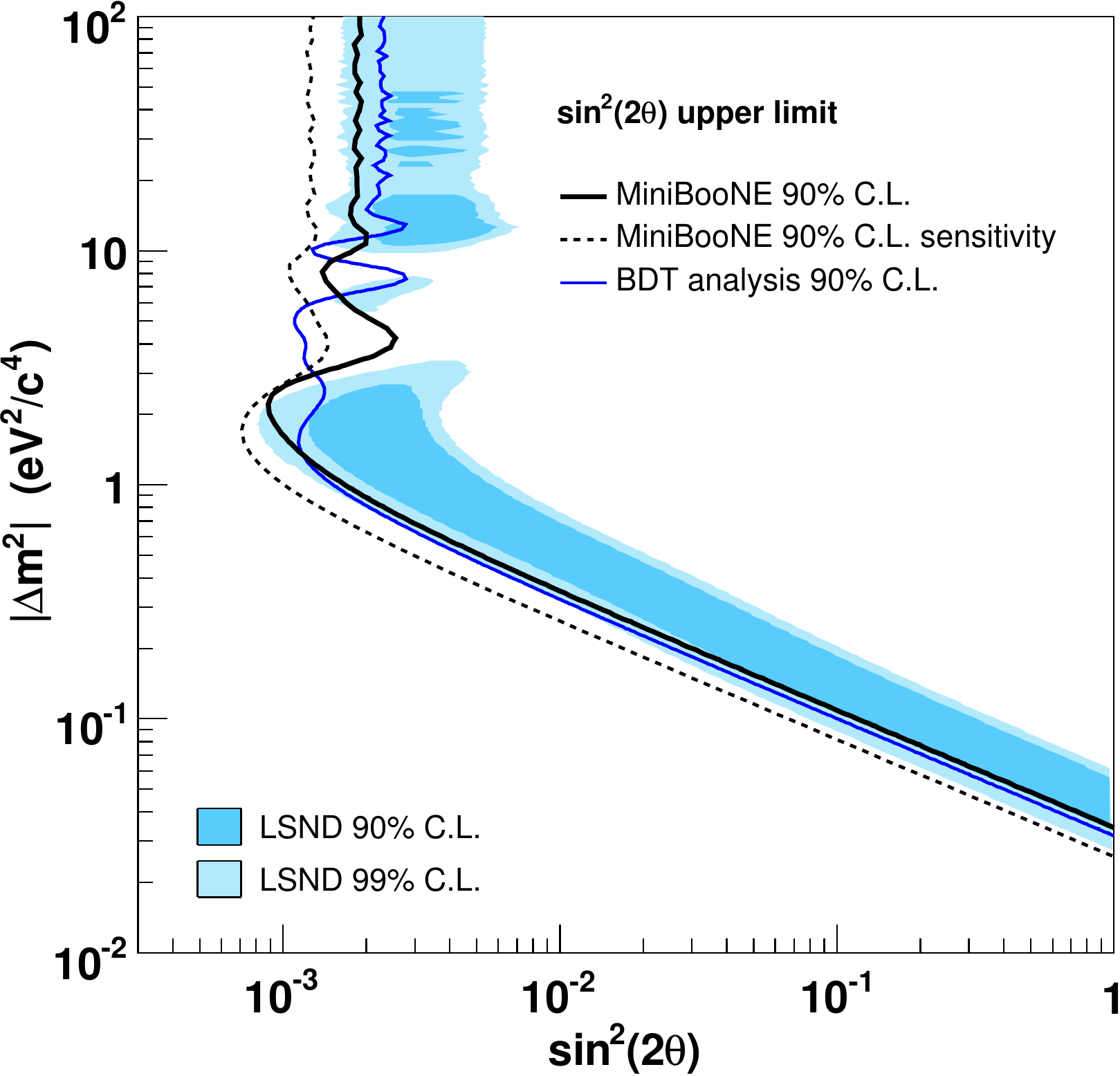}
\caption{(left) Neutrino energy distribution for $\nue$ CCQE data and
  background and (right) 90\% CL limit (thick curve) and sensitivity (dashed curve)
  for events with energy $> 475$~MeV within a two-neutrino
  $\numu \rightarrow \nue$ oscillation model (from Ref.~\cite{miniboone_nue}).}
\label{fig:miniboone_nue}
\end{center}
\end{figure}

Figure~\ref{fig:miniboone_nue} shows the MiniBooNE results for
$\numu$--$\nue$ oscillations in terms of the reconstructed energy
distribution of $\nue$ candidates~\cite{miniboone_nue}. Points are data
with the statistical error and the histogram is the background prediction
with systematic errors. For the analysis region between 475~MeV
and 1.25~GeV, there is no evidence of oscillations. Data are consistent
with background. MiniBooNE has excluded two neutrino oscillations in
the LSND region at 98\% CL. However, in the low-energy part of the
spectrum (between 200 and 475~MeV) they have found a sizeable excess of
data. The excess at low energy has a significance of 1.7$\sigma$ or
3.4$\sigma$ and is incompatible with LSND-type oscillations. The
source of this excess remains unknown.

\begin{figure}[ht]
\begin{center}
\includegraphics[width=0.47\linewidth]{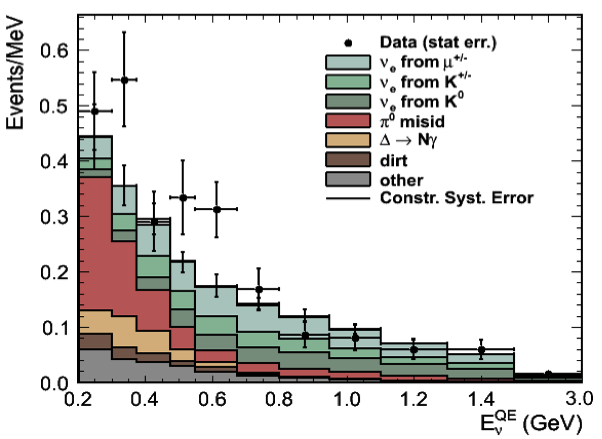}
\includegraphics[width=0.37\linewidth]{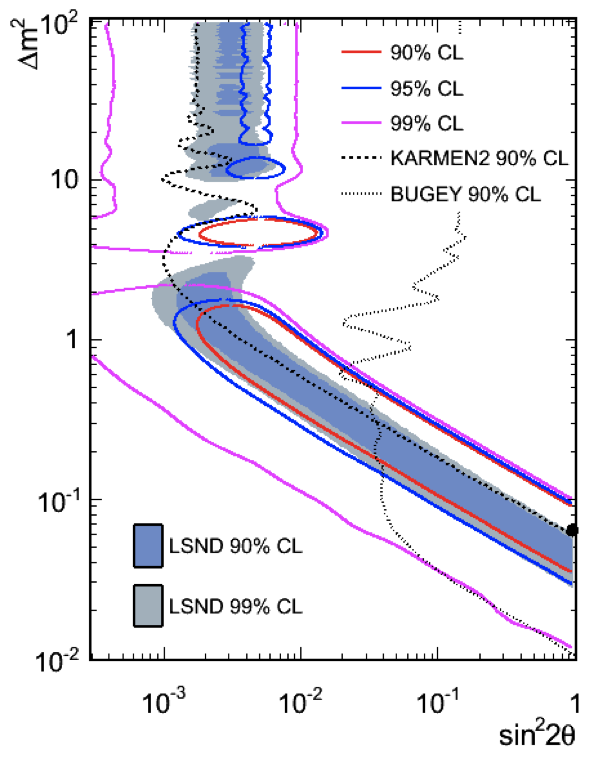}
\caption{(left) Neutrino energy distribution for $\anue$ CCQE data and
  background and (right) the 90, 95 and 99\% CL allowed regions
  for events with energy $> 475$~MeV within a two-neutrino
  $\anumu \rightarrow \anue$ oscillation model (from Ref.~\cite{miniboone_anue}).}
\label{fig:miniboone_anue}
\end{center}
\end{figure}

MiniBooNE has also reported results from the search for
$\anumu$--$\anue$ oscillations~\cite{miniboone_anue}. For the oscillation
study, no contribution from the low-energy neutrino mode excess has
been accounted for in the $\bar{\nu}$ prediction. In
Fig.~\ref{fig:miniboone_anue} the $\anue$ \ced{charged-current quasi-elastic (CCQE)}\aq{Given in full here. OK?} energy for data and
background events is shown. From 200 to 3000~MeV there is a
total excess of $43.2 \pm 22.5$ events. The excess is present in the low
($< 475$~MeV) and high ($> 475$~MeV) energy regions.

Many checks have been performed on the data to ensure that backgrounds
are correctly estimated. Any single background would have to be
increased by more than 3$\sigma$ to explain the observed excess of
events. On the right-hand plot of Fig.~\ref{fig:miniboone_anue} the 90, 95 and 99\% CL
contours for $\anumu \rightarrow \anue$ oscillations in the energy
range $> 475$~MeV are shown. The allowed regions are in agreement with LSND
allowed regions. The probability of background-only fit relative to
the best oscillation fit is 0.5\%. Comparison between MiniBooNE and
LSND as a function of $L/E$ also shows consistency between both results.

The source of the excess observed by MiniBooNE at low energy and the
difference found between neutrino and antineutrino results are still
under study.

CNGS~\cite{cngs} is the neutrino beam facility in Europe. It is mainly
a $\numu$ beam from the CERN SPS with a mean energy of $\sim$17~GeV, a 4\%
$\anumu$ contamination and a 0.9\% ($\nue + \anue$) contamination. Two
experiments, OPERA and ICARUS, are located in the LNGS laboratory in
Italy, 730~km away from the neutrino source.
Their main goal is to detect $\numu \rightarrow \nutau$ transitions in
appearance mode.
Physics operations started in 2007 and they have provided neutrino
beams in 2008, 2009 and 2010.

OPERA looks for $\nutau$ CC interactions through the
measurement of $\tau$ decay kinks in different channels.
The detector consists of a large set of emulsion--lead targets combined
with electronic detectors and a magnetic spectrometer. This technique
provides a very good spatial resolution of the order of the micrometres.

In August 2009 the OPERA collaboration presented the first
$\nutau$ neutrino candidate~\cite{opera}. With their statistics, they expected 0.5
$\nutau$ candidates. The statistical significance of the measurement of
a first $\nutau$ candidate is 2.36$\sigma$. For five years of data taking
at the nominal CERN performance of $4.5 \times 10^{19}$~POT, they
expect to detect 10~$\tau$ events.

The ICARUS T600 detector at LNGS~\cite{icarus} is a 600~ton LAr TPC providing 3D
imaging of any ionizing event. T600 is presently taking data and has
smoothly reached the optimal working conditions. They have already
observed neutrino interactions. The first data analysis is ongoing,
together with the development of fully automated reconstruction
software. They expect to measure one or two $\tau$ events in the next two
years.

\subsection{Global analysis of oscillation data}

In previous sections, all the current neutrino oscillation results
have been summarized. The three pieces of neutrino oscillation
evidence have been explained corresponding to three different values
of mass-squared differences. The mixing of three standard neutrinos
can only explain two of the anomalies. The explanation of the three
sets of data would require the existence of sterile $\nu$ species,
since only three light neutrinos can couple to the Z$^0$ boson.

In the case of the solar and atmospheric neutrino indications, several
experiments agree on the existence of the effect and they have been
confirmed by terrestrial reactor and accelerator
experiments. Therefore, the standard scenario is to consider
three-neutrino mixing without the LSND result. Several attempts have
been made in the literature to accommodate also LSND data (include a
fourth sterile neutrino, break CPT symmetry, make neutrinos and
antineutrinos have different masses). However, the present
phenomenological situation is that none of these explanations can
successfully describe all the neutrino data.

Table \ref{tab:global} summarizes the present values of the oscillation
parameters from a recent global three-flavour neutrino oscillation
analysis of the experimental data~\cite{schwetz}. The upper (lower)
row corresponds to normal (inverted) mass hierarchy.

\begin{table}[htbp]
\begin{center}
\caption{Neutrino oscillation parameter summary (from Ref.~\cite{schwetz}).}
\label{tab:global}
\begin{tabular}{cccc}
\hline\hline
{\bf Parameter} & {\bf Best fit $\pm 1\sigma$} & {\bf $2\sigma$} &
{\bf $3\sigma$} \\
\hline
\Rule{1.5em}
$\Delta m^2_{21}$ ($10^{-5}$~eV$^2$) & 7.59$^{+0.20}_{-0.18}$ &
7.24--7.99 & 7.09--8.19 \\
\Rule{2em}
$\Delta m^2_{32}$ ($10^{-3}$~eV$^2$) &
\begin{tabular}{c}
2.45 $\pm$ 0.09 \\
-- (2.34$^{+0.10}_{-0.09}$) \\
\end{tabular} &
\begin{tabular}{c}
2.28--2.64 \\
-- (2.17--2.54) \\
\end{tabular}
&
\begin{tabular}{c}
2.18--2.73 \\
-- (2.08--2.64) \\
\end{tabular}
\\
\Rule{2em}
$\sin^2\theta_{12}$ & 0.312$^{+0.017}_{-0.015}$ & 0.28--0.35 &
0.27--0.36 \\
\Rule{2em}
$\sin^2\theta_{23}$ &
\begin{tabular}{c}
0.51 $\pm$ 0.06 \\
0.52 $\pm$ 0.06 \\
\end{tabular} &
\begin{tabular}{c}
0.41--0.61 \\
0.42--0.61 \\
\end{tabular} &
0.39--0.64 \\
\Rule{2em}
$\sin^2\theta_{13}$ &
\begin{tabular}{c}
0.0$10^{+0.009}_{-0.006}$ \\[3pt]
0.013$^{+0.009}_{-0.007}$ \\
\end{tabular} &
\begin{tabular}{c}
$\leq$ 0.027 \\[3pt]
$\leq$ 0.031 \\
\end{tabular} &
\begin{tabular}{c}
$\leq$ 0.035 \\[3pt]
$\leq$ 0.039 \\
\end{tabular} \\
\vspace{-0.3cm} \\
\hline \hline
\end{tabular}
\end{center}
\end{table}


There are two possible mass orderings, which we denote as normal
($\Delta m^2_{32}$ > 0) and inverted ($\Delta m^2_{32}$ < 0). The two
orderings are often referred to in terms of sgn($\Delta m^2_{32}$).

As you may see, not all of the neutrino oscillation parameters have
been measured: the value of the $\theta_{13}$ angle, the sign of the
$\Delta m^2_{32}$ (mass hierarchy) and the CP violation phase are
still unknown.

Past and present experiments tried to measure the $\theta_{13}$ mixing angle
without success. We only have an upper limit on its value, indicating
that this angle must be very small. However, the best-fit point of
this parameter is not zero. There are independent hints for
$\theta_{13} > 0$ computed using different data ranging between
1.4$\sigma$ and 2.8$\sigma$.

\begin{figure}[ht]
\begin{center}
\includegraphics[width=10cm]{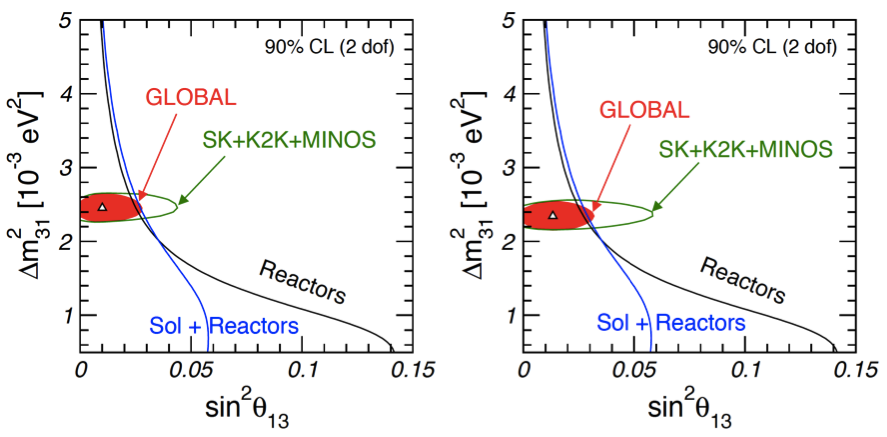}
\caption{Bound on $\sin^2\theta_{13}$ using global data, corresponding to (left) normal hierarchy and (right) inverted hierarchy (from Ref.~\cite{schwetz}).}
\label{fig:limit_th13}
\end{center}
\end{figure}

Figure~\ref{fig:limit_th13} illustrates the interplay of the various
datasets in the plane of $\sin^2\theta_{13}$ and $\Delta
m^2_{32}$. The latest T2K results have not been considered in this
analysis.

\section{Current and future neutrino oscillation experiments}

In the previous section I provided a summary of the present situation in
terms of experimental neutrino oscillation results and data
analysis. As pointed out, there are still many questions
to be answered by \ced{forthcoming neutrino oscillation experiments in the next
few years and further in the future}.

The main topics that will be addressed by the current and near-future
experiments are:
\begin{itemize}
\item
measurement of $\theta_{13}$ mixing angle;
\item
accurate measurements of other oscillation parameters ($\Delta
m^2_{32}$, $\theta_{23}$, is $\theta_{23}$ maximal?);
\item
understanding of LSND/MiniBooNE anomalies;
\item
understanding of differences observed between neutrinos and
antineutrinos in accelerator experiments;
\item
searching for CP violation in the leptonic sector; and
\item
searching for the sign of $\Delta m^2_{32}$.
\end{itemize}

Current experiments that will try to study these questions are
the accelerator experiments MINOS, OPERA, ICARUS and MiniBOONE, which
will continue their operation to accumulate more statistics, and the
new ones T2K in Japan and NO$\nu$A in the USA. Concerning reactor neutrinos,
there are essentially three new reactor experiments that would like to
measure $\theta_{13}$: Double Chooz in France is already taking data,
RENO in Korea is coming soon, and Daya Bay in China a bit later. In
addition, more news on SK and Borexino concerning natural sources will
be reported.

The first goal of the near-future experiments is to measure the
$\theta_{13}$ mixing angle. There are essentially two ways of studying
this parameter: with neutrino accelerator long-baseline experiments or
reactor experiments.

The long-baseline accelerator experiments will try to measure the
$\theta_{13}$ mixing angle by looking for the appearance of electron
neutrinos in a muon neutrino beam generated at great distance from the
detector.
The main problem to measure $\theta_{13}$ with accelerator experiments
is that the oscillation probability depends on several parameters in
such a way that the measurement of $\theta_{13}$ will be affected by
correlations and degeneracies between parameters (their sensitivity
will be reduced). Owing to the long baseline, they can also be sensitive
to matter effects.

However, reactor neutrino experiments are unique for providing an
unambiguous determination of $\theta_{13}$. The electron antineutrino
disappearance probability does not depend on the CP phase nor on
the sign of $\Delta m^2_{32}$. It depends essentially on $\theta_{13}$
(only has a weak dependence on $\Delta m^2_{21}$). Therefore, unlike appearance
experiments, they do not suffer from parameter degeneracies. Moreover,
the matter effects are negligible due to the small distances. So, they
will provide a clean measurement of the mixing angle.
In addition to this, they can help to solve the $\theta_{23}$
degeneracy (the octant of $\theta_{23}$ if not maximal, $\theta_{23}
> \pi/4$ or $< \pi/4$) combined with accelerator experiments.

 The experimental challenges of neutrino accelerator experiments are
 related to the neutrino beam intensity, the contamination of other
 flavours in the neutrino beam, the uncertainties on the neutrino flux
 properties and the neutrino--nucleus interactions. On the other hand,
 reactor neutrino experiments have a pure antineutrino flux without
 flavour contamination, the flux is known at few per cent level, and the cross-section is high, so the needed detectors are smaller (and
 cheaper compared to accelerator experiments). However, they need to
 deal with backgrounds and reduce the systematic uncertainties to
 provide a precise measurement. On the other hand, accelerator
 experiments are able to provide other measurements like CP-violation.
 Anyway, both kinds of experiments are necessary. They
 provide independent and complementary information.


\subsection{New reactor experiments}

The main goal of the new reactor experiments is to measure the
$\theta_{13}$ mixing angle. In order to achieve this, several
improvements with respect to previous reactor measurements are
needed. It will be necessary to increase the statistics. More powerful
reactors are desired, longer exposure and larger detector mass. On the
other hand, backgrounds should be further reduced with a better
detector design: using veto detectors and external shields against
muons and external radioactivity. Finally, an important reduction of
the systematic uncertainties is fundamental to reach the high
precision needed. It could be achieved by performing relative
measurements using two identical detectors and comparing them to
minimize the reactor errors. A detailed calibration programme will be
needed.

Reactor experiments will look for the disappearance of electron
antineutrinos coming from nuclear reactors. The corresponding
oscillation probability (Eq.~\eqref{eq:posc_react}) essentially depends
on $\Delta m^2_{32}$ and $\sin^2 2\theta_{13}$. The second term
corresponds to a second oscillation amplitude dominated by solar
parameters and has been measured with the KamLAND experiment:
\begin{equation}
P(\nu_\rme{\rightarrow}\nu_\rme)= 1 - \sin^2 2\theta_{13} \sin^2\left(\frac{\Delta
    m_{32}^2 L}{4 E_\nu}\right) - \cos^4\theta_{13} \sin^2
2\theta_{12} \sin^2\left(\frac{\Delta m_{21}^2 L}{4 E_\nu}\right).
\label{eq:posc_react}
\end{equation}

\begin{figure}[ht]
\begin{center}
\includegraphics[width=10cm]{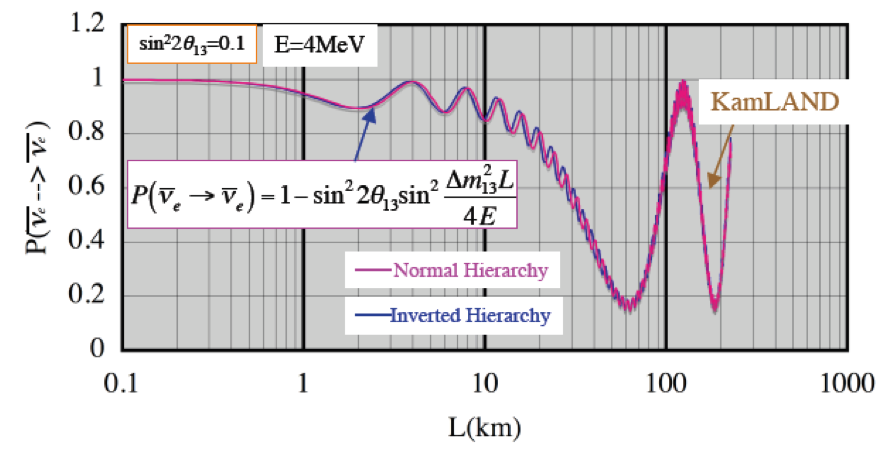}
\caption{Survival oscillation probability for a typical reactor
  neutrino experiment.}
\label{fig:oscprob_react}
\end{center}
\end{figure}

Figure~\ref{fig:oscprob_react} shows the survival probability as a
function of the distance of detection for a typical reactor neutrino
energy of 4~MeV. Owing to the low neutrino energies, reactor neutrino
experiments are disappearance experiments located at short distances
in order to maximize the disappearance probability. At $\sim$1--2~km from
the neutrino source, a small antineutrino deficit is expected over a
large neutrino flux. High precision will be necessary to measure the
mixing angle.

The reactor antineutrinos are detected through the inverse beta decay
reaction (Eq.~\eqref{eq:ibd}) giving a prompt signal due to the e$^+$ annihilation and a delayed signal from the neutron capture ($\sim$30~$\mu$s later) giving
photons of $\sim$8~MeV in the case of capture in Gd. In the case of H, the
delayed signal happens 200~$\mu$s later and the photons are of 2~MeV. The spectrum peaks around 3.6~MeV and neutrino energy threshold is 1.8~MeV.

The signature of a neutrino interaction can be mimicked by two types
of background events: accidental or correlated. All backgrounds are
linked to the cosmic muon rate and detector radiopurity. Compared to
CHOOZ, backgrounds can be reduced with a better detector design and
\textit{in situ} measurements.

The accidental events occur when a neutron-like event by chance falls
in the time window of $\sim$100~$\mu$s after an event in the
scintillator with an energy above 0.5--0.7~MeV. The positron-like
signal comes from natural radioactivity of the rock or of the detector
materials, in general, dominated by the PMT radioactivity. The delayed
background (neutron-like signal) comes from neutron captures on
Gd. They are energy deposits over 6~MeV isolated in time from other
deposits.

The correlated background are events that mimic both parts of the
coincidence signal: one single process induces both a fake positron
and a neutron signal. They come from fast neutrons induced by cosmic
muons, which slow down by scattering in the scintillator, deposit more
than 0.5~MeV visible energy and are captured on Gd. Correlated
background can also be produced by long-lived isotopes like $^8$He,
$^9$Li or $^{11}$Li, which undergo beta decay with neutron emission.

There are several reactor neutrino experiments that \ced{are looking to measure}
the $\theta_{13}$ angle with sensitivities on $\sin^2 2\theta_{13}$ up to 0.01:
Double Chooz in France, RENO in Korea and Daya Bay in China. Table
\ref{tab:reactors} summarizes the three reactor neutrino experiments
in progress.\aq{I've assumed ``y'' stands for ``year'', for which the preferred abbreviation is ``yr''. If that is not the case please explain what it is.}
Double Chooz~\cite{dchooz} is the most advanced of the three reactor experiments,
since it is already taking data. The three detectors are quite similar
with slight variations between them.

\begin{table}[htbp]
\begin{center}
\caption{Comparison between reactor neutrino experiments.}
\label{tab:reactors}
\begin{tabular}{ccccccc}
\hline\hline
{\bf Experiment} & {\bf Location} & {\bf Th. power} & {\bf Distances } & {\bf Depth near} & {\bf Target} & {\bf Expect.} \\
 & & {\bf (GW)} & {\bf near/far (m)} & {\bf /far (mwe)} & {\bf
   mass (ton)} & {\bf sensit. (3 yr)} \\
\hline
Double Chooz & France & 8.5 & 400/1050 & 115/300 & 10/10 & 0.03 \\
RENO & Korea & 16.4 & 290/1380 & 120/450 & 15/15 & 0.02 \\
Daya Bay & China & 11.6 & 360(500)/ & 260/910 &
40$\times$2/80 & 0.01 \\
& & (17.4) & 1985(1613) &  &  &  \\
\hline\hline
\end{tabular}
\end{center}
\end{table}

The antineutrinos used in Double Chooz are produced by the pair of
reactors (type N4) located at the Chooz-B nuclear power station in
France. The maximum operating thermal power of each core amounts to
4.27~GW. The idea of Double Chooz is to use two almost identical
neutrino detectors of medium size, containing 10.3~m$^3$ of liquid
scintillator target doped with 0.1\% of gadolinium. The neutrino
laboratory of the first CHOOZ experiment is located 1.050~km from
the two cores. The far detector is already installed at this site. The
far site is shielded by about 300~mwe of rocks. In order to cancel the
systematic errors originating from the lack of knowledge of the
$\anue$ flux and spectrum, as well as to reduce the set
of systematic errors related to the detector and event selection
procedure, a second detector will be installed close to the nuclear
cores, at $\sim$400~m.

\begin{figure}[ht]
\begin{center}
\includegraphics[width=6cm]{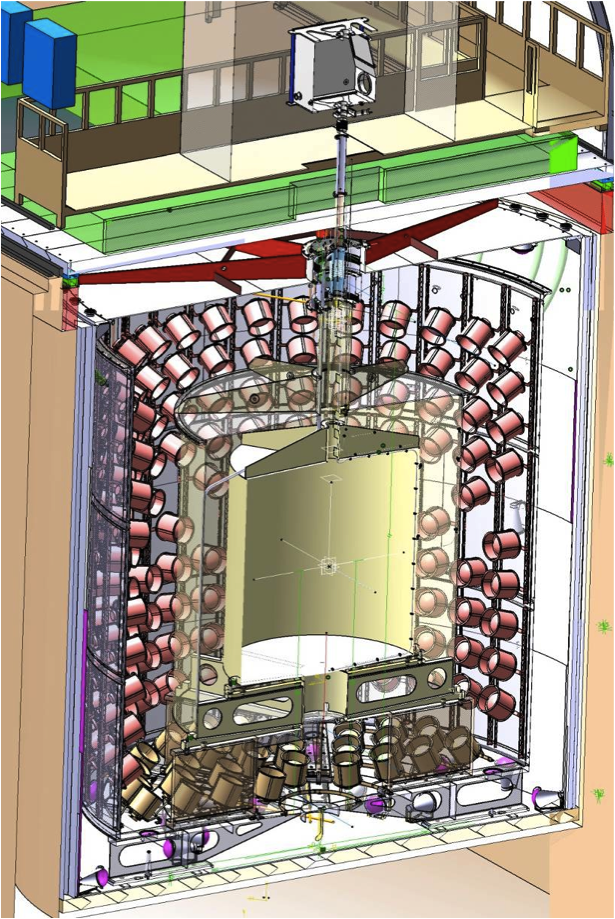}
\caption{Sketch of the Double Chooz detector.}
\label{fig:DC_detect}
\end{center}
\end{figure}

The Double Chooz detector consists of concentric cylinders
(Fig.~\ref{fig:DC_detect}). A target cylinder of 1.2~m radius and 2.5~m
height, providing a volume of 10.3~m$^3$, is filled
with a liquid scintillator doped with gadolinium (1~g/l). This is the
volume for neutrino interactions. Surrounding the target we have the
gamma-catcher region of 22.6~m$^3$ containing non-loaded liquid
scintillator with the same optical properties as the $\anue$ target
(light yield, attenuation length). This is an extra volume for gamma
interaction. This region is needed to measure the gammas from the
neutron capture on Gd, to measure the positron annihilation and to
reject the background from fast neutrons. Surrounding the
gamma-catcher acrylic tank there is a 1~m thick non-scintillating
(oil) buffer contained in a stainless-steel tank. The goal of this
region is to decrease the level of accidental background mainly from
the contribution from PMT radioactivity. The photomultiplier tubes are
mounted from the interior surface of the buffer vessel and they
collect the light from the target volume and the gamma-catcher. They
are 390 10-inch PMTs per detector to cover $\sim$13\%. Then a 50~cm thick
inner veto region is filled with liquid scintillator to tag the muon-related
background events. Finally, a 15~cm thick steel shielding will
protect the detector from natural radioactivity of the rocks around
the pit with a significant gamma reduction. An additional muon outer
veto (plastic scintillator planes) will be required to help identify
muons, which could cause neutrons or other cosmogenic backgrounds.

The statistical error in CHOOZ was 2.8\% while in Double Chooz in
three years it is expected to be $\sim$0.5\%. The fiducial volume has been
increased with respect to CHOOZ and longer data taking is
expected. Concerning the systematic errors, in CHOOZ the total
systematic error was 2.7\%, dominated by the reactor antineutrino flux
and spectrum uncertainties (1.9\%). In Double Chooz the uncertainty
related to the reactor is cancelled by using two identical
detectors. The relative normalization between the two detectors is the
most important source of error. The goal of Double Chooz is to reduce
the overall systematic uncertainty to 0.6\%. Table \ref{tab:DC_sys}
summarizes the expected systematic errors in Double Chooz.

\begin{table}[h]
\begin{center}
\caption{Summary of Double Chooz systematic errors.}
\label{tab:DC_sys}
\begin{tabular}{lcc}
\hline\hline
 & \textbf{CHOOZ} & \textbf{Double Chooz}\\
\hline
Reactor uncertainties  & 2.1\% & < 0.1\% \\
($\nu$ flux and reactor power) &  & \\
Number of protons & 0.8\% & < 0.2\% \\
Detector efficiency & 1.5\% & < 0.5\% \\
{\bf Systematic error} & {\bf 2.7\%} & {\bf < 0.5\%} \\
\hline\hline
\end{tabular}
\end{center}
\end{table}

\begin{figure}[ht]
\begin{center}
\includegraphics[width=7cm]{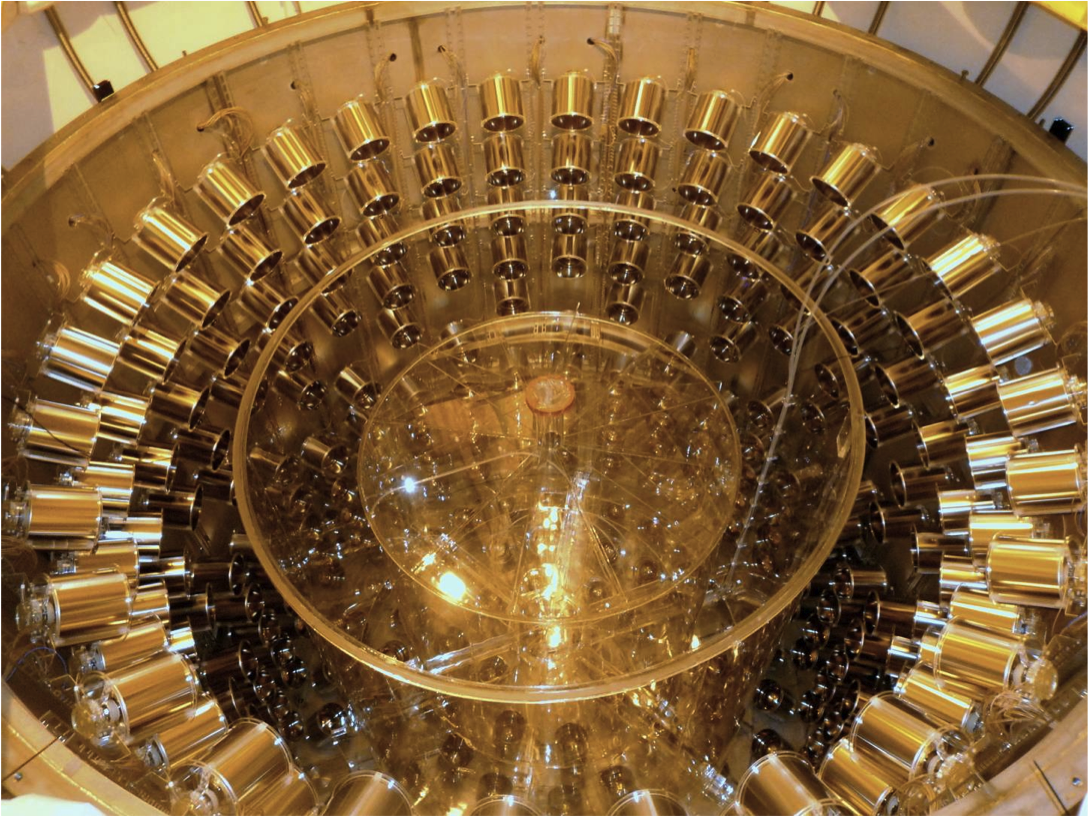}
\includegraphics[width=5cm]{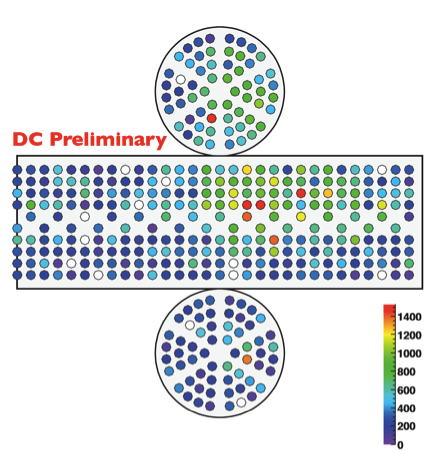}
\includegraphics[width=3.5cm]{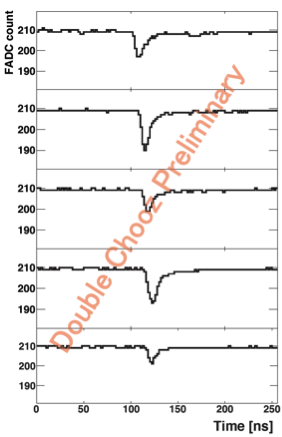}
\caption{(left) Internal view of the Double Chooz detector with the PMTs
  installed and (right) one event of a few photoelectrons contained in
  the inner detector.}
\label{fig:first_DC}
\end{center}
\end{figure}

The Double Chooz far detector started to take data at the end of
2010. Figure~\ref{fig:first_DC} shows an internal view of the detector with all
PMTs installed inside the buffer volume and, on the right, the first
signals of a few photoelectrons contained in the inner detector.

\begin{figure}[ht]
\begin{center}
\includegraphics[width=8cm]{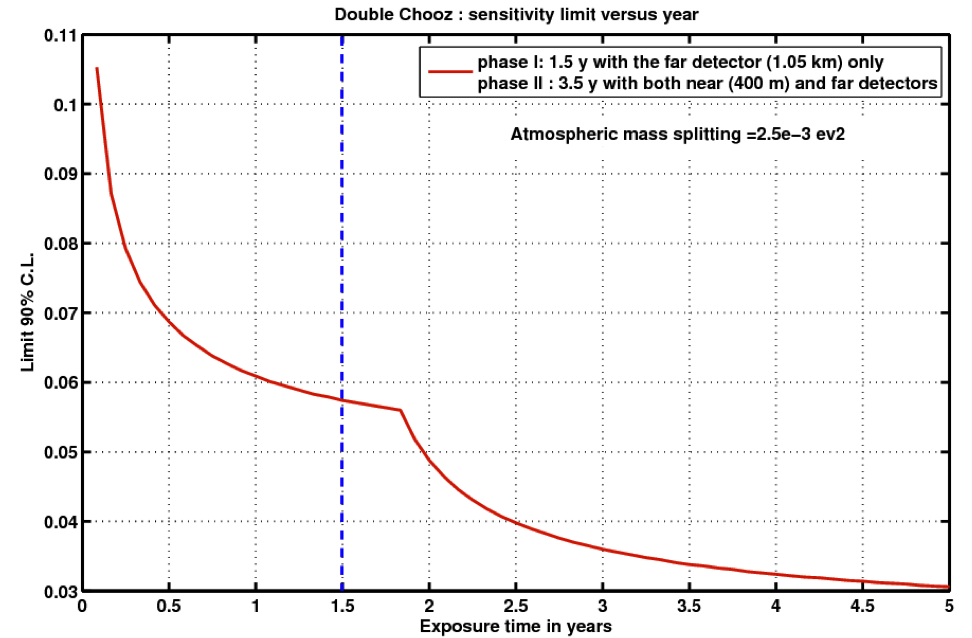}
\caption{Expected Double Chooz sensitivity on $\sin^2 2\theta_{13}$.}
\label{fig:dc_sensi}
\end{center}
\end{figure}

The expected Double Chooz $\sin^2 2\theta_{13}$ sensitivity as a
function of time is shown in Fig.~\ref{fig:dc_sensi} in the case when no
signal is observed. Double Chooz will operate in two phases. In the
first one, after 1.5 years of data taking with the far detector, a
limit of $\sin^2 2\theta_{13} < 0.06$ at 90\% CL can be reached. Using both
detectors, it is possible to measure $\sin^2 2\theta_{13}$ up to 0.05 at
3$\sigma$ or to obtain a limit down to 0.03 at 90\% CL after three
years of data taking.

The RENO~\cite{reno} reactor neutrino experiment is under construction at
YongGwang in South Korea. The plant consists of six equally spaced
reactors in line spanning $\sim$1.3~km. The total average thermal
power is $\sim$16.4~GWth. One near detector and one far detector are
placed in the iso-flux line from the reactors. The near detector is
located $\sim$290~m from the cores' barycentre with an overburden
$\sim$120~mwe. The far detector is at 1380~m surrounded by 450~mwe.
The design of the RENO detectors is quite similar to the Double
Chooz one. The inner detector is bigger (16~ton) and the main
difference is the muon veto system, which is a 30~cm concrete vessel
filled with water observed by 60~PMTs.

The goal of the experiment is to have a systematic error of the order
of 0.5\% and a statistical error $\sim$0.3\%. The expected limit is $<0.02$
and the discovery reach at 3$\sigma$ up to 0.04 after three years of
data taking. Both detectors are being commissioned. They expect to
start data taking with the two detectors at the end of 2011.

A third reactor neutrino experiment, Daya Bay~\cite{dayabay}, is under construction
in China, at the Ling Ao and Daya Bay nuclear power plants. The power
plant complex is now composed of two pairs of reactors, Daya Bay and
Ling Ao-I. Other two reactors named Ling Ao-II are under construction
and should be operational in 2011. Each core yields 2.9~GW, thus the
site is 11.6~GW and will be 17.4~GW. Near detectors are needed in near
sites to monitor the different reactors. Two detectors will be
installed at $\sim$360~m from Daya Bay, two detectors at $\sim$500~m
from Ling Ao, and four far detectors at 1.6~km from the barycentre of
Ling Ao sites and $\sim$2~km from Daya Bay. Each detector contains 20~ton
of liquid scintillator doped with Gd. Horizontal tunnels connect
the detector halls for cross-calibration. The design of the detector
is very similar to the Double Chooz one except for the shielding. The
detectors of each site are submerged into a swimming pool filled with
purified water giving protection against radiation and fast
neutrons. The water pool is instrumented with PMTs to read the
Cerenkov light to tag muons together with \ced{resistive plate chambers
(RPCs)}\aq{Given in full OK?} placed on top of
the water pool. This system is under production. The excavation of
access tunnels and experimental halls is nearing completion. Two near
detectors are completed. The expected systematic error is 0.38\%. They have
the ambitious idea of swapping the detectors of different sites, moving
them through the tunnels to reduce the systematic errors from relative
detector normalization to 0.12\%. The expected sensitivity at 90\%
CL with three years of data taking (assuming a systematic error of
0.4\%) is 0.01. Daya Bay plans to start taking data with the first
near site in summer 2011 and with the three sites operational at the
end of 2012.

\subsection{New accelerator experiments}

Complementary to reactors, new accelerator experiments will measure
neutrino oscillations in the next few years. Their main goal is to look for
$\nue$ appearance in a muon neutrino beam. The approximate formula for
the oscillation probability can be written as
\begin{eqnarray}
P{(\nu_\rme{\rightarrow}\nu_\mu )}  & \approx & s_{23}^2
\sin^22\theta_{13} \sin^2\!\left(\frac{\Delta m^2_{32}L}{4E_\nu}\right ) + P_{\text{sol}}(\theta_{12}, \Delta m^2_{21}) \nonumber \\
&&{} \pm \sin2\theta_{13} {F}_{\text{solar}} {F}(\sin2\theta_{23}, |\Delta
m^2_{32}|) {F}(\delta_\text{CP}, \Delta m^2_{32}).
\end{eqnarray}
The first term corresponds to atmospheric oscillations, the second one
is the solar one and there is an interference term, which has the
information on the  $\delta_{\rm CP}$ phase and also dependence on the sign
of $\Delta m^2_{32}$. The +~($-$) sign applies to neutrinos
(antineutrinos), respectively.

Accelerator experiments will try to measure the $\theta_{13}$ mixing
angle, provide more precise measurements of the atmospheric parameters
and in principle look for CP-violation,
\begin{equation}
P(\nu_{\alpha}{\rightarrow}\nu_{\beta})-P(\bar{\nu}_{\alpha}{\rightarrow}\bar{\nu}_{\beta}) \neq 0 \qquad\quad (\alpha \neq \beta),
\end{equation}
and matter effects.

There are two effects, one  from the CP phase and another from matter effects, that produce differences between neutrinos
and antineutrinos. We need to disentangle these two effects using
different experimental set-ups. At short distances, CP-violating
effects dominate, while at long distances, matter effects completely
hide CP-violating effects. They can be distinguished by the
different neutrino energy dependence. In order to achieve this, an
improvement of the present beams is needed: with much higher
intensities and almost monochromatic beams.

New detectors at accelerators are located off-axis in order to reduce
the beam energy and have a more monochromatic beam. This technique
allows experiments to pick the energy corresponding to the maximum
oscillation signal and, at the same time, to get rid of the high-energy
part \ced{contributing most of the background}. The $\nue$ contamination
from the beam could be reduced below the 1\% level.

\begin{figure}[ht]
\begin{center}
\includegraphics[width=6cm]{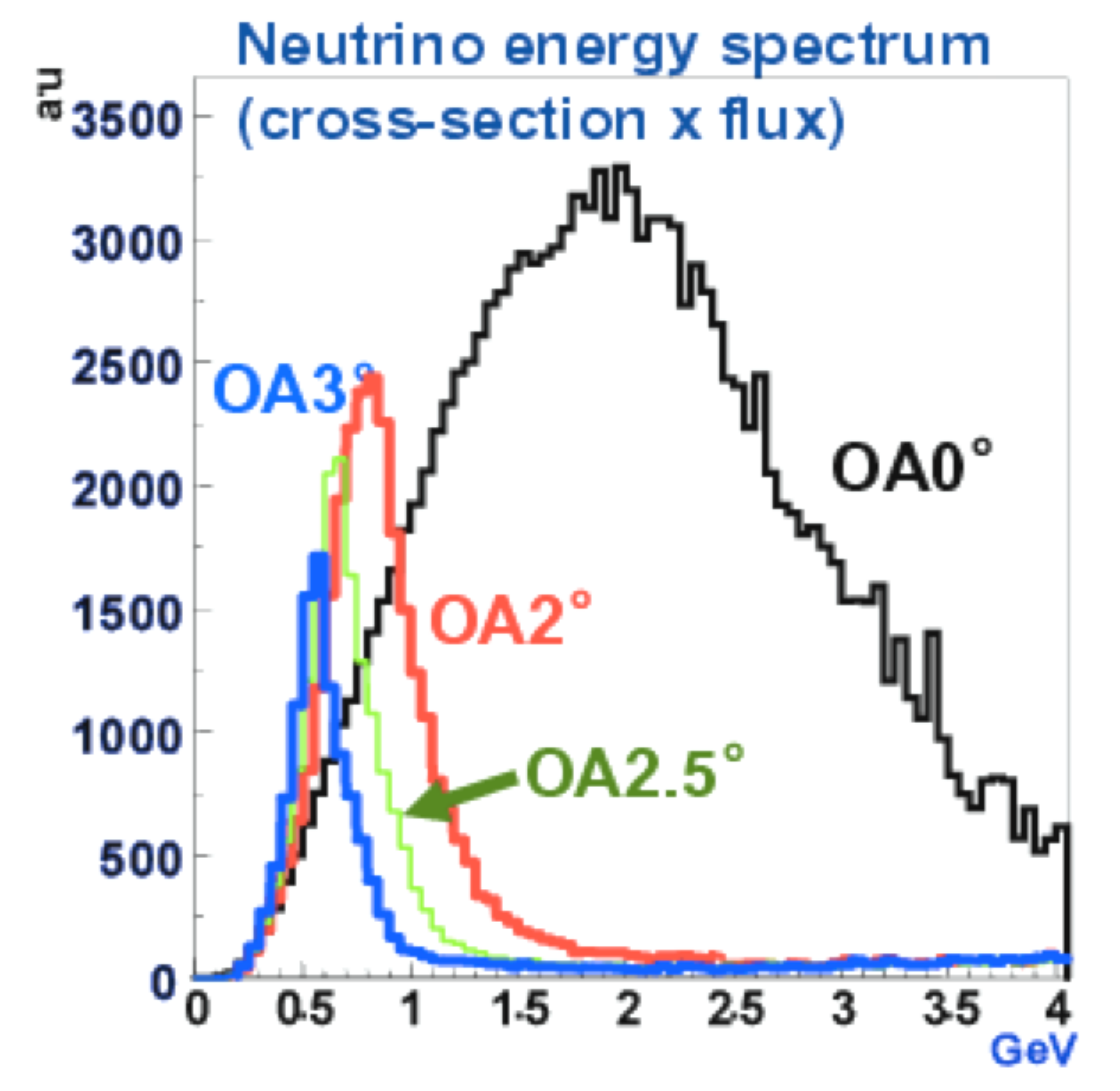}
\caption{Neutrino energy spectrum variation as a function of the
  off-axis angle.}
\label{fig:off-axis}
\end{center}
\end{figure}

In Fig.~\ref{fig:off-axis} the neutrino energy spectrum is shown for
different off-axis degrees. The energy peak is reduced and becomes
narrower by increasing the off-axis \ced{angle}. In addition, the contamination of
$\nue$ from the beam is greatly reduced. The problem with this technique
is the reduced rate. Thus, large detectors and \ced{intense}\aq{I don't think ``high'' was quite the right word here. Is ``intense'' OK?} proton sources are
needed.

In accelerator experiments, the main neutrino signal will be CCQE
interactions. They will look for the muon or electrons coming from
these reactions. They have to deal with backgrounds coming essentially
from the $\nue$ contamination of the beam and from $\pi^0$ production in
neutral currents.

The main long-baseline project that has begun operation this
year is T2K~\cite{t2k}. The neutrino beam is produced in the
accelerator complex of J-PARC in Japan and it will travel 295~km to
Kamioka, where the SK detector is located. They will use near and far
detectors to control the beam and measure the oscillations. Both
detectors are 2.5$^\degree$ off neutrino beam axis. This gives a
neutrino beam energy of $\sim$600~MeV. Owing to the short distance, this
experiment is not sensitive to matter effects but can provide
information on CP if $\theta_{13}$ is not too small.

T2K will have different detectors along the beam line. A muon monitor
is located after the beam dump and measures the direction and intensity of
the beam. It is used as a proton beam detector, target monitor and horn
monitor. The INGRID on-axis detector, at 280~m from the target, is made of steel
and scintillator layers and measures the intensity and direction of
the neutrino beam. It monitors the beam using muons from CC neutrino
interactions. The ND280 off-axis near detector is a magnetized detector inside
the former UA1 magnet donated by CERN to this experiment (0.2~T). It
is composed of several subdetectors: a $\pi^0$ detector, a tracker
made of fine-grain detectors and TPCs to detect charged particles and
measure their momentum, and an electromagnetic calorimeter. The Side
Muon Range Detector will detect muons and measure their momenta. This
detector measures the neutrino flux and spectrum before oscillations,
different interaction cross-sections and also the backgrounds for
$\nue$ appearance. Finally, the SK detector expects 10 $\numu$ events per
day at full beam power.

T2K completed its first run in the first half of 2010. A total of $3.23
\times 10^{19}$ protons were delivered at 30~GeV. The beam was
working at 50~kW of power. T2K has analysed both $\numu$ and
$\nue$ samples. For the $\numu$ disappearance analysis, $\numu$ is
consistent with previous disappearance experiments. In the appearance
$\nue$ channel, they have observed six $\nue$ candidates, and the expected
number of events in a three-flavour neutrino oscillation scenario with
$|\Delta m^2_{32}| = 2.4 \times 10^{-3}$~eV$^2$,
$\sin^22\theta_{23} = 1$ and $\sin^2 2\theta_{13} = 0$ was $1.5 \pm
0.3$ (syst.)~\cite{t2k_nue}. At 90\% CL, the data are consistent with
$0.03~(0.04) < \sin^2 2\theta_{13} < 0.28~(0.34)$ for $\delta_{\rm CP} = 0$
and normal (inverted) hierarchy.

Their goal for 2011 was to accumulate 150~kW $\times$ $10^7$~s by
July and increase the beam power. However, due to the March 2011
earthquake, the experiment has been somewhat delayed. More data are required to
firmly establish $\nue$ appearance and to determine the $\theta_{13}$
angle.

\begin{figure}[ht]
\begin{center}
\includegraphics[width=7cm]{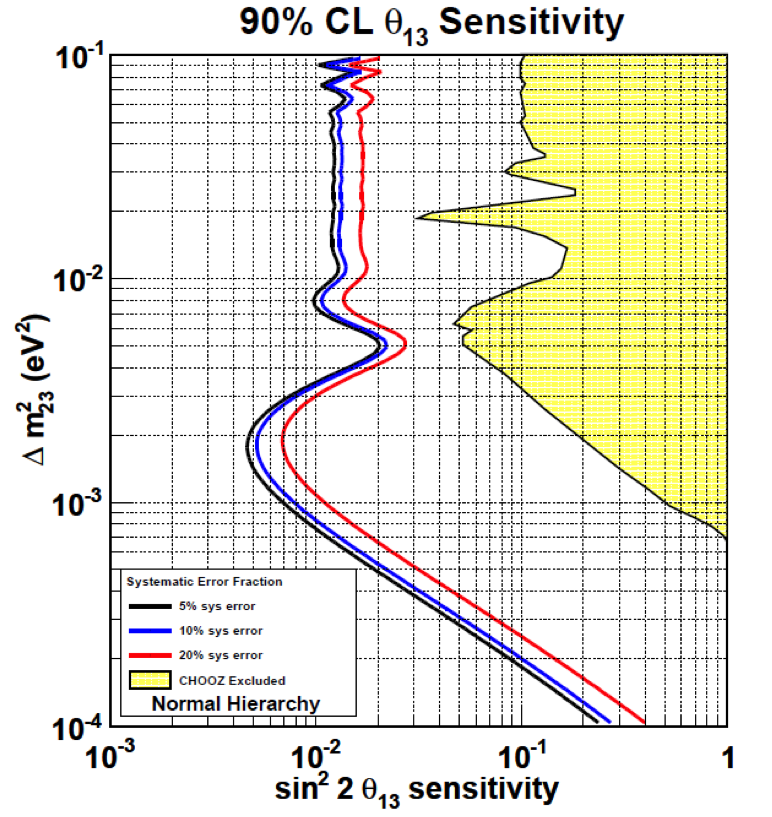}
\caption{Expected sensitivity of the T2K experiment to
  $\sin^2 2\theta_{13}$ for five years of data taking with a 750~kW beam
  assuming $\delta_{\rm CP}=0$ and normal mass hierarchy (from
  Ref.~\cite{t2k_sensit}).}
\label{fig:t2k_sensit}
\end{center}
\end{figure}

Assuming the beam running at 750~kW for five years,
Fig.~\ref{fig:t2k_sensit} shows the expected sensitivity that T2K plans
to reach for $\sin^2 2\theta_{13}$ as a function of $\Delta m^2_{32}$ at
90\% CL and for different systematic errors, assuming $\delta_{\rm CP} =
0$. They could be sensitive down to 0.01 at 90\% CL. The final
sensitivity will depend on the value of $\delta_{\rm CP}$.

There is another approved experiment that will start its operation in
2013: the NO$\nu$A experiment (NUMI Off-axis Neutrino
Appearance)~\cite{nova}. They will search for $\nue$ appearance using an upgraded
version of the NUMI beam at 700~kW with two identical detectors: a 220~ton near
detector located close to the source at 1~km and a 15~ton far detector
at 810~km away at Ash River, Minnesota, USA. They will \ced{use} active tracking
liquid scintillator calorimeters with very good electron
identification capability.

The unique feature of this experiment is that, depending on
the value of $\theta_{13}$, NO$\nu$A could be the only approved
experiment with sensitivity to determine the neutrino mass
hierarchy. The detectors will be placed off-axis at 0.8$^\degree$
to tune the neutrino energy to 2~GeV and maximize the
$\nue$ appearance. They can run in the neutrino and antineutrino
modes. The NuMI beam will be upgraded from 320~kW to 700~kW during the
shutdown of 2012.

\begin{figure}[ht]
\begin{center}
\includegraphics[width=7cm]{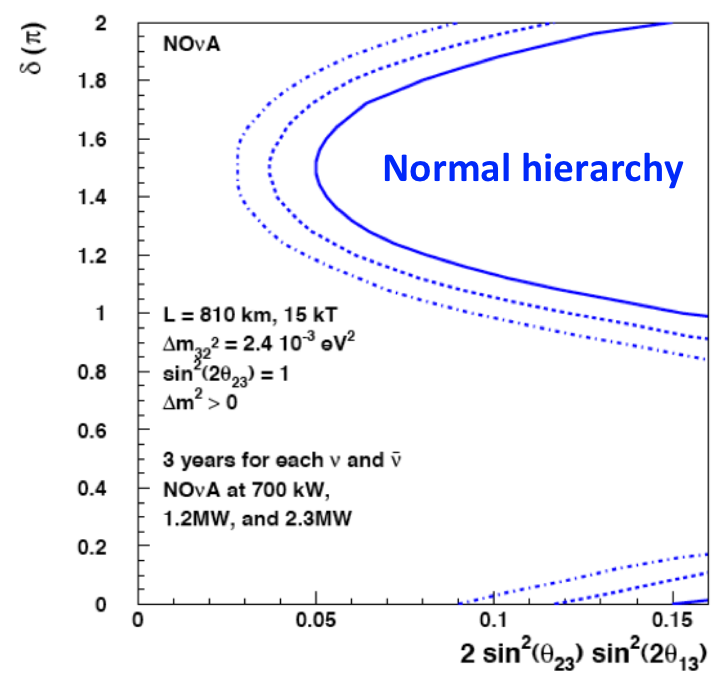}
\caption{The 95\% resolution of the mass ordering as a function of
  $\delta_{\rm CP}$ for six years of NO$\nu$A running split evenly between
  neutrinos and antineutrinos for different beam powers in the case
  of normal mass ordering (from Ref.~\cite{nova}).}
\label{fig:nova}
\end{center}
\end{figure}

NO$\nu$A plans to run for three years in neutrino mode and three years in
antineutrino mode. They plan to take advantage of the large matter
effects.
Figure~\ref{fig:nova} shows the NO$\nu$A sensitivity to matter effects
depending on the $\delta_{\rm CP}$ value. The mass ordering can only be
solved by NO$\nu$A alone in this region of the parameter space, if the
hierarchy is normal. For the rest of the parameter space, we need to
combine these measurements with other experiments, like T2K, which
will improve the sensitivity a bit. NO$\nu$A plans to have the far
detector completed in October 2013.

In the next few years, it is possible that these experiments will provide a
measurement of $\theta_{13}$ mixing angle, if $\sin^2 2\theta_{13} >
0.01$ and solve the $\theta_{23}$ degeneracy. However, they will have
limited sensitivity to CP-violation and matter effects. More than 70\%
of the parameter space will not be accessible.

The ultimate goals of the future generation will depend on these
measurements, but in principle they will focus on CP-violation (new
measurements are needed to solve degeneracies) and on the mass
hierarchy. To achieve these measurements, many improvements are needed
from the experimental point of view: We will need upgraded beams that are more
energetic, more powerful and more pure. We will need huge detectors (one
order of magnitude bigger), with more granularity and energy
resolution. And, to solve the degeneracies, we will need different
energies, baselines (longer baselines to enhance matter effects)
and detection channels.

New facilities and experiments are being proposed that can realize
some (or all) of the pending issues:

\begin{itemize}
\item[(a)]
{\it Superbeams} are more powerful versions of conventional pion decay-based
beams. They could be obtained with new megawatt proton sources. They
will need to be coupled with huge detectors at longer distances to
explore matter effects. In these facilities, the main
beam consists of $\numu$ and the experiments will search for both
$\numu$ disappearance and $\nue$ appearance. Several possibilities are
under study: a CERN upgraded beam to large detectors located in
European underground laboratories (LAGUNA), a new beamline from an
upgraded accelerator complex (2.3~MW beam power) to be sent to a
large detector located in the DUSEL underground laboratory (1300~km),
and an upgraded version of the J-PARC beam (1.66~MW) to T2HK in Japan
or another detector in Korea.
\item[(b)]
{\it Beta-beams} are very pure $\nue$ or $\anue$ beams made
by allowing accelerated radioactive ions to decay in a storage
ring. Both $\nue$ disappearance and $\numu$ appearance are sought.
However, $\numu$ disappearance cannot be studied.
\item[(c)]
{\it Neutrino factories} are facilities where muons are produced by pion
decay, cooled, injected into a storage ring and allowed to decay in
straight sections. This provides a very clean $\numu$ and $\anue$ beams
(or vice versa) with well-known energy spectrum. The dominant search is
the appearance of ``wrong-sign'' muons from the oscillation of
$\anue$. Other oscillation channels can also be observed. They will need
detectors with capability to distinguish between $\mu^+$ and
$\mu^-$.
\end{itemize}
\noindent
Figure~{\ref{fig:mezzetto} (from Ref.~\cite{mezzetto}) compares the
  $\sin^2 2\theta_{13}$ discovery reach at 3$\sigma$ for different
  future facilities.

\begin{figure}[ht]
\begin{center}
\includegraphics[width=9cm]{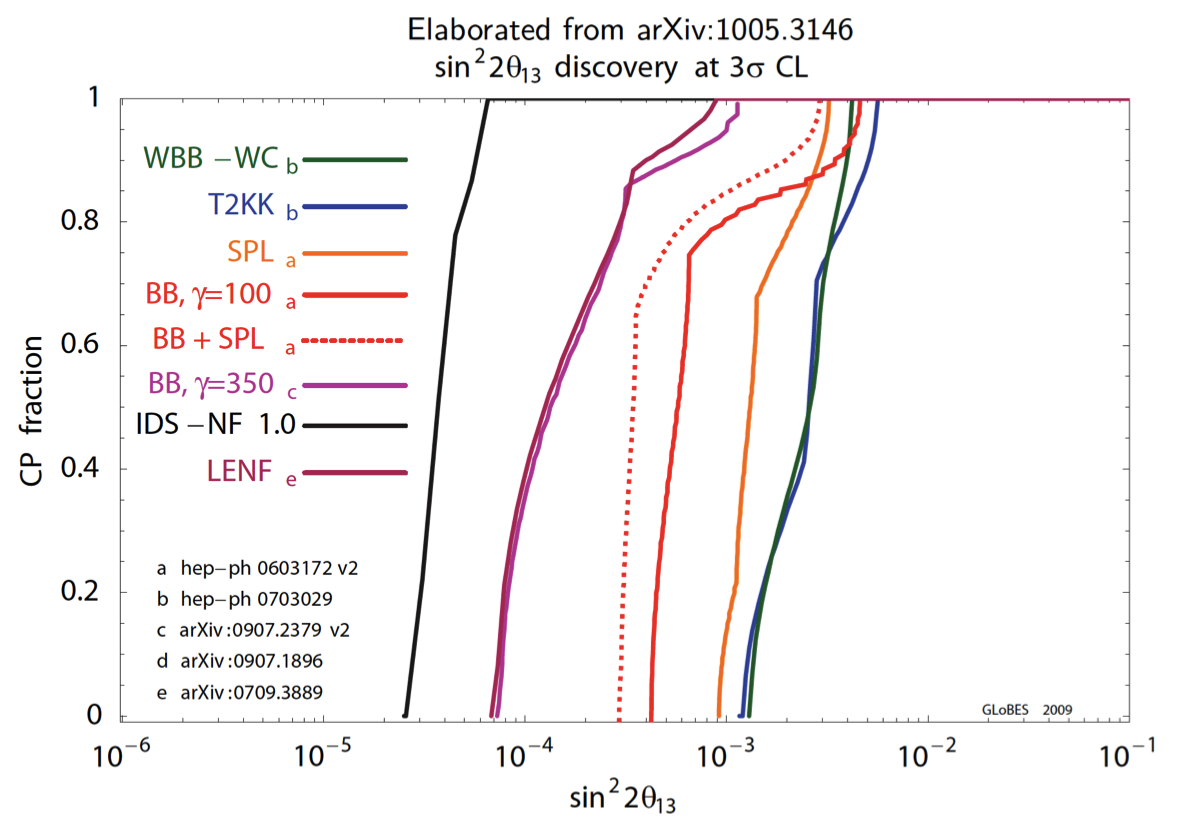}
\caption{Physics reach of different future facilities in
  $\sin^2 2\theta_{13}$ (from Ref.~\cite{mezzetto}).}
\label{fig:mezzetto}
\end{center}
\end{figure}

\section{Direct measurements of neutrino mass}

 The properties of neutrinos and especially their rest mass play an
 important role in cosmology, particle physics and astroparticle
 physics. Neutrino oscillation experiments provide compelling evidence
 that neutrinos are massive but they cannot provide the absolute mass
 value.

There are two complementary approaches for measuring the
neutrino mass in laboratory experiments: one is the precise
spectroscopy of beta decay at its kinematic endpoint, and the other
is the search for neutrinoless double-beta decay ($0\nu\beta\beta$).

The $0\nu\beta\beta$ process requires the neutrino to be a Majorana
particle and the effective Majorana mass $m_{\beta\beta}$ can be
determined as
\begin{equation}
m_{\beta\beta} = \left|\sum_i U_{\rme i}^2  m_{i}\right| .
\label{eq:mbb}
\end{equation}
This is the coherent sum of all mass eigenstates $m_i$ with
respect to the PMNS mixing matrix $U_{\rme i}$; $m_{\beta\beta}$ depends
on complex CP phases with the possibility of cancellations. Therefore,
this implies a model-dependent determination of the Majorana mass.

Experiments investigating single-beta decay offer a direct and
model-independent method to determine the absolute neutrino
mass; $m_{\nu_\rme}$ is determined as an incoherent sum of all mass
eigenstates according to the PMNS matrix:
\begin{equation}
m_{\nu_\rme}^2 = \sum_i |U_{\rme i}|^2 m_{i}^2 .
\end{equation}

The experiments looking for $0\nu\beta\beta$ decay have the potential
to probe $m_{\beta\beta}$ in the 20--50~meV region. New single-$\beta$
experiments will increase the sensitivity on $m_{\nu_\rme}$ by one order of
magnitude to 200~meV.

The basic principle applied in the single-beta decay
model-independent method is based on kinematics and energy
conservation. The idea is to measure the spectral shape of
beta decay electrons close to their kinematic endpoint
(Eq.~\eqref{eq:lepton_spect}), where $E_0 - E$ is small and the mass
term $m_i$ becomes significant. A non-zero neutrino mass will not only
shift the endpoint but also change the spectral shape:
\begin{equation}
\frac{\rmd\Lambda_i}{\rmd E} = C p (E + m_\rme) (E_0 - E)
\sqrt{(E_0 - E)^2 - m_i^2} \, F(E, Z) \Theta({E_0 - E - m_i^2}) .
\label{eq:lepton_spect}
\end{equation}
Here $E_0$ is the maximum energy, $F(E, Z)$ is the Fermi function and
$m_i$ is the neutrino mass.

The experimental requirements for doing this measurement are having a
low-endpoint $\beta$ source for a large fraction of electrons in the
endpoint region, high energy resolution and very low
background. Figure~\ref{fig:limits_mnu} shows the evolution of the
experimental bounds on neutrino masses with time.

\begin{figure}[ht]
\begin{center}
\includegraphics[width=8cm]{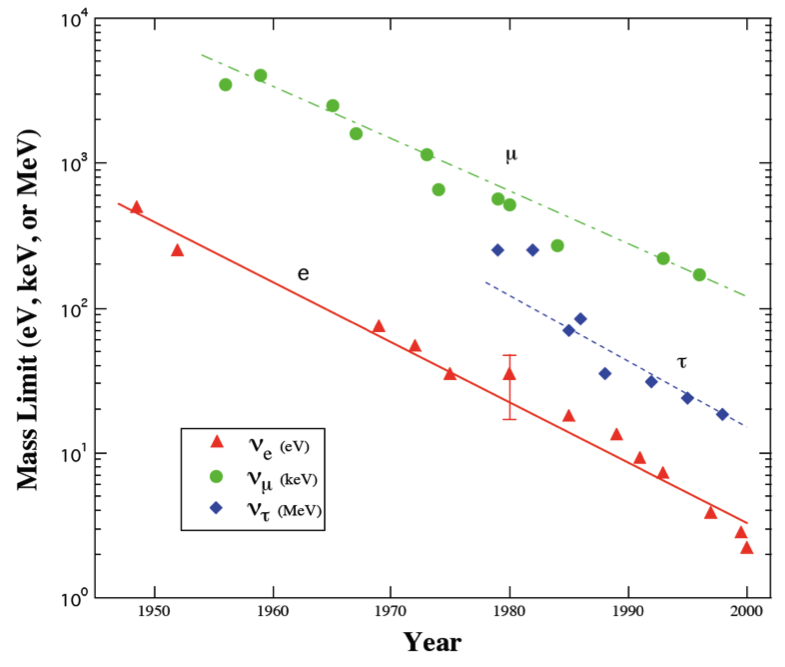}
\caption{Limits on neutrino masses versus year (from
  Ref.~\cite{PDG}).}
\label{fig:limits_mnu}
\end{center}
\end{figure}

At present the best experimental limits from single-beta decay have
been determined by the Mainz and Troitsk experiments~\cite{mainz}
through the tritium beta decay:
\begin{equation}
{}^3{\rm H} \rightarrow {}^3{\rm He} + \rme^- + \bar{\nu}_\rme \qquad\quad  (m_{\nu_\rme} < 2.2~\text{eV at 95\% CL}).
\end{equation}
The direct limits on the other two neutrino masses are much
weaker. The muon neutrino mass limit ($m_{\nu_\mu} < 170$~keV) has
been determined from the endpoint spectrum of the pion decay $\pi^+
\rightarrow \mu^+ \nu_\mu$. The tau neutrino mass ($m_{\nu_\tau} <
18.2$~MeV) has been measured using the tau hadron decay $\tau
\rightarrow 5 \pi \nu_\tau$.

There are two complementary experimental approaches (calorimetry and
spectroscopy) for measuring the neutrino mass from single-$\beta$
decays with different systematics. In the calorimeter approach, the
source is identical to the detector. The best choice for the source is
$^{187}$Re crystal bolometers and the entire beta decay energy is
measured as a differential energy spectrum. $^{187}$Re has the lowest
endpoint (2.47~keV) but, due to its rather long half-life ($4.3 \times
10^{10}$~yr), the activity is rather low. Since bolometers are
modular, their number can be scaled in order to increase the
sensitivity. This approach is being followed in the MARE experiment
\cite{mare}. In the spectrometer approach, an external tritium source is
used. Electrons are magnetically or electrostatically selected and
transported to the counter. The kinetic energy of the beta electrons
is analysed as an integral spectrum by an electrostatic spectrometer. The
material is a high-purity molecular tritium source with a low endpoint at
18.6~keV and a short half-life providing high activity. This approach
has reached its ultimate size and precision in the KATRIN
experiment~\cite{katrin}.

The KATRIN set-up (Fig.~\ref{fig:katrin}) \ced{extends} over 70~m. KATRIN
uses a molecular gaseous tritium source. Electrons emitted by the
T$_2$ decay are guided by strong magnetic fields (3.6~T in the source
and 5.6~T in the transport section) to the transport section and
finally to the spectrometer section. The gas flow is retained by 14
orders of magnitude by active and cryogenic pumping. The
pre-spectrometer can be used to transmit only electrons with energies
close to the T$_2$ endpoint. Only electrons of the endpoint region
would enter the main spectrometer for precise energy analysis. The
low-energy part of the spectrum is filtered. Then, when electrons
enter the spectrometer, the magnetic field drops by several orders of
magnitude. Only electrons able to cross the potential in the
spectrometer are counted. The main spectrometer offers a resolution of
0.93~eV for 18.6~keV electrons by applying a magnetic field ratio of
1/20\,000. The selected electrons are counted in a final detector
(Si PIN diodes with energy resolution of 1~keV). The main inconvenience
here is that the source is external and results suffer from many
systematic uncertainties, since the final energy of the electrons needs to
be corrected for the energy lost in the different steps.

\begin{figure}[ht]
\begin{center}
\includegraphics[width=10cm]{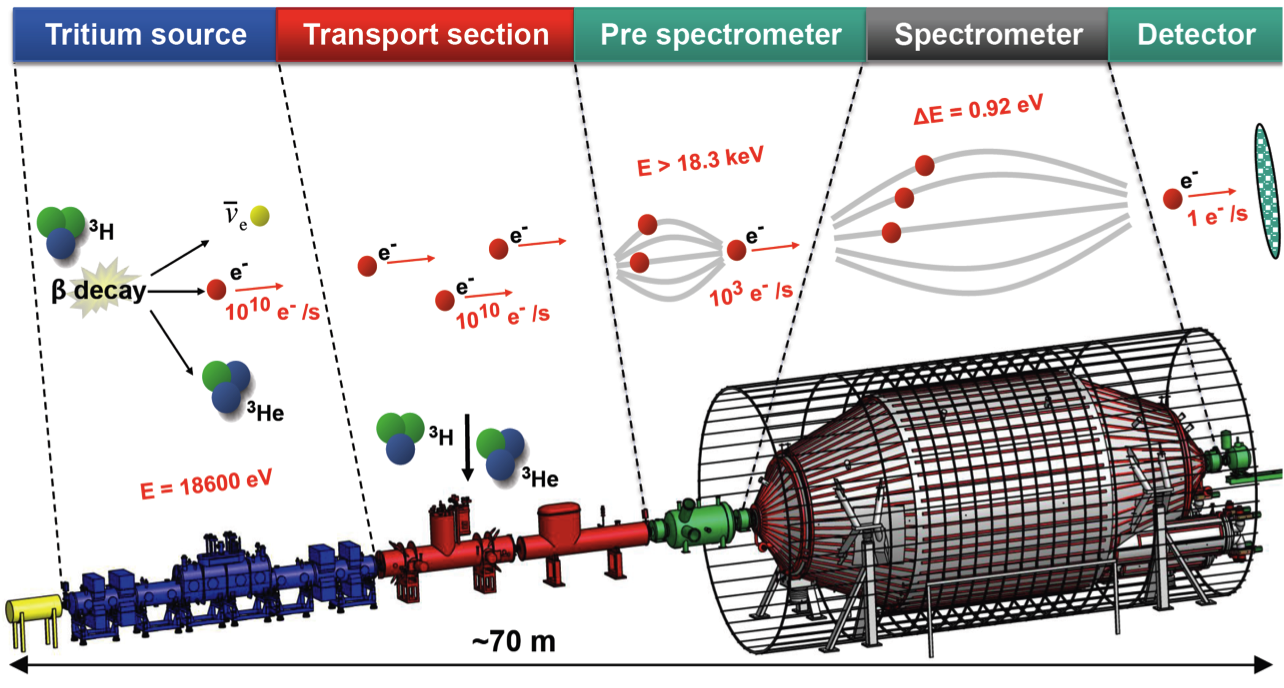}
\caption{Set-up of the KATRIN experiment.}
\label{fig:katrin}
\end{center}
\end{figure}

The main spectrometer is going to follow a test programme in 2011 and
they plan to have the system integrated for late 2012. After three
years of data taking, they plan to arrive at a sensitivity for the
neutrino mass $< 0.2$~eV at 90\% CL or they could be able to detect a
neutrino mass up to 0.35~eV at 5$\sigma$ significance.

The MARE experiment wants to make a direct and calorimetric
measurement of the $\nue$ mass with sub-eV sensitivity. They plan to use
$^{187}$Re (or $^{163}$Ho) as beta emitter and they will measure all
the energy released in the decay except the $\nue$ energy. The
systematic uncertainties from the external electron source are
eliminated. On the other hand, because they detect all the decays
occurring over the entire beta decay spectrum, the source activity
must be limited to avoid pulse pile-up at the endpoint. Thus, the
statistics at the endpoint will be limited. They use thermal
microcalorimeters whose absorbers contain the beta decay isotope with
a low $Q$-value ($\sim$2.5~keV). They plan to improve the energy resolution
to 1--2~eV. The MARE project is subdivided into two phases: MARE~I is an
R\&D phase focused on the choice of the best isotope and the best detector
technology for the final experiment. They will use 300 bolometers and
plan to take data for three years to investigate masses between 2 and 4~eV.
MARE~II will be the final large scale of the detector with sub-eV
sensitivity (improving the mass sensitivity by one order of magnitude)
and investigate the KATRIN region. They plan to use 50\,000 bolometers
and five years of data taking.

New ideas have recently come up to measure the neutrino mass. Project~8
\cite{project8} aims to make use of radio-frequency techniques to
measure the kinetic energy of electrons from a gaseous tritium
source. When a relativistic electron moves in a uniform magnetic
field, cyclotron radiation is emitted. The characteristic frequency is
inversely proportional to the energy of the electron. An array of
antennas would capture the cyclotron radiation emitted by the
electrons when moving and, by measuring the frequency, the energy of the
electron could be obtained. The authors claim that they can obtain
sensitivities of 0.1~eV. They are now preparing a proof-of-principle
experiment to show the feasibility of detecting electrons and determining
their kinetic energy.

\section{Neutrinoless double-beta decay}

Direct information on neutrino masses can also be obtained from
neutrinoless double-beta decay ($0\nu\beta\beta$) searches. This
process violates the total lepton number and requires Majorana
neutrinos. Therefore, the detection of such a process would prove that
neutrinos are their own antiparticles.

The double-beta ($\beta\beta$) decay process is allowed when
single-beta decay is energetically forbidden or strongly
suppressed. The double-beta decay is characterized by a nuclear
process that changes the charge $Z$ in two units while leaving the total
mass $A$ unchanged:
\begin{equation}
(A, Z) \rightarrow (A, Z+2) + 2 \rme^- + 2 \bar{\nu}_\rme .
\end{equation}
For this, it is necessary that the mass $m(A,Z) > m(A,
Z+2)$. This condition is fulfilled in several nuclei, with lifetimes
between $10^{18}$ and $10^{21}$ years.

In the $0\nu\beta\beta$ decay process, only two electrons are
emitted:
\begin{equation}
(A, Z) \rightarrow (A, Z+2) + 2 \rme^- .
\end{equation}
The process can be mediated by the exchange of a light
Majorana neutrino or other particles. The existence of
$0\nu\beta\beta$ decay requires Majorana neutrino mass, no matter what
the actual mechanism is and a violation of the total lepton number
conservation.  A limit on the half-life of this process implies a limit
on the effective Majorana neutrino mass.

In the case of $\beta\beta$ decay, we should observe a continuous
energy spectrum corresponding to the two electrons up to the endpoint
of the decay (Fig.~\ref{fig:spect_0nbb}). In the case of 0$\nu\beta\beta$ decay, we should only
see a line coming from the two electron energies since no neutrinos
are carrying away part of the energy of the process. In that sense, to
observe and be sensitive to $0\nu\beta\beta$, we need good energy
resolution to separate this line from the possible background
(including the possible $\beta\beta$ decay up to the $Q$-value).

\begin{figure}[ht]
\begin{center}
\includegraphics[width=8cm]{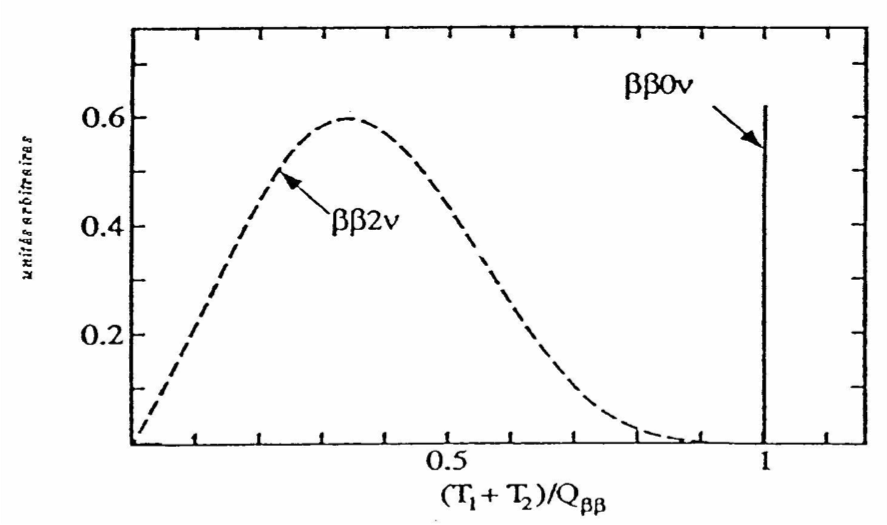}
\caption{Energy spectrum of $\beta\beta$ and $0\nu\beta\beta$ processes.}
\label{fig:spect_0nbb}
\end{center}
\end{figure}

The inverse half-life ($T^{-1}_{1/2}$) of the neutrinoless double-beta decay rate is
proportional to the square of the effective Majorana mass and also
depends on the phase space factor ($G^{0\nu}$) and the nuclear matrix
elements ($M^{0\nu}$), which are difficult to evaluate.
While the phase space can be calculated reliably, the
computation of the nuclear matrix is subject to uncertainty. This
would give a factor $\sim$3 uncertainty in the derived $m_{\beta\beta}$
values:
\begin{eqnarray}
T^{-1}_{1/2} \simeq G^{0\nu} |M^{0\nu}|^2 \langle m_{\beta\beta} \rangle^2 .
\end{eqnarray}

The effective neutrino mass $m_{\beta\beta}$ depends directly on the
assumed form of lepton number-violating interactions. The simplest one
is a light Majorana neutrino exchange. Assuming this, the effective Majorana
neutrino mass can be written as the sum of the mass eigenvalues
multiplied by the mixing matrix elements and the CP phases:
\begin{equation}
m_{\beta\beta} = | m_1 c_{12}^2 c_{13}^2 + m_2 s_{12}^2 c_{13}^2 \, \rme^{\rmi \alpha_1} +  m_3 s_{13}^2 \, \rme^{\rmi \alpha_2} | .
\end{equation}


The individual neutrino masses can be expressed in terms of the
smallest neutrino mass and the mass-squared differences.
For the normal mass hierarchy (NH),
\begin{equation}
m_3 \simeq \sqrt{\Delta m_{\rm atm}^2} \gg m_2 \simeq \sqrt{\Delta m_{\rm sun}^2} \gg m_1 ,
\end{equation}
the effective mass is
\begin{equation}
\left<m_{\beta\beta}\right>^{\rm NH} \simeq \big| (m_1 c_{12}^2 +
  \sqrt{\Delta m_{\rm sun}^2} \, s_{12}^2 \,\rme^{\rmi \alpha_1}) c_{13}^2 +
  \sqrt{\Delta m_{\rm atm}^2} \, s_{13}^2 \,\rme^{\rmi \alpha_2} \big|
\end{equation}
For the inverted mass hierarchy (IH), the smallest neutrino mass is $m_3$,
\begin{equation}
m_2 \simeq m_1 \simeq \sqrt{\Delta m_{\rm atm}^2} \gg m_3 ,
\end{equation}
and the effective mass can be written as
\begin{equation}
\left<m_{\beta\beta}\right>^{\rm IH} \approx \sqrt{\Delta m_{\rm atm}^2} \,
c_{13}^2 \, \big| c_{12}^2 + s_{12}^2 \, \rme^{\rmi \alpha_1} \big| .
\end{equation}
In the quasi-degenerate case (QD),
\begin{equation}
m_0^2 \equiv m_1^2 \simeq m_2^2 \simeq m_3^2 \gg \Delta m_{\rm atm}^2 ,
\end{equation}
the effective mass is
\begin{equation}
\left<m_{\beta\beta}\right>^{\rm QD} \approx m_0 \big| (c_{12}^2 +
  s_{12}^2 \,\rme^{\rmi \alpha_1}) c_{13}^2 + s_{13}^2 \,\rme^{\rmi \alpha_2} \big| .
\end{equation}

Given our present knowledge of the neutrino
oscillation parameters, one can derive the relation between the
effective Majorana mass and the mass of the lightest neutrino, as
shown in Fig.~\ref{fig:mbb_plot}.

\begin{figure}[ht]
\begin{center}
\includegraphics[width=8cm]{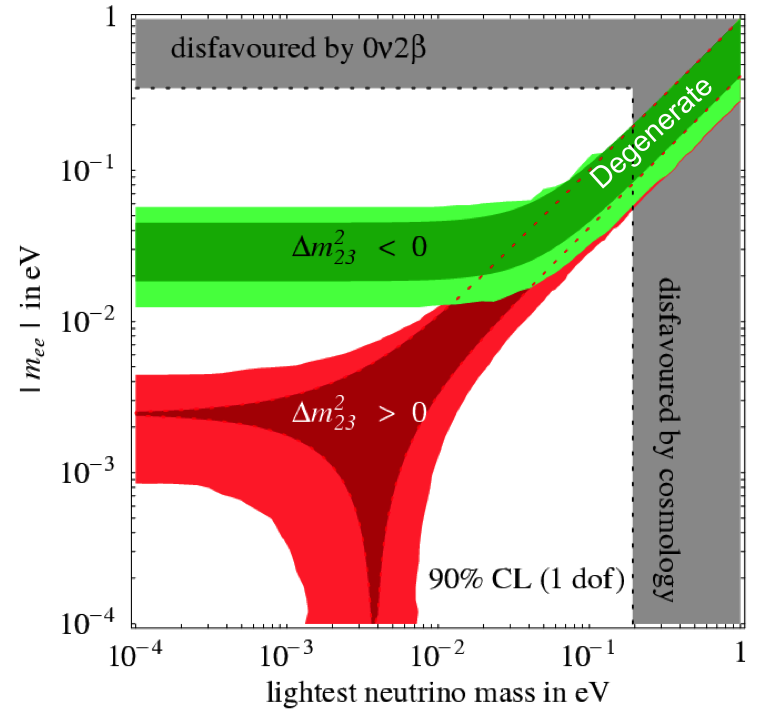}
\caption{Effective Majorana neutrino mass as a function of the
  smallest neutrino mass.}
\label{fig:mbb_plot}
\end{center}
\end{figure}

In principle, a determination of the Majorana mass would allow us to
distinguish between these regions. The three different mass
hierarchies allowed by the oscillation data result in different
projections. The width of the innermost dark bands reflects the
uncertainty introduced by the unknown Majorana phases. Because of the
overlap of the different mass scenarios, a measurement of
$m_{\beta\beta}$ in the degenerate or hierarchical ranges would not
determine the hierarchy. Naturally, if $m_{\beta\beta} < 0.01$~eV,
normal hierarchy becomes the only possible scenario.

The most sensitive double-beta experiments, Heidelberg--Moscow HM-1 and
IGEX, have used $^{76}$Ge as source and detector, and reach
sensitivities around 0.3~eV in the effective neutrino mass. Both
collaborations have reported almost the same upper limit on the
half-life of $1.6 \times 10^{25}$~yr, corresponding to a mass range
of 0.33 to 1.3~eV~\cite{hm_igex}.

However, part of the Heidelberg--Moscow collaboration claimed in 2001
the observation of the $0\nu\beta\beta$ process~\cite{klapdor1} with five enriched
high-purity $^{76}$Ge detectors (10.96~kg of active volume). New
results were presented in 2004 with collected statistics of 71.7~kg~yr~\cite{klapdor2}. The background achieved in the energy region of $0\nu\beta\beta$
decay is very low (0.11~events/kg~yr~keV). The confidence level for the
neutrinoless signal was improved to 4.2$\sigma$ with a
$T_{1/2} = 0.69\mbox{--}4.18 \times 10^{25}$~yr corresponding
to $\left< m_{\beta\beta} \right> = 0.24\mbox{--}0.58$~eV. This would imply
a degenerate neutrino mass hierarchy.

This result has been much criticized and remains controversial (in
contradiction with HM-1 and IGEX experiments, only part of the
collaboration agrees with the result, not all the background peaks are
explained) and needs to be confirmed or refuted by other experiments.

The latest reanalysis of data from 1990 to 2003 shows a 6$\sigma$ excess
of counts at the decay energy, which corresponds to a Majorana neutrino
mass of $0.32 \pm 0.03$~eV at 68\% CL (Fig.~\ref{fig:klapdor}).

\begin{figure}[ht]
\begin{center}
\includegraphics[width=8cm]{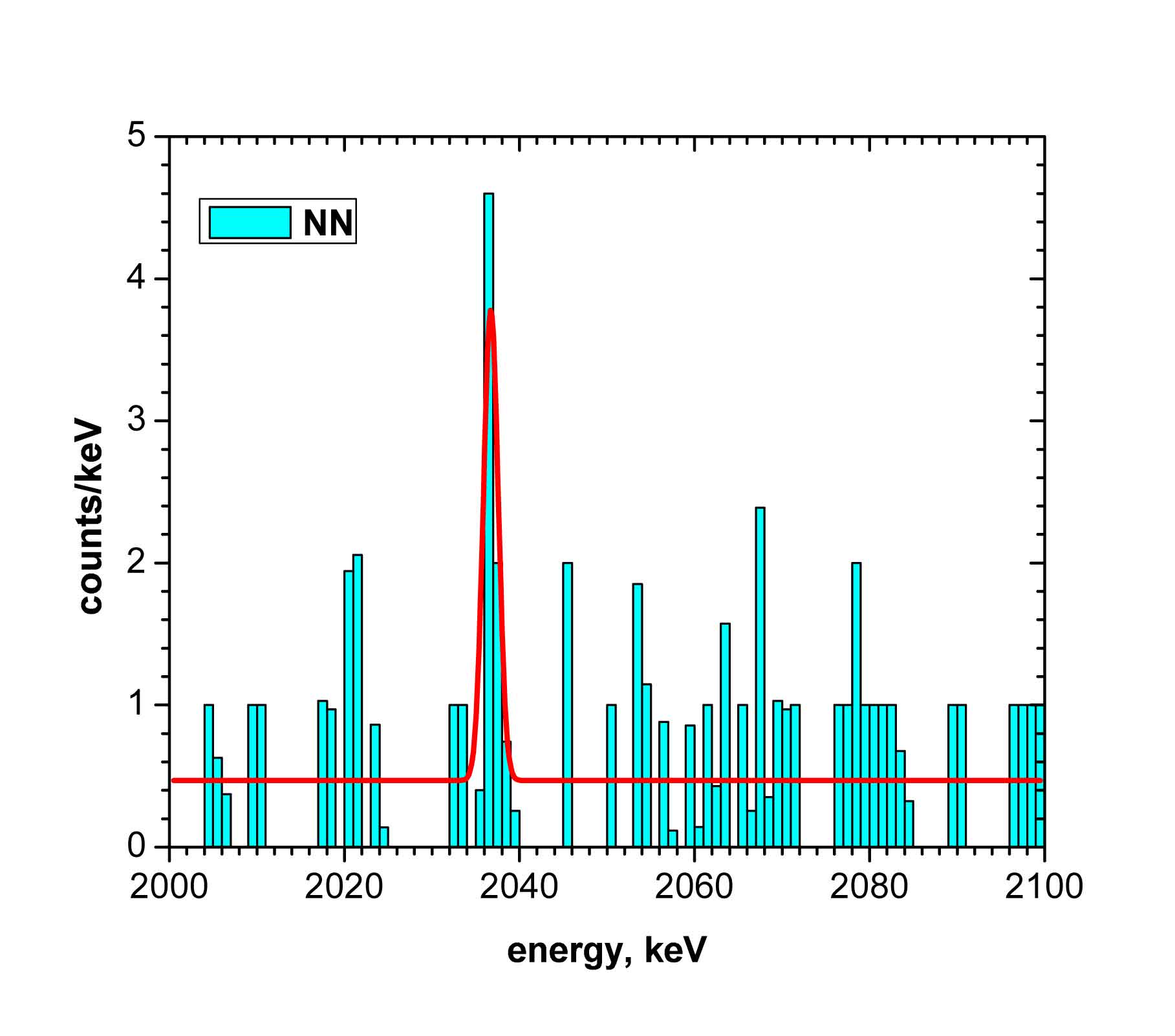}
\caption{The claim by the Heidelberg--Moscow experiment to have
  observed the $0\nu\beta\beta$ process at $>6\sigma$ (from
  Ref.~\cite{klapdor3}).}
\label{fig:klapdor}
\end{center}
\end{figure}

\subsection{Experimental detection}

Neutrinoless double-beta decay is a very rare process.
The half-life sensitivity of this process depends on whether
there is background or not. The sensitivity (without background) is
proportional to the exposure (mass $M$ $\times$ time of measurement
$t$) and the isotopic abundance $a$; with background, it is inversely
proportional to the background rate $B$ and the energy resolution
$\Delta E$:
\begin{equation}
\begin{array}{rcl@{\qquad\quad}l}
T_{1/2} & \propto & a M \epsilon \, t & \text{(background free),} \\
T_{1/2} & \propto & a \epsilon \sqrt{\displaystyle\frac{M t}{\Delta E \, B}} & \text{(background limited).}
\end{array}
\end{equation}
Therefore, the basic experimental requirements for detecting this
process are a large and highly efficient source mass, excellent energy
resolution and an extremely low background in the $0\nu\beta\beta$
peak region.

The neutrinoless double-beta decay experiments can be classified into two
types, depending on whether or not the source is \ced{the same as} the detector. The first experimental approach is calorimetric detectors, where the source
is the detector. They have good energy resolution and good scaling-up,
but modest background discrimination. Thus, strong requirements on
radiopurity and shielding are needed. The semiconductors, cryogenic
bolometers, scintillators and liquid and gaseous Xe TPCs are in this
category. The second approach involves detectors where the source is different
from the detector. This is the case for the \ced{combined tracking and calorimetry (tracko-calo)}\aq{Expanded on first occurrence. OK?} experiments,
where foils of $\beta\beta$ source are surrounded by a tracking
detector that provides direct detection of the two electron tracks
emitted in the decay. They have a moderate energy resolution and are
difficult to scale-up. However, they can provide information on the
event topology.

The main goals of the future $0\nu\beta\beta$ experiments will be to
reach sensitivities of the order of $\left< m_{\beta\beta} \right>
\sim 0.01\mbox{--}0.1$~eV (IH mass region) using different isotopes and
different experimental techniques. Table \ref{tab:0nbb_exp} shows a
summary of the \ced{forthcoming} $0\nu\beta\beta$ experiments.

\begin{table}[h]
\begin{center}
\caption{Overview of upcoming 0$\nu\beta\beta$ experiments.}
\label{tab:0nbb_exp}
\begin{tabular}{cccccc}
\hline\hline
\textbf{Experiment} & \textbf{Isotope} & \textbf{Mass (kg)} &
\textbf{Technique} & \textbf{Sensit. $T^{0\nu}_{1/2}$ (yr)} & \textbf{Status}  \\
\hline
GERDA & $^{76}$Ge & 40 & ionization & 2 $\times$ $10^{26}$ & in
progress \\
Majorana & $^{76}$Ge & 30 & ionization & 1 $\times$ $10^{26}$ & in
progress \\
COBRA & $^{116}$Cd, $^{130}$Te & t.b.d. & ionization & t.b.d. & R\&D \\
CUORE & $^{130}$Te & 200 & bolometers & 6.5 $\times$ $10^{26}$ & in
progress \\
EXO & $^{136}$Xe & 200 & liquid TPC & 6.4 $\times$ $10^{25}$ & in progress \\
NEXT & $^{136}$Xe & 100 & gas TPC & 1.8 $\times$ $10^{26}$ & in
progress \\
SNO+ & $^{150}$Nd & liquid scintillator & 56 & 4.5 $\times$ $10^{24}$
& in progress\\
KamLAND-Zen & $^{136}$Xe & liquid scintillator & 400 & 4 $\times$
$10^{26}$ & in progress \\
SuperNEMO & $^{82}$Se, $^{150}$Nd & 100 & tracko-calo & 1--2 $\times$
$10^{26}$ & in progress \\
\hline\hline
\end{tabular}
\end{center}
\end{table}

The GERDA and Majorana experiments will search for $0\nu\beta\beta$ in
$^{76}$Ge using arrays of high-purity germanium detectors. This is a
well-established technique offering outstanding energy resolution
(better than 0.2\% full width at half-maximum (FWHM) at the $Q_{\beta\beta}$ value) and high efficiency ($\sim$80\%) but limited methods to reject backgrounds.

The GERDA detector~\cite{gerda} is made of an 86\% enriched pure naked $^{76}$Ge
crystal array immersed in LAr and surrounded by 10~cm of lead and 2~m
of water. In phase~I, they will operate the refurbished HM and
IGEX enriched detectors ($\sim$18~kg). They will verify or reject the
Heidelberg--Moscow claim with the same detectors. They expect a
background rate of the order of 0.01 counts/keV~kg~yr. In phase~II,
they will add 20~kg of segmented detectors to arrive at a background
level of $\sim$0.001 counts/keV~kg~yr. Depending on the outcome,
there could be a phase~III, merging GERDA and Majorana detectors, to
reach a mass of the order of 1~ton and test the IH mass region.

The Majorana experiment~\cite{majorana} is located in Sanford Laboratory and will be
composed of 30~kg of enriched $^{76}$Ge crystals with a passive Cu and
Pb shielding providing a low background. They anticipate a background
rate of 0.001 counts/keV~kg~yr.

 The COBRA experiment~\cite{cobra} aims to search for the $0\nu\beta\beta$ decay of
 $^{116}$Cd and $^{130}$Te with CdZnTe semiconductors. It is currently
 in the R\&D phase and they have a test set-up working at the Gran Sasso
 laboratory. The idea is to use an array of CdZnTe room-temperature
 semiconductors. The exploration of pixellated detectors will add
 tracking capabilities to the pure energy measurements and even
 further background reduction by particle identification. A scientific
 proposal is foreseen by the end of 2012.

CUORICINO was an experiment at the Gran Sasso laboratory working from 2003 to
2008. It was composed by cryogenic bolometers of TeO$_2$ crystals. The
$0\nu\beta\beta$ decay was not observed and the experiment has been
able to set the world's most stringent lower limit for the half-life
for  $0\nu\beta\beta$ in $^{130}$Te, namely, $T_{1/2} \ge 2.8 \times
10^{24}$~yr at 90\% CL~\cite{cuoricino}.

The CUORE detector~\cite{cuore} will consist of an array of 988 TeO$_2$ crystals
that contain 27\% $^{130}$Te as the source of $0\nu\beta\beta$ with
$\sim$200~kg of $^{130}$Te for a total detector mass of about 740~kg. The
crystals will be cooled inside a specially built dilution refrigerator
-- one of the world's largest -- to a temperature of $\sim$10~mK, at which
point they have such a small heat capacity that the energy deposited
by individual particles or gamma rays in a crystal produces a
temporary, measurable rise of its temperature. The measured
temperature pulses will be used to construct an energy spectrum of the
interactions occurring inside the crystals, and the spectrum is then
inspected for a small peak at 2527~keV. The next project goal for
CUORE will be the construction and operation of CUORE-0, the first
52-crystal tower produced by the CUORE detector assembly line. The
CUORE-0 tower will be installed in the existing CUORICINO cryostat,
and it will take data for the next two years while the 19 CUORE towers
are assembled. CUORE-0 is primarily intended to serve as a test of the
CUORE detector assembly protocols and to verify the functionality of
the experimental components, but it will nevertheless represent a
significant measurement: it will be comparable in
size to CUORICINO, yet its energy spectrum will have a lower
background due to improvements in materials and assembly
procedures. The advantages and disadvantages of the technique are
similar to those of germanium experiments, with about the same energy
resolution and efficiency for the signal. The expected sensitivity for
a background of 0.001 counts/keV~kg~yr and $\Delta E = 5$~keV is
$\sim 6.5 \times 10^{26}$~yr.

SNO+~\cite{sno+} proposes to fill the Sudbury Neutrino Observatory (SNO) with
liquid scintillator. A mass of several tens of kilograms of
$\beta\beta$ decaying material can be added to the experiment by
dissolving a neodymium salt in the scintillator. The natural abundance
in the $^{150}$Nd isotope is 5.6\%. Given the liquid scintillator light
yield and photocathode coverage of the experiment, a modest energy
resolution performance (about 6\% FWHM at $Q_{\beta\beta}$) is
expected. This could be compensated by large quantities of isotope
and low backgrounds. They plan to use enriched Nd to increase the mass.

KamLAND-Zen~\cite{kamland-zen} plans to dissolve 400~kg of $^{136}$Xe in the liquid
scintillator of KamLAND in the first phase of the experiment, and up to
1~ton in a projected second phase. Xenon is relatively easy to
dissolve (with a mass fraction of more than 3\% being possible) and
also easy to extract. The major modification to the existing KamLAND
experiment is the construction of an inner, very radiopure and very
transparent balloon to hold the dissolved xenon. The balloon, 1.7~m 
in radius, would be shielded from external backgrounds by a
large, very radiopure liquid scintillator volume. While the energy
resolution at $Q_{\beta\beta}$ (about 10\%) is inferior to that of
SNO+, the detection efficiency is much better (80\%) due to its double
envelope.

The NEMO-3 experiment~\cite{nemo3} combines calorimetry and tracking
techniques. The foils of the source are surrounded by a tracking detector
that provides a direct detection of the two electron tracks emitted in the
decay. NEMO-3 is installed in the Frejus underground laboratory and is
searching for neutrinoless double-beta decay for two main isotopes ($^{100}$Mo and
$^{82}$Se) and studying the two-neutrino double-beta decay of seven
isotopes. The experiment has been taking data since 2003 and, up to
the end of 2009, showed no evidence for neutrinoless double-beta decay.

SuperNEMO~\cite{supernemo} uses the NEMO-3 approach with series of modules, each one
consisting of a tracker and a calorimeter that surround a thin foil of
the isotope. In SuperNEMO the target will likely be $^{82}$Se, although other isotopes such as $^{150}$Nd or $^{48}$Ca are also being considered. The mass of the
target is limited to a few kilograms (typically 5--7~kg) by the need to build
it foil-like, and to minimize multiple scattering and energy loss. The
tracker and calorimeter can record the trajectory of the charged
particles and measure their energies independently. This technique,
which maximally exploits the topological signature of the events,
leads to excellent background rejection. However, the selection efficiency is
relatively low (about 30\%), and the resolution rather modest (4\%
FWHM at $Q_{\beta\beta}$). Moreover, this technique is very hard to extrapolate
to large masses due to the size, complexity and cost of each module.

Another technique used in $0\nu\beta\beta$ experiments is the xenon
time projection chambers. Xenon is a suitable detection medium,
providing both scintillation and ionization signals. It has a decaying
isotope, $^{136}$Xe, with a natural abundance of about 10\%. Compared
to other sources, xenon is easy (thus relatively cheap) to enrich in
the candidate isotope.

When an event occurs, the energetic electrons produced interact with
the liquid xenon (LXe) to create scintillation light that is detected, for example,
with avalanche photodiodes (APDs). The electrons also ionize some of
the xenon and the ionized electrons drift to charge collection wires
at the ends of the vessel in an electric field. The time between the
light pulse and the electrons reaching the wires tell us how far in
the event occurred, since we know the drift time.

There are two possibilities for a xenon TPC: a cryogenic liquid xenon time
projection chamber (LXe TPC), or a (high-pressure) xenon (HPXe) gas chamber.

EXO~\cite{exo} is a LXe TPC with a modest energy resolution (3.3\% FWHM at
$Q_{\beta\beta}$) through ionization and scintillation readout. A 200~kg detector of 80\% enriched $^{136}$Xe is currently being installed at the Waste Isolation Pilot
Plant (WIPP) in New Mexico, USA. This experiment aims to measure the -- as yet
unobserved -- two-neutrino mode of double-beta decay of $^{136}$Xe and
provide a competitive limit on neutrinoless double-beta
decay. Background rates of order 0.001 counts/keV~kg~yr are expected
in EXO-200. The improvement with respect to the high-resolution
calorimeters comes from the event topological information.
The collaboration is undergoing extensive R\&D to develop the xenon
detector and a way to ``tag'' the products of the decay
($^{136}\mathrm{Ba}^{2+}$ tagging) in order to eliminate all backgrounds.

The NEXT experiment~\cite{next} proposes to build a 100~kg high-pressure gaseous
xenon (enriched at 90\% in $^{136}$Xe) TPC. The experiment aims to
take advantage of both good energy resolution ($\leq 1$\% FWHM at
$Q_{\beta\beta}$) and the presence of a $0\nu\beta\beta$ topological
signature for further background suppression. NEXT plans to rely on
electroluminescence to amplify the ionization signal, using two
separate photodetection schemes for an optimal measurement of both
calorimetry and tracking.

Figure~\ref{fig:jj} shows the background rate in the region of interest
(1~FHWM around $Q_{\beta\beta}$) versus the energy resolution (FWHM)
for different past and present experiments~\cite{pau}. The (green) circles correspond to measured data, while the (blue) squares and (red) diamonds correspond, respectively, to the R (reference) and O (optimistic) background assumptions of the
experiments, according to Ref.~\cite{pau}. The results for the $m_{\beta\beta}$ sensitivity (90\% CL) of the
proposals as a function of exposure are also shown. The filled circles
indicate 10 years of run-time according to the reference scenario.

\begin{figure}[ht]
\begin{center}
\includegraphics[width=7.5cm]{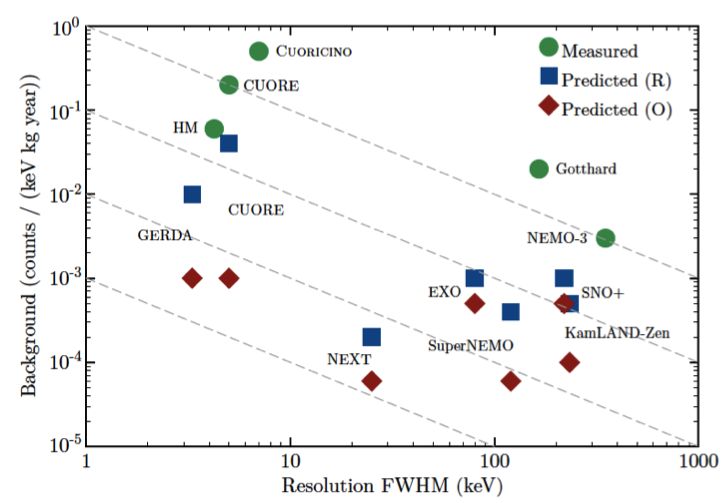}
\includegraphics[width=6cm]{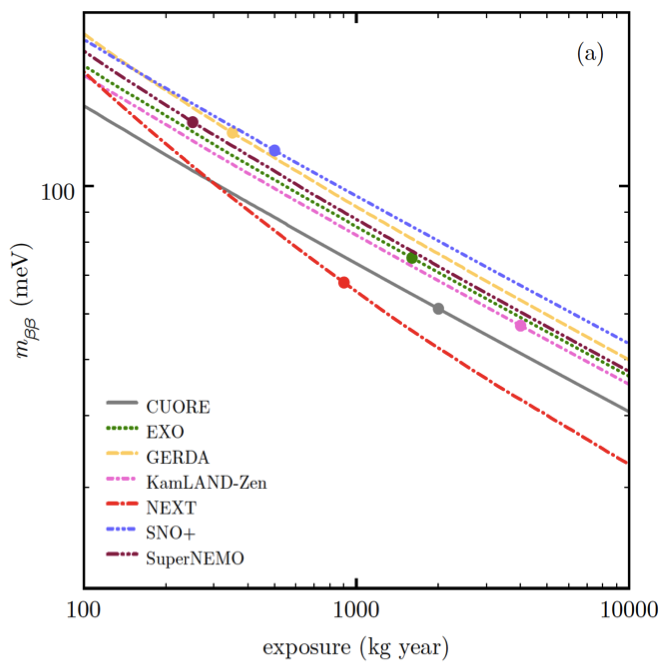}
\caption{(left) Background rate as a function of the energy resolution (FWHM)
for different past and present experiments and (right) $m_{\beta\beta}$
sensitivity (at 90\% CL) as a function of the exposure (from Ref.~\cite{pau}).}
\label{fig:jj}
\end{center}
\end{figure}

NEXT and CUORE have the best sensitivities, reaching 66 and 73~meV at
90\% CL, respectively. KamLAND-Zen, EXO and SuperNEMO follow, with
sensitivities in the 82--87~meV range. GERDA and SNO+ reach
sensitivities of 94 and 96~meV, respectively. In the optimistic
scenario, the lower background regime for all experiments allows 
significantly better sensitivities to be obtained.

In summary, the goals of the next generation of $0\nu\beta\beta$
experiments are to push the $m_{\beta\beta}$ limit down to 100~meV to
confirm or discard the Heidelberg--Moscow claim. In a second and more
ambitious step, they should reach $m_{\beta\beta} \sim 50$~meV
to fully explore the degenerate spectrum. Finally, depending on their
capability to scale their technology to larger masses ($\sim$~ton scale), they will try
to partially explore the inverse hierarchy down to $\sim 10$~meV.

\section{Supernova neutrinos}

Type II supernovae (SNe) are massive stars that begin their lives made out of
hydrogen. Hydrogen starts nuclear fusion in the core, and when all H is
converted into He, the star starts to collapse until He is hot enough
to fuse. Then, He will begin the same process as H. The same happens
for the rest of the elements up to Fe. The Fe fusion reaction absorbs more
energy than it releases, and then the core shrinks, heats up and
produces no new, more massive elements. The star cannot resist the
pressure of its internal gravitational force and then collapses. The
collapse leads to an explosion, that is known as a type~II~SN.

In the core-collapse mechanism, three stages are important from the
point of view of neutrino emission:
\begin{itemize}
\item[(1)]
The collapse of the core: a first electron neutrino burst is emitted
since the high density of matter enhances the electron capture by
protons.
\item[(2)]
Then, neutrinos are trapped and an elastic bounce of the core is
produced, which results in a shock wave. When the shock crosses the
electron neutrino sphere, an intense burst of $\nue$ is produced, called
the {\it shock breakout} or {\it neutronization burst}, and a total
energy of $3 \times 10^{51}$~erg is radiated \ced{in milliseconds}.
\item[(3)]
The process will finish in an explosion. Then, the external layers of
the star are expelled into space. After this, the star loses
energy by emitting neutrinos of all flavours and the {\it cooling
process} ($\sim 10$~s) starts until a neutron star or a black hole is formed.
\end{itemize}
\noindent
The total energy released during this process is enormous: $\approx 3
\times 10^{53}$~erg. Some 99\% of the gravitational binding energy of the star ($E_B$) is released in the form of neutrinos of all flavours: 1\% are produced during the
neutronization process while the rest are $\nu$--$\bar{\nu}$ pairs from
later cooling reactions. The expected supernovae rate in our Galaxy is
about three per century.

In 1987 astrophysics entered a new era with the detection of the
neutrinos from the SN1987A~\cite{sn1987}, which exploded in the Large Magellanic
Cloud at a distance of $\sim 50$~kpc. The burst of light was visible to
the naked eye. Around three hours before the observation of the SN, an increase of
neutrinos was detected by three water Cerenkov detectors: Kamiokande,
IMB and Baksan. This observation confirmed important parts of the
neutrino supernova theory such as total energy, mean temperature and
time duration. However, limited quantitative information on the
neutrino spectrum was obtained due to the small statistics (only
about 20 events) recorded.

The flavour composition, energy spectrum and time structure of the
neutrino burst from a supernova can give information about the
explosion mechanism and the mechanisms of proto-neutron star
cooling. In addition, the intrinsic properties of the neutrino such as
flavour oscillations can also be studied.

The neutrinos in the cooling stage are in equilibrium with their
surrounding matter density and their energy spectra can be described
by a function close to a Fermi--Dirac distribution. The flux of an
emitted neutrino $\nu_{\alpha}$ can then be written as
\cite{Lunardini:2003eh}
\begin{equation}
\phi_{\alpha}(E_{\alpha},  L_{\alpha}, D, T_{\alpha}, \eta_{\alpha}) =
\frac{L_{\alpha}}{4\pi D^2 F_3(\eta_\alpha) T^4_{\alpha}} \,
\frac{E_{\alpha}^2}{\rme^{E_{\alpha}/T_{\alpha}-\eta_{\alpha}}+1} ,
\end{equation}
\noindent
where $L_{\alpha}$ is the luminosity of the flavour $\nu_{\alpha}$ ($E_B
= \sum L_{\alpha}$), $D$ is the distance to the supernova,
$E_{\alpha}$ is the energy of the $\nu_{\alpha}$ neutrino,
$T_{\alpha}$ is the neutrino temperature inside the neutrinosphere and
$\eta_{\alpha}$ is the ``pinching'' factor.

The original $\numu$, $\nutau$, $\anumu$ and $\anutau$ fluxes are
approximately equal and therefore we treat them as $\nux$. An energy
hierarchy between the different neutrino flavours is generally believed
to hold and implies $\langle E_{\nu_\rme} \rangle < \langle E_{\bar{\nu}_\rme}
\rangle < \langle E_{\nu_x} \rangle$. However, the specific neutrino
spectra remain a matter of detailed calculations. In particular,
recent simulations seem to indicate that the energy differences
between flavours could be very small and possible collective neutrino
flavour conversions could arise for either mass hierarchy depending on
the primary fluxes~\cite{Choubey:2010up}.

Neutrino oscillations and matter effects in the supernova will change
the neutrino fluxes significantly and, therefore, the number of events
expected in the detectors.

If the neutrino energy spectra are
different, then $\theta_{13}$ and the mass hierarchy can be
probed. For small mixing angle ($\sin^2 \theta_{13} < 2 \times
10^{-6}$), there are no effects on $\theta_{13}$ and we cannot
distinguish among mass hierarchies. Only an upper bound on
$\sin^2\theta_{13}$ can be set. For intermediate $\theta_{13}$ ($2
\times 10^{-6} < \sin^2\theta_{13} < 3 \times 10^{-4}$),
maximal sensitivity to the angle is achieved and measurements of the
angle are possible in this region. For large mixing angle
($\sin^2\theta_{13} > 3 \times 10^{-4}$), maximal conversions
occur. The mass hierarchy can be probed but only a lower bound on
$\theta_{13}$ can be established.

In addition to matter effects in the SN matter, when neutrinos
traverse the Earth, regeneration effects can produce a distortion of
the neutrino energy spectrum. If we compare the signals from different
detectors in different locations, we could probe such an effect.

\subsection{Supernova neutrino detection in terrestrial experiments}

Most of the current and near-future supernova neutrino experiments
\cite{KateTAUP} are water Cerenkov or liquid scintillator detectors
and, therefore, primarily sensitive to the $\anue$ component of the
signal, via inverse beta decay $\bar{\nu}_\rme + \rmp \rightarrow  \rmn + \rme^+$. For
supernova burst detection, not only statistics but also diversity of
flavour sensitivity is needed: neutral current sensitivity, which gives
access to the $\numu$ and $\nutau$ components of the flux, and
$\nue$ sensitivity are particularly valuable.

Only two near-future experiments will be mainly sensitive to
the $\nue$. The HALO detector~\cite{HALO} is under construction at SNOlab
and it uses 80~tons of lead blocks instrumented with the unused SNO
NCD counters to record neutrons and electromagnetic signals. However,
this technique has some limitations since no energy or pointing
information can be obtained and only rates are provided. The ICARUS
detector at Gran Sasso~\cite{ICARUS} is a 600~ton LAr TPC with
excellent $\nue$ sensitivity via $^{40}$Ar CC interactions, for which
de-excitation gammas will be visible.

All current supernova neutrino experiments participate in the
Supernova Early Warning System (SNEWS)~\cite{SNEWS}, the network of SN
neutrino observatories whose main goal is to provide the astronomical
community with a prompt alert for the next galactic core-collapse
supernova explosion.

Very promising for the future are a number of planned mega-detectors
exploring essentially three technologies: megaton-scale water Cerenkov
detectors, like LBNE in DUSEL~\cite{LBNE}, Hyper-K in Japan~\cite{HK}
and Memphys in Europe~\cite{Memphys}; 100~kton-scale LAr TPC detectors,
like GLACIER in Europe~\cite{Glacier} or LAr LBNE in DUSEL
\cite{LBNE-LAr}; and 50~kton-scale liquid scintillator detectors, like
LENA in Europe~\cite{LENA} or Hanohano in Hawaii~\cite{Hano-Hano}. Some
such detectors can hope to collect individual neutrino events every
few years from beyond the Local Group of galaxies (a few megaparsecs), assuming
that background can be reduced sufficiently.

The LAGUNA~\cite{LAGUNA} project in Europe is studying the
performance of these three technologies for detecting supernova
neutrinos. The three proposed large-volume detector neutrino
observatories can guarantee continuous exposure for several decades, so
that a high statistics supernova neutrino signal could eventually be
observed. The expected numbers of events for GLACIER, LENA and MEMPHYS
are reported in Ref.~\cite{LAGUNA_detect}, including the neutronization
burst rates and diffuse supernova neutrino background.

\subsection{Diffuse supernova neutrino background}

The diffuse supernova neutrino background (DSNB) is the flux of
neutrinos and antineutrinos emitted by all core-collapse supernovae
that have occurred so far in the Universe. It will appear isotropic and 
time-independent in feasible observations. The DSNB has not been detected
yet, but discovery prospects are excellent.

The Super-Kamiokande experiment established an upper limit on the
$\anue$ flux of $\Phi (\anue) < 1.2~\mathrm{cm^{-2}~s^{-1}}$ for neutrino
energies higher than 19.3~MeV~\cite{sk-dsnb}, close to the
predictions.

Figure~\ref{fig:sk-dsnb} shows the energy spectrum of DSNB
candidates. Points are data and the expected total atmospheric
neutrino background is shown by the thick solid line. The largest
allowed DSNB signal is shown by the shaded region added to the
atmospheric background.\aq{What is the boxed ``GADZOOKS!'' at the top of the right-hand panel? Can it be deleted?}

\begin{figure}[ht]
\begin{center}
\includegraphics[width=6.2cm]{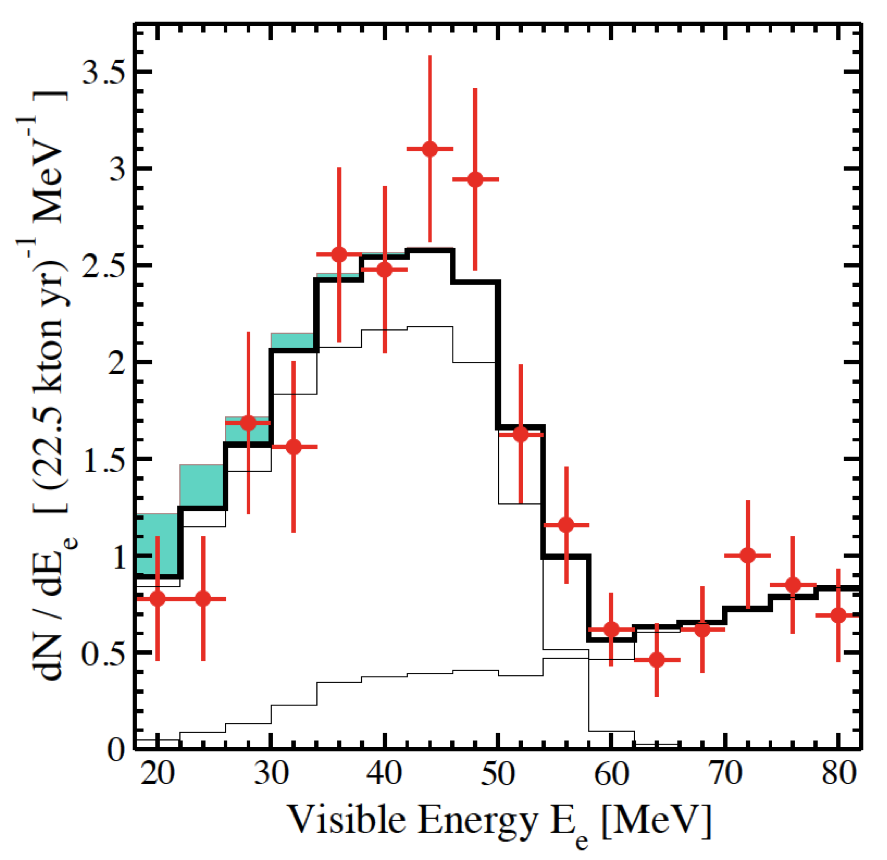}
\includegraphics[width=6.5cm]{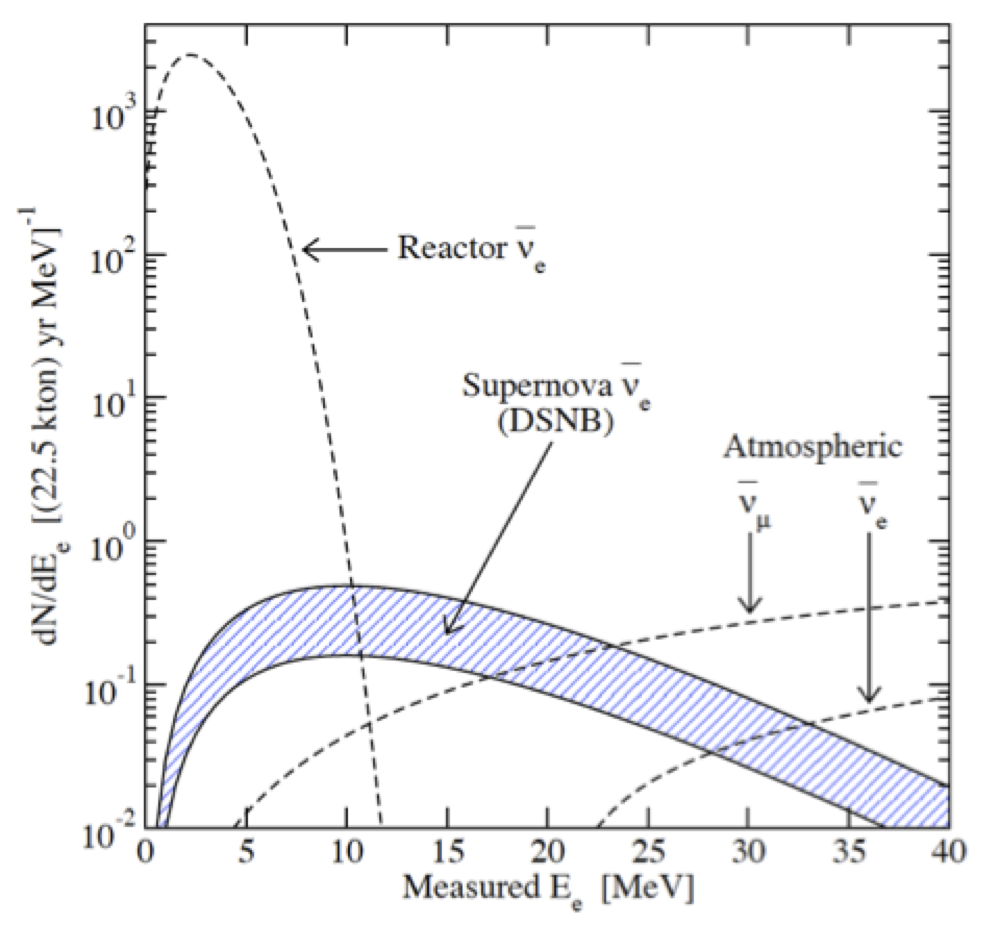}
\caption{(left) Energy spectrum of DSNB candidates measured by SK and
  (right) expected detection rates in SK with dissolved gadolinium (from
  Ref.~\cite{beacom}).}
\label{fig:sk-dsnb}
\end{center}
\end{figure}

If Super-Kamiokande is modified with dissolved gadolinium
to reduce detector backgrounds and increase the energy range for
analysis, then the DSNB could be detected at a rate of a few events
per year~\cite{beacom}.

LAr TPCs would be able to detect mainly the $\nue$ component of the DSNB
signal, providing complementary information with respect to
Super-Kamiokande. The main background sources for these events in the
relevant neutrino energy range of 10--50~MeV are solar and low-energy
atmospheric neutrinos. Depending on the theoretical predictions for
the DSNB flux, a 100~kton LAr detector running for five years would
get more than 4$\sigma$ measurement of the DSNB flux~\cite{cocco}.



\section{Conclusions}

Neutrinos are responsible for one of the most important discoveries in the past
few years in particle and astroparticle physics. Experimental data have
proved that neutrinos oscillate and, therefore, they are massive particles.

Nevertheless, fundamental questions regarding neutrinos remain
unsolved, and present and future neutrino experiments will try to
provide an answer to them. The main goals of such a research programme include
the measurement of the unknown $\theta_{13}$ mixing angle, the sign
of $\Delta m^2_{32}$ (type of mass hierarchy), the determination of
the existence or not of CP-violation in the leptonic sector, the value
of the neutrino masses, and the Majorana or Dirac nature of neutrinos,
among others. New facilities and detectors are being proposed to
answer these questions, using both oscillation and non-oscillation
experiments.

Neutrinos still have surprises for us, and the near future is going to
be very exciting. We will have a better understanding of the neutrino
physics thanks to the experimental programme in the coming years.

\section*{Acknowledgements}

I would like to thank the organizers for inviting me to this great
School and in particular I am very grateful to the discussion leaders
and students for creating a friendly environment and for very
interesting discussions.

\section*{Bibliography}

\noindent
C. Giunti and C.W. Kim, \emph{Fundamentals of Neutrino Physics and
Astrophysics} (Oxford University Press, Oxford, 2007).

\noindent
M.C. Gonzalez-Garcia and M. Maltoni, \emph{Phenomenology
  with massive neutrinos}, 
    {\it Phys.\ Rept.}  {\bf 460} (2008)~1.

\noindent
K. Zuber, \emph{Neutrino Physics}, Series in High Energy Physics, Cosmology
and Gravitation, 2nd edn. (CRC Press, Boca Raton, FL, 2010).


\end{document}